 \documentclass[twocolumn]{IEEEtran}

\usepackage{amsmath,epsfig,amssymb,verbatim,amsopn,cite,subfigure,multirow}
\usepackage{balance}
\usepackage{footnote}
\usepackage{algorithm,algorithmic}
\usepackage[usenames,dvipsnames]{color}
\usepackage[all]{xy}  
\usepackage{url}
\usepackage{adjustbox}
\usepackage{enumitem}
\usepackage{cite,algorithm,algorithmic,amsmath,amssymb,amsthm,empheq,mhsetup}
\usepackage{subfigure,amsfonts,balance}
\usepackage{epstopdf}
\usepackage{setspace}

\usepackage{tikz}
\newcommand{\tikzxmark}{%
\tikz[scale=0.23] {
    \draw[line width=0.7,line cap=round] (0,0) to [bend left=6] (1,1);
    \draw[line width=0.7,line cap=round] (0.2,0.95) to [bend right=3] (0.8,0.05);
}}

\usepackage{pifont}
\newcommand{\cmark}{\checkmark}
\newcommand{\xmark}{\tikzxmark}

\usepackage{array}
\newcolumntype{P}[1]{>{\centering\arraybackslash}p{#1}}
\newcolumntype{M}[1]{>{\centering\arraybackslash}m{#1}}

\newtheorem{Lemma}{Lemma}

\newtheorem{Definition}[Lemma]{Definition}

  {\proof}{\proofend}

\newcounter{mytempeqcounter}

\newcommand{\qa}{{\bf a}}
\newcommand{\qb}{{\bf b}}

\newcommand{\qe}{{\bf e}}
\newcommand{\qf}{{\bf f}}
\newcommand{\qg}{{\bf g}}
\newcommand{\qh}{{\bf h}}

\newcommand{\qn}{{\bf n}}

\newcommand{\qs}{{\bf s}}

\newcommand{\qu}{{\bf u}}
\newcommand{\qv}{{\bf v}}
\newcommand{\qw}{{\bf w}}
\newcommand{\qx}{{\bf x}}
\newcommand{\qy}{{\bf y}}
\newcommand{\qz}{{\bf z}}

\newcommand{\qA}{{\bf A}}
\newcommand{\qB}{{\bf B}}
\newcommand{\qC}{{\bf C}}
\newcommand{\qD}{{\bf D}}

\newcommand{\qF}{{\bf F}}
\newcommand{\qG}{{\bf G}}
\newcommand{\qH}{{\bf H}}
\newcommand{\qI}{{\bf I}}

\newcommand{\qM}{{\bf M}}
\newcommand{\qN}{{\bf N}}

\newcommand{\qP}{{\bf P}}

\newcommand{\qR}{{\bf R}}
\newcommand{\qS}{{\bf S}}

\newcommand{\qV}{{\bf V}}
\newcommand{\qW}{{\bf W}}
\newcommand{\qX}{{\bf X}}
\newcommand{\qY}{{\bf Y}}

\newcommand{\ettall}{\emph{et al.}}

\newcommand{\Sn}{\sigma_n^2}

\newcommand{\UE}{\mathtt{I}}

\newcommand{\FD}{\mathsf{FD}}
\newcommand{\HD}{\mathsf{HD}}

\newcommand{\Ntx}{N}
\newcommand{\Nrx}{N}

\newcommand{\dl}{\mathtt{dl}}
\newcommand{\ul}{\mathtt{ul}}

\newcommand{\gamdmk}{\gamma_{mk}^{\dl}}
\newcommand{\gamdmkp}{\gamma_{mk'}^{\dl}}
\newcommand{\gamuml}{\gamma_{m\ell}^{\ul}}

\newcommand{\vsl}{\varsigma_\ell}

\newcommand{\betamkd}{\beta_{mk}^{\dl}}

\newcommand{\betakldu}{\beta_{k\ell}^{\mathtt{du}}}

\newcommand{\betamlu}{\beta_{m\ell}^{\ul}}

\newcommand{\bsHz}{\text{[bit/s/Hz]}}

\newcommand{\etamkI}{\eta_{mk}^{\mathtt{I}}}

\newcommand{\alphml}{\alpha_{m\ell}}

\newcommand{\betamkue}{\beta_{mk}^{\UE}}
\newcommand{\Bmu}{\boldsymbol{\mu}}

\newcommand{\xik}{x_{\mathtt{I},k}}

\newcommand{\xel}{x_{\mathtt{E},\ell}}

\DeclareMathOperator{\VARSIGMA}{\boldsymbol{\varsigma}}

\newcommand{\vFD}{\mathsf{NAFD}}

\newcommand{\ECB}{\mathtt{ECB}}
\newcommand{\NCB}{\mathtt{NCB}}
\newcommand{\CB}{\mathtt{CB}}

\newcommand{\FZF}{\mathtt{FZF}}

\newcommand{\MR}{\mathtt{MR}}
\newcommand{\PZF}{\mathtt{PZF}}
\newcommand{\CZF}{\mathtt{CZF}}

\newcommand{\PPZF}{\mathtt{PPZF}}
\newcommand{\SE}{\mathtt{SE}}
\newcommand{\LPMMSE}{\mathtt{LP-MMSE}}
\newcommand{\LMMSE}{\mathtt{L-MMSE}}
\newcommand{\CMMSE}{\mathtt{C-MMSE}}

\newcommand{\tauup}{\tau_{u,p}}

\newcommand{\SINRE}{\mathrm{SINR_{E}}}
\newcommand{\SINRone}{\mathrm{SINR}_{1}}
\newcommand{\SINRk}{\mathrm{SINR_{k}}}
\newcommand{\ps}{\mathrm{psc}}

\newcommand{\SINR}{\mathrm{SINR}}

\newcommand{\Zone}{\mathcal{Z}_{1}}

\newcommand{\betalE}{\beta_{\ell E}}

\newcommand{\Mk}{\mathcal{M}_k}
\newcommand{\Nris}{N_{\mathtt{RIS}}}

\newcommand{\SEQoS}{\mathcal{S}_\dl^o}

\newcommand{\betalk}{\beta_{\ell k}}

\newcommand{\betalone}{\beta_{\ell 1}}

\newcommand{\hhatlmk}{\hat {\qh}_{\ell k}}

\newcommand{\hhatlmE}{\hat{\qh}_{\ell E}}

\newcommand{\gamalmk}{\gamma_{\ell k}}

\newcommand{\gamalmE}{\gamma_{\ell E}}

\newcommand{\rholk}{\eta_{\ell k}}

\newcommand{\rholt}{\eta_{\ell t}}

\newcommand{\tausl}{|\mathcal{S}_{\ell}|}

\newcommand{\etamkpI}{\eta_{mk'}^{\mathtt{I}}}

\newcommand{\wimk}{\qw_{\mathrm{I},mk}}
\newcommand{\weml}{\qw_{\mathrm{E},m\ell}}

\DeclareMathOperator{\ETAI}{\boldsymbol{\eta}^{\mathtt{I}}}
\DeclareMathOperator{\ETAE}{\boldsymbol{\eta}^{\mathtt{E}}}
\DeclareMathOperator{\aaa}{\mathbf{a}}

\DeclareMathOperator{\MM}{\mathcal{M}}

\DeclareMathOperator{\K}{\mathcal{K}}

\DeclareMathOperator{\C}{\mathbb{C}}

\newcommand{\etamlE}{\eta_{m\ell}^{\mathtt{E}}}

\newcommand{\Bpsi}{\boldsymbol{\psi}}
\newcommand{\BBPsi}{\boldsymbol{\Psi}}

\newcommand{\gamuemk}{\gamma_{mk}^{\UE}}
\newcommand{\wlk }{\qw_{\ell k}}
\newcommand{\wlkp }{\qw_{\ell k'}}
\newcommand{\mtc }{\mathbb{C}}
\newcommand{\Ex}{\mathbb{E}}

\newcommand{\etalk}{\eta_{\ell k}}
\newcommand{\etalkp}{\eta_{\ell k'}}

\DeclareMathOperator{\THeta}{\boldsymbol{\theta}}

\newcommand{\diag}{\mathrm{diag}}
\newcommand{\trace}{\mathrm{tr}}

\title{Next Generation Multiple Access with Cell-Free Massive MIMO}

\author{Mohammadali Mohammadi,~\IEEEmembership{Senior Member,~IEEE,} Zahra Mobini,~\IEEEmembership{Member,~IEEE,}\\
Hien Quoc Ngo,~\IEEEmembership{Senior Member,~IEEE,} and Michail Matthaiou~\IEEEmembership{Fellow, IEEE}
\thanks{The authors are with the Centre for Wireless Innovation (CWI), Queen's University Belfast, U.K.
email:\{m.mohammadi, zahra.mobini, hien.ngo, m.matthaiou\}@qub.ac.uk.

This work is a contribution by Project REASON, a UK Government funded project under the Future Open Networks Research Challenge (FONRC) sponsored by the Department of Science Innovation and Technology (DSIT). This work was supported in part by the U.K. Engineering and Physical Sciences Research
Council (EPSRC) under Grant EP/X04047X/1. The work of M. Mohamamdi and M. Matthaiou was
supported by the European Research Council
(ERC) under the European Union’s Horizon 2020 research
and innovation programme (grant agreement No. 101001331). The work of Z.~Mobini and  H.~Q.~Ngo
 was supported by the U.K. Research and Innovation Future
Leaders Fellowships under Grant MR/X010635/1, and a research grant from the Department for the Economy Northern Ireland under the US-Ireland R\&D Partnership Programme.
}}

\allowdisplaybreaks

\begin{document}

\bstctlcite{IEEEexample:BSTcontrol}
\maketitle
\thispagestyle{empty}

\begin{abstract}
To meet the unprecedented mobile traffic demands of future wireless networks, a paradigm shift from conventional cellular networks to distributed communication systems is imperative. Cell-free massive multiple-input multiple-output (CF-mMIMO) represents a practical and scalable embodiment of distributed/network MIMO systems. It inherits not only the key benefits of co-located massive MIMO systems but also the macro-diversity gains from distributed systems. This innovative architecture has demonstrated significant potential in enhancing network performance from various perspectives, outperforming co-located mMIMO and conventional small-cell systems. Moreover, CF-mMIMO offers flexibility in integration with emerging wireless technologies such as full-duplex (FD), non-orthogonal transmission schemes, millimeter-wave (mmWave) communications, ultra-reliable low-latency communication (URLLC),  unmanned aerial vehicle (UAV)-aided communication, and  reconfigurable intelligent surfaces (RISs). In this paper, we provide an overview of current research efforts on CF-mMIMO systems and their promising future application scenarios. We then elaborate on new requirements for CF-mMIMO networks in the context of these technological breakthroughs. We also present several current open challenges and outline future research directions aimed at fully realizing the potential of CF mMIMO systems in meeting the evolving demands of future wireless networks.

\end{abstract}


\section*{Main Nomenclature}
\addcontentsline{toc}{section}{Nomenclature}
\begin{IEEEdescription}[\IEEEusemathlabelsep\IEEEsetlabelwidth{$V_1,V_2,V_3,V_4$}]
\fontsize{0.33cm}{0.4cm}\selectfont
\item[$5$G]    Fifth generation
\item[$6$G]    Sixth generation 
\item[ADC]     Analog-to-digital converter 
\item[AI]     Artificial intelligence
\item[AN]      Artificial noise 
\item[AP]      Access point
\item[AWGN]    Additive white Gaussian noise
\item[BS]      Base station

\item[CB]     Conjugate beamforming
\item[CLI]      Cross-link interference
\item[COMP-JT]     Coordinated multi-point with joint transmission 
\item[CPU]     Central processing unit
\item[CSI]     Channel state information
\item[DAC]        Digital-to-analog  converter
\item[DL]      Downlink
\item[DRL]     Deep reinforcement learning
\item[ECB]       Enhanced conjugate beamforming
\item[EE]      Energy efficiency
\item[EH]      Energy harvesting
\item[EU]       Energy user
\item[FD]      Full-duplex
\item[FDD]     Frequency-division duplexing
\item[FZF]     Full-pilot zero-forcing
\item[GP]     Geometric programming
\item[HD]      Half-duplex
\item[IoE]     Internet-of-Everything 
\item[IoT]     Internet-of-Things
\item[IU]       Information user
\item[ISAC]       Integrated sensing and communication 

\item[L-MMSE]   Local minimum mean-squared error
\item[LoS]     Line-of-sight
\item [LSF]   Large-scale fading
\item [LSFD]   Large-scale fading decoding
\item[MIMO]    Multiple-input multiple-output
\item[ML]      Machine learning
\item[mmWave]  Millimeter wave
\item[mMIMO]    Massive multiple-input multiple-output
\item[MMSE]    Minimum mean square error  
\item [MR]    Maximum ratio 
\item [MRC]    Maximum ratio combining
\item [MRT]    Maximum ratio transmission
\item [NAFD]    Network-assisted full-duplex
\item[NCB]     Normalized conjugate beamforming
\item[NOMA]    Non-orthogonal multiple access
\item [OTFS]   Orthogonal time-frequency space modulation 
\item[OMA]     Orthogonal multiple access
\item[PLS]     Physical layer security 
\item[PPZF]     Protective partial zero-forcing
\item[PZF]      Partial zero-forcing 
\item[QoS]     Quality-of-service
\item[RF]      Radio-frequency
\item[RIS]     Reconfigurable intelligent surface
\item[RL]      Reinforcement learning
\item[RSMA]    Rate splitting multiple access 
\item[RZF]     Regularized zero-forcing
\item[SE]      Spectral efficiency
\item[SI]      Self interference
\item[SIC]     Successive interference cancellation
\item[SINR]    Signal-to-interference-plus-noise ratio
\item[SNR]     Signal-to-noise ratio
\item[STAR-RIS] Simultaneous transmitting and reflecting RIS
\item[SWIPT]   Simultaneous wireless information and power transfer
\item[TDD]     Time-division duplexing 
\item[TDMA]     Time-division-multiple-access
\item[UAV]     Unmanned aerial vehicle
\item[UC]      User-centric 
\item[UL]      Uplink
\item[URLLC]   Ultra-reliable low-latency communication
\item[WIPT]    Wireless information and power transmission
\item[WIT]     Wireless information transfer
\item [WMMSE]   Weighted sum-minimum mean square error
\item[WPCN]    Wireless power communication network
\item[WPT]     Wireless power transfer
\item[ZF]      Zero forcing
\end{IEEEdescription}

\section{Introduction}
The emergence of revolutionary applications, such as the Internet-of-Everything (IoE), e-Health, pervasive connectivity in smart environments, augmented/virtual reality, and high-definition video streaming necessitates fundamental changes in wireless communication networks and multiple access techniques from one generation to  the next. More specifically, earlier generations of cellular networks mainly relied on orthogonal multiple access (OMA) techniques, such as time division multiple access (TDMA),  frequency multiple access (FDMA), and orthogonal frequency division multiple access (OFDMA). However, these traditional methods are constrained by the limited radio spectrum and cannot meet the high quality of service (QoS) demands of  next-generation wireless networks with massive device access. Consequently, new access methods that leverage user, space, and frequency domain resources are essential, as time-domain resources are insufficient due to strict latency requirements~\cite{Xiaoming:JSAC:2021}. As a remedy, non-orthogonal multiple access (NOMA)~\cite{Ding:JSAC:2017} and massive multiple-input multiple-output (mMIMO)~\cite{Larsson:CMG:2014} have gained significant attention in recent years as promising approaches for next-generation multiple access~\cite{Xiaoming:JSAC:2021}.  In particular, by utilizing a large number of antennas, mMIMO employs the space division multiple access (SDMA) principle to simultaneously serve multiple users (UEs) on each time-frequency slot, thereby greatly enhancing resource utilization efficiency to support massive access. From an architectural perspective, the concept of ultra-dense networking has   been harnessed in fifth-generation (5G) wireless networks relying on the dense deployment of small cells in high-traffic hotspots and mMIMO technology~\cite{Lu:JSTSP:2014,Matthaiou:COMMag:2021} at macro base stations (BSs)~\cite{Wang:Tut:2023}. 
In cellular 5G mMIMO systems, however, each UE is connected to only one access point (AP)/BS in only one of the many cells. Therefore, these networks suffer from performance degradation for cell-edge UEs, primarily due to reduced channel gain from the serving cell and  strong interference caused by neighboring cells, termed as inter-cell interference~\cite{interdonato2019ubiquitous,demir2021foundations}.

Sixth generation (6G) wireless networks will address the shortcomings of 5G systems by incorporating of innovative technologies. To address the cell-edge issue and inter-cell interference, distributed communication~\cite{Shidong:MCOM:2003} and network MIMO~\cite{Venkatesan:ACSSC:2007} overlaid with the signal co-processing concept, can be leveraged to coherently serve each UE through multiple APs. Network MIMO encompasses various forms, such as distributed antenna systems~\cite{Choi:TWC:2007}, multi-cell MIMO cooperative network~\cite{Gesbert:JSAC:2010}, coordinated beamforming, coordinated multi-point with joint transmission (COMP-JT)~\cite{Irmer:COMG:2011}, and virtual MIMO~\cite{Hong:JSAC:2013}. Nevertheless, such systems are often implemented in a network-centric fashion, wherein the APs are divided into disjoint clusters and serve the UEs residing in their joint coverage area~\cite{interdonato2019ubiquitous}. 

In addition, these systems require precise synchronization, complex signal co-processing, increased deployment cost, and backhaul infrastructure to share  channel state information (CSI) between the APs. Another main drawback is their tendency to utilize network resources less efficiently. In particular, the advantages of network MIMO techniques, whether through coordinated beamforming or CoMP-JT, are particularly pronounced for UEs located near the cell boundaries but less so for those closer to the APs. Accordingly, the widespread deployment of these techniques in 5G networks have been hindered by challenges in achieving high-quality CSI and sufficiently fast synchronization among coordinating APs.

On the other hand, if the co-processing is implemented in a \textit{user-centric (UC)} fashion, i.e., each UE is coherently served by  its selected subset of APs, interference becomes more manageable with less CSI exchange overhead between the APs. The combination of the UC concept with ultra-dense networks and distributed mMIMO concept can be effectively accomplished through a new paradigm of mMIMO systems known as \textit{cell-free mMIMO (CF-mMIMO)}. This approach utilizes the fusion of distributed communication systems~\cite{Shidong:MCOM:2003} and mMIMO, grounded in the concept of signal co-processing. This network architecture essentially consists of a large number of distributed APs connected to one or several central processing units (CPUs), often referred to as the cloud radio access network data centers. The CPUs operate in a network MIMO fashion, without traditional cell boundaries, to provide coordination and computational assistance to the APs. The APs coherently serve UEs over the same time-frequency resources through joint transmission and reception techniques. CF-mMIMO offers several unique advantages, including the following: 
\vspace{-1em}
\begin{itemize}
    \item It can provide stronger macro-diversity gains leading to an increased coverage probability compared to conventional colocated networks. This advantage is achieved through the use of distributed APs~\cite{Hien:cellfree}.
     \item Unlike conventional network MIMO, CF-mMIMO systems utilize an excess of service antennas compared to the number of UEs, which significantly enhances the system's ability to manage interference, particularly the boundary effect, effectively.

    \item Channel estimation and signal processing can be performed locally at each AP by leveraging the channel reciprocity, while there is no need for instantaneous CSI sharing among the APs. 
    \item Performance analysis under various setups is supported by rigorous closed-form expressions for the spectral efficiency (SE), accounting for channel estimation errors and interference from pilot contamination. These results facilitate power control design.  
\end{itemize}

Despite the aforementioned advantages, CF-mMIMO faces some challenges in practice. Specifically, the persistent increase in data throughput demands and network densification raises critical questions about the  development of practical and scalable system architectures. In addition, efficient signal processing, robust and flexible fronthaul/backhaul designs, accurate channel state information acquisition, effective resource allocation such as power control and user scheduling, and synchronization are the major challenges. To cope with these challenges and also to achieve the promised/potential prospects of the 6G wireless networks, including \textit{i)} higher SE for massive connectivity, e.g., for Internet of Things (IoT) devices; \textit{ii)} ultra reliable and low-latency communication (URLLC), e.g., in an industrial environment; \textit{iii)} supporting distributed compute and artificial intelligence (AI)-powered applications in integrated AI and communication networks,  it is essential to incorporate CF-mMIMO into the framework of emerging technologies.  Considering these challenges and opportunities, we present a comprehensive survey of the latest contributions that combine CF-mMIMO with the emerging technologies, emphasizing the interplay and mutual benefits that arise from their integration and co-existence.

\vspace{-1em}
\subsection{Existing Surveys and Tutorials}
In recent years, a few  surveys and tutorials on CF-mMIMO systems have been published~\cite{Elhoushy:Tuts:2022,HeLJCOM:2022,Zhang:Survey:2022,kassam2023review}, covering topics such as signal processing and resource allocation strategies for CF-mMIMO networks, as well as the integration of various 5G communication techniques into CF-mMIMO networks. These techniques include non-orthogonal transmissions, physical layer security (PLS), millimeter-wave (mmWave) communication, reconfigurable intelligent surfaces (RISs), unmanned aerial vehicle (UAV) communications, and AI.

\begin{table*}
	\centering
	\caption{\label{tabel:Survey} Contrasting our contributions to the literature}
	\vspace{-0.6em}
	\small
\begin{tabular}{|p{5.5cm}|p{1.6cm}|p{0.5cm}|p{0.5cm}|p{0.5cm}|p{0.5cm}|p{0.5cm}|p{0.5cm}|}
	\hline
        \textbf{Contributions} 
        & \textbf{This paper}
        &~\cite{Elhoushy:Tuts:2022} 
        &~\cite{HeLJCOM:2022} 
        &~\cite{Zhang:Survey:2022} 
        &~\cite{kassam2023review} 
        &~\cite{Ngo:PROC:2024}
        &~\cite{Zheng:WC:2024}
        \\
        \hline
	\hline

        Signal processing and resource allocation     
        &\centering\cmark 
        &\centering\cmark 
        & \centering\cmark 
        & \centering\cmark    
        & \centering\cmark
        & \centering\cmark    
        & \centering\xmark 
        \cr
        
        \hline
        Network-assisted full duplexing          
        &\centering\cmark 
        &\centering \xmark 
        &\centering \xmark 
        &\centering \xmark
        & \centering\xmark 
        & \centering\xmark    
        & \centering\xmark
        \cr

        \hline
        Non-orthogonal multiple access         
        &\centering\cmark 
        &\centering\cmark  
        & \centering \xmark
        & \centering\cmark
        & \centering\xmark
        & \centering\xmark    
        & \centering\xmark
        \cr

           \hline
        Rate-splitting multiple access         
        &\centering\cmark 
        &\centering\cmark  
        & \centering \xmark
        &\centering \xmark 
        &\centering\xmark 
        & \centering\xmark    
        & \centering\cmark
        \cr
        
        \hline
        Physical layer security                   
        &\centering\cmark 
        &\centering\cmark  
        & \centering \xmark
        &\centering\cmark  
        &\centering\xmark 
        & \centering\xmark    
        & \centering\xmark 
        \cr

         \hline         
         Energy harvesting    
         &\centering\cmark 
         &\centering \xmark 
         &\centering \xmark 
         &\centering \xmark 
        &\centering\xmark 
        & \centering\xmark    
        & \centering\xmark 
        \cr
        
        \hline
        Millimeter wave communication                
        &\centering\cmark 
        &\centering\cmark 
        & \centering \xmark
        &\centering\cmark  
        &\centering\xmark 
        & \centering\xmark    
        & \centering\xmark 
        \cr
        
        \hline
        Reconfigurable intelligent surfaces                   
        &\centering\cmark 
        &\centering\cmark 
        & \centering\cmark
        &\centering \xmark 
        &\centering\cmark
        & \centering\xmark    
        & \centering\xmark 
        \cr
       
        \hline
        Ultra-reliable low-latency communication                
        &\centering\cmark 
        &\centering \xmark 
        &\centering \xmark 
        &\centering \xmark 
        &\centering\xmark 
        & \centering\xmark    
        & \centering\xmark 
        \cr

        \hline
        Unmanned aerial vehicle                   
        &\centering\cmark 
        &\centering\cmark 
        & \centering \xmark
        &\centering \xmark 
        &\centering\cmark
        & \centering\xmark    
        & \centering\xmark 
        \cr

        \hline
        Artificial intelligence and machine learning       
        &\centering\cmark 
        &\centering\cmark 
        &\centering\cmark
        &\centering \xmark        
        &\centering\cmark 
        & \centering\xmark    
        & \centering\xmark 
        \cr
        \hline

        Integrated sensing and communication       
        &\centering\cmark 
        &\centering\cmark 
        &\centering\xmark
        &\centering \xmark        
        &\centering\xmark 
        & \centering\xmark    
        & \centering\cmark 
        \cr
        \hline
     
\end{tabular}
\vspace{-0.5em}
\end{table*}

More specifically, Elhoushy~\ettall~\cite{Elhoushy:Tuts:2022}  reviewed different aspects of CF-mMIMO systems, with a focus on precoding and detection techniques for downlink (DL) and uplink (UL) transmission under various channel fading models. They also emphasized the potential of integrating CF-mMIMO systems with various enabling techniques and technologies for 5G and B5G networks. A review paper by He~\ettall~\cite{HeLJCOM:2022} discussed the enabling physical layer technologies for CF-mMIMO, such as UE association, pilot assignment, transmitter, and receiver design, as well as power control. Zhang \ettall~\cite{Zhang:Survey:2022} presented a comprehensive survey and quantified the advantages of CF-mMIMO systems in terms of their energy- and cost-efficiency. Furthermore, they analyzed the signal processing techniques used to reduce the fronthaul burden for joint channel estimation and transmit precoding. Kassam~\ettall~\cite{kassam2023review} provided an overview of the current state-of-the-art in CF-mMIMO systems, focusing on the challenges related to the limited capacity of the fronthaul links and the connection between APs and UEs. They also discussed prospective cell-free technologies, including RIS, UAV, and AI-enabled CF mMIMO systems. Ngo~\ettall~\cite{Ngo:PROC:2024}  provided a contemporary overview of ultra-dense CF-mMIMO networks and addressed important unresolved questions on their future deployment. Zheng~\ettall~\cite{Zheng:WC:2024} outlined an overview of the mobile CF-mMIMO communication system, focusing on four deployment structures and four application scenarios. The above-mentioned surveys related to CF-mMIMO systems are outlined at a glance in Table~\ref{tabel:Survey}, which allows the readers to capture the main contributions of each of the existing surveys.

\vspace{-0.5em}
\subsection{Paper Contributions and Organization}
This survey paper differs from prior magazines, tutorials, and surveys by elaborating on  fundamental concepts and integration challenges. Specifically, for certain emerging technologies, we conduct an in-depth examination of the potential and use case scenarios for integrating them into CF-mMIMO systems to complement and enhance the network performance. In other instances, we explore the potential of CF-mMIMO for their distributed implementation. We highlight state-of-the-art advancements and research progress in each area and outline compelling directions for future research. By exploring recent CF-mMIMO research discoveries, we create a broader knowledge base which will stimulate further exploration in this rapidly evolving area of research. The contributions of this paper are summarized as follows:

\begin{enumerate}
    \item We discuss the fundamentals of CF-mMIMO networks and provide a general overview of the signal processing requirements, including channel estimation, precoding and decoding designs, resource allocation algorithms, and practical challenges.
    \item We investigate the compatibility of CF-mMIMO with emerging technological breakthroughs. Specifically, we assess the feasibility of full-duplex (FD) transmissions in CF-mMIMO systems. Additionally, we examine the coexistence of CF-mMIMO with non-orthogonal transmissions, including NOMA and rate splitting multiple access (RSMA). Furthermore, we delve into topics such as PLS, wireless energy harvesting (EH), mmWave communication, and RIS. We also provide case studies and outline future research directions for each of these aspects in the CF-mMIMO systems.   
    \item We further articulate the remaining challenges and open issues related to URLLC and UAV-aided communication in CF-mMIMO systems. Finally, we explore the applications of machine learning (ML) for the advancement of CF-mMIMO systems. 

\end{enumerate}

The rest of this paper is outlined  as follows: In Section~\ref{sec:fundamentals}, we overview the principles and discuss the relevant challenges in CF-mMIMO systems. Section~\ref{sec:SignalProcess} provides a comprehensive survey of the signal processing techniques and resource allocation algorithms. Section~\ref{sec:NAFD} integrates FD into CF-mMIMO networks. Sections~\ref{sec:NOMA} elaborates on the interaction of CF-mMIMO with non-orthogonal transmission techniques. Section~\ref{sec:PLS} discusses PLS approaches for CF-mMIMO networks. The state-of-the-art EH CF-mMIMO systems are detailed in Section~\ref{sec:EH}. Section~\ref{sec:mmwave} describes the application of CF-mMIMO in mmWave communications, followed by Section~\ref{sec:RIS}, where we discuss the potential application of RIS for CF-mMIMO networks. Finally, Section~\ref{sec:other} lists potential future research directions, followed by conclusions in Section~\ref{sec:conc}.

\textit{Notation:} We use bold upper case letters to denote matrices, and lower case letters to denote vectors. The superscripts $(\cdot)^*$, $(\cdot)^T$ and $(\cdot)^H$ stand for the conjugate, transpose, and conjugate-transpose (Hermitian), respectively;  
$\qI_M$  represents the $M\times M$ identity matrix;  $\boldsymbol{0}_N$ denotes the all-zero vector of size $N\times 1$, while $\trace(\cdot)$ returns the trace of a matrix. A circular symmetric complex Gaussian distribution having variance $\sigma^2$ is denoted by $\mathcal{CN}(0,\sigma^2)$. Finally, $\Ex\{\cdot\}$ and $\mathtt{Var}(\cdot)$ denote the statistical expectation and variance.

\section{Cell-free Massive MIMO: Fundamentals}~\label{sec:fundamentals}
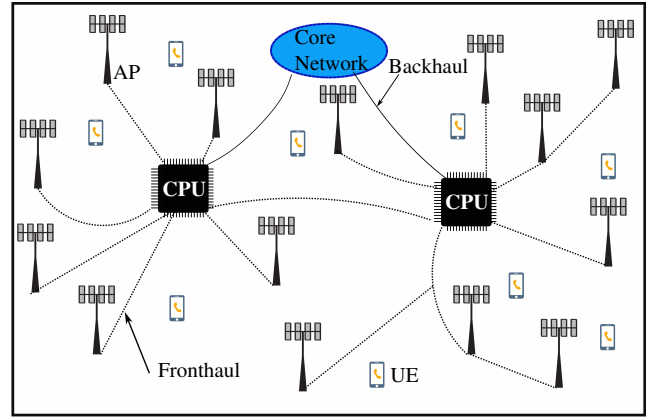
\begin{figure}[!t]\centering \vspace{0em}
    \def\svgwidth{240pt} 
    \fontsize{8}{7}\selectfont 
\begingroup%
  \makeatletter%
  \providecommand\color[2][]{%
    \errmessage{(Inkscape) Color is used for the text in Inkscape, but the package 'color.sty' is not loaded}%
    \renewcommand\color[2][]{}%
  }%
  \providecommand\transparent[1]{%
    \errmessage{(Inkscape) Transparency is used (non-zero) for the text in Inkscape, but the package 'transparent.sty' is not loaded}%
    \renewcommand\transparent[1]{}%
  }%
  \providecommand\rotatebox[2]{#2}%
  \newcommand*\fsize{\dimexpr\f@size pt\relax}%
  \newcommand*\lineheight[1]{\fontsize{\fsize}{#1\fsize}\selectfont}%
  \ifx\svgwidth\undefined%
    \setlength{\unitlength}{1029.97503662bp}%
    \ifx\svgscale\undefined%
      \relax%
    \else%
      \setlength{\unitlength}{\unitlength * \real{\svgscale}}%
    \fi%
  \else%
    \setlength{\unitlength}{\svgwidth}%
  \fi%
  \global\let\svgwidth\undefined%
  \global\let\svgscale\undefined%
  \makeatother%
  \begin{picture}(1,0.652443)%
    \lineheight{1}%
    \setlength\tabcolsep{0pt}%
    \put(0,0){\includegraphics[width=\unitlength]{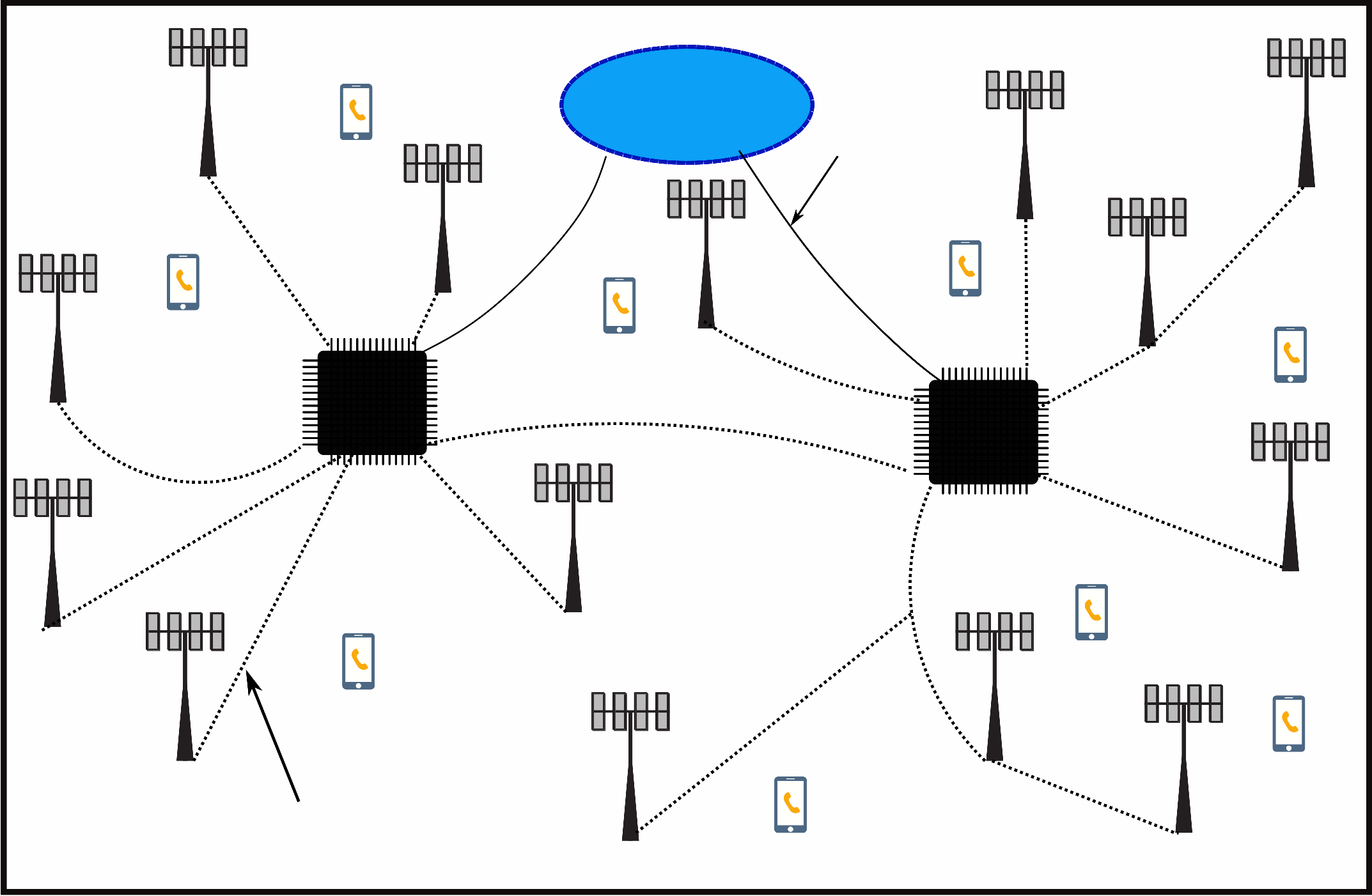}}%
    \put(0.23180379,0.0618499){\color[rgb]{0,0,0}\makebox(0,0)[lt]{\lineheight{1.25}\smash{\begin{tabular}[t]{l}Fronthaul\end{tabular}}}}%
    \put(0.59512449,0.54112633){\color[rgb]{0,0,0}\makebox(0,0)[lt]{\lineheight{1.25}\smash{\begin{tabular}[t]{l}Backhaul\end{tabular}}}}%
    \put(0.59792349,0.05423431){\color[rgb]{0,0,0}\makebox(0,0)[lt]{\lineheight{1.25}\smash{\begin{tabular}[t]{l}UE\end{tabular}}}}%
    \put(0.16095556,0.53198756){\color[rgb]{0,0,0}\makebox(0,0)[lt]{\lineheight{1.25}\smash{\begin{tabular}[t]{l}AP\end{tabular}}}}%
    \put(0.44716026,0.58773012){\color[rgb]{0,0,0}\makebox(0,0)[lt]{\lineheight{1.25}\smash{\begin{tabular}[t]{l}   Core\\Network\end{tabular}}}}%
    \put(0.68431565,0.32688347){\color[rgb]{1,1,1}\makebox(0,0)[lt]{\lineheight{1.25}\smash{\begin{tabular}[t]{l}\textbf{CPU}\end{tabular}}}}%
    \put(0.23874752,0.34823589){\color[rgb]{1,1,1}\makebox(0,0)[lt]{\lineheight{1.25}\smash{\begin{tabular}[t]{l}\textbf{CPU}\end{tabular}}}}%
  \end{picture}%
\endgroup%
 \vspace{0mm}
    \caption{Illustration of a CF-mMIMO network with many
distributed APs connected to CPUs.}\vspace{0mm} \label{fig:CFmMIMO}
\vspace{2em}
\end{figure}

A CF-mMIMO system consists of $M$ geographically distributed N-antenna APs that serve $K$ distributed single-antenna UEs, where $M N\gg K$. 
The channel coefficient vector between AP $\ell$ and UE $k$, $\qh_{\ell k}\in \mtc^{N\times 1}$, is modeled as  ${\qh}_{\ell k}\sim\mathcal{CN}(\boldsymbol{0}, \qR_{\ell,k})$ and $\qR_{\ell,k}\in\mathbb{C}^{N\times N}$ is the spatial correlation matrix; $\beta_{\ell k}=\frac{1}{N}\trace(\qR_{\ell,k})$ accounts for the average channel gain from an antenna at AP $\ell$ to UE $k$. In the case of uncorrelated fading channels, $\qh_{\ell k}=\sqrt{\beta_{\ell k}} \qg_{\ell k}$, where $\beta_{\ell k}$ denotes the large-scale fading (LSF) and $\qg_{\ell k}\sim \mathcal{CN}(\boldsymbol{0}, \qI_N)$ represents the small-scale fading.  
Each AP is connected via a fronthaul link to the CPU, which is responsible for AP cooperation, allowing all APs to communicate with all UEs over the same time-frequency resources and through spatial multiplexing. Specifically, the fronthaul links are used for payload-data, power control coefficient, CSI exchange between the CPU and APs, and phase-synchronization between the APs~\cite{demir2021foundations}. As shown in Fig.~\ref{fig:CFmMIMO}, there can be multiple CPUs, which are connected to the network core via backhaul links.  The CF-mMIMO system could operate either in frequency-division duplex (FDD) mode or in time-division duplex (TDD) mode, which is further discussed in Section~\ref{sec:FDD} and Section~\ref{sec:TDD}, respectively.

\subsection{FDD Cell-free Massive MIMO}~\label{sec:FDD}
In FDD CF-mMIMO systems, UL and DL transmissions occur simultaneously and over different frequency bands. Therefore, the DL and UL channels are not reciprocal and, hence, the APs must acquire the DL channel CSI to perform DL precoding. This can be performed through the DL channel estimation phase followed by a CSI feedback from UEs towards APs. However, the amount of this CSI feedback scales linearly with the number of APs~\cite{Abdallah:TCOM:2020}, which can make FDD inefficient for CF-mMIMO systems. To circumvent this issue, the angular reciprocity feature of FDD systems can be exploited to reduce the required feedback overhead. Angular reciprocity indicates that the angle of departures are the same in both UL and DL, as long as the UL and DL carrier frequencies are not too far from each other, i.e., less than several GHz~\cite{Xie:TVT:2017}. Therefore,  we only need to feed back the estimate of the path gain information to the APs. When the propagation channels are sufficiently sparse, the feedback overhead is considerably reduced~\cite{Kim:TWC:2020}. 

\subsection{TDD Cell-free Massive MIMO}~\label{sec:TDD}
When CF-mMIMO system operates under the TDD mode, the DL and UL data transmission occur over the same frequency band. The time-frequency resources are structured in coherence blocks, wherein the channel is approximately static and frequency flat. Each TDD coherence block is $\tau_c$ samples long and is determined by the shortest UE's coherence time, $T_c$, and bandwidth, $B_c$, in the system as $\tau_c = T_c B_c$. Each coherence interval accommodates three main phases, namely 1) UL training of duration $\tauup$ samples, 2) DL transmission of duration $\tau_{d}$ samples, and 3) UL transmission with $\tau_u$ samples. During the UL training phase, all UEs transmit pilot sequences of length $\tauup < \tau_c$ and the APs estimate their channels towards all UEs. Subsequently, the APs exploit the estimated UL CSI to combine the received signals in the UL, and to perform DL data precoding, leveraging the inherent channel reciprocity  in TDD protocol. Therefore, the estimation overhead scales with the number of served UEs and is independent of the total number of  AP antennas.

\subsubsection{Downlink Transmission}
Let $\wlk \in \mtc^{N\times 1}$ denote the precoding vector at AP $\ell$ assigned to UE $k$. Different precoding designs will be described in Section~\ref{sec:DL precod}.  Then, the transmit signal by AP $\ell$ can be expressed as
\begin{align}~\label{eq:sl:dl}
  \qs_{\ell} = \sum\nolimits_{k=1}^{K} \sqrt{\rho_d \etalk}\wlk x_k, 
\end{align}
where $x_k$ is the unit-power data signal intended for  UE $k$, i.e., $\Ex\{\vert x_k\vert^2\}=1$ and $\Ex\{x_kx_j^*\}=0$ for $k \neq j$; $\rho_d \triangleq p_d/\Sn$ denotes the normalized transmit signal-to-noise ratio (SNR) related to the data symbol, where $p_d$ is the DL transmit power and $\Sn$ denotes the noise power; $\etalk$, $\ell=1,\ldots,M$, $k=1,\ldots,K$, are the power control coefficients at AP $\ell$ for UE $k$, satisfying the following average power constraint at each AP
\begin{align}~\label{eq:sk}
 \Ex\{\Vert \qs_{\ell}\Vert^2\} \leq \rho_d.  
\end{align}
The $k$th UE receives a linear combination of the signals transmitted by all APs given by
\begin{align}~\label{eq:rk}
 r_{k}^{\dl} 
 &= \sqrt{\rho_d} a_{kk} x_k 
 +
  \sum\nolimits_{k'\neq k}^{K}\sqrt{\rho_d} a_{kk'} x_{k'} + n_k, 
\end{align}
where $a_{kk'} \triangleq \sum_{\ell=1}^{M} \sqrt{ \etalkp} \qh_{\ell k}^H \wlkp $ describes the effective channel gain. Moreover, in~\eqref{eq:rk} the first term is the desired signal, the second term describes the multi-user interference, and the third term is the independent and identically distributed (i.i.d.) Gaussian noise at the receiver, $n_k\sim \mathcal{CN}(0,1)$. To decode $x_k$ reliably, UE $k$ must have sufficient knowledge of $a_{kk}$. There are different decoding approaches in the literature for decoding $x_k$ at the UE, which are summarized as follows.

\textbf{Channel statistics-based data decoding:}
When the effective channel is not available at the UE, channel statistics can be employed for data decoding. In this case, an achievable DL SE for UE $k$ can be obtained by using the popular \textit{hardening} bound or \textit{use-and-forget} capacity-bounding technique~\cite{ngo16}. This approach is widely used in the literature of mMIMO and assumes that $a_{kk}\approx \Ex\{a_{kk}\}$. Channel hardening indicates that after the precoding/combining, the fading channel between the AP and UE is transformed into an almost deterministic scalar channel~\cite{Ngo:TWC:2017,Sutton:TVT:2021}.

\textbf{Coherent decoding via downlink training:}
In CF-mMIMO systems with a low and moderate number of APs, only a few dominant APs, located close to a specific UE, effectively contribute to its received signal. Therefore, the channel hardening phenomenon is less pronounced than in cellular mMIMO. As a result, in these cases, relying on the channel statistics at the UE side for decoding may be no longer considered valid. The lower degree of channel hardening in CF-mMIMO compared to co-located mMIMO was pointed  out in~\cite{Interdonato:TWC:2019}, and also analytically demonstrated in~\cite{Chen:TCOM:2018,Polegre:ACESs:2020}
under different channel model assumptions. To tackle this issue, DL training is performed to acquire the estimate of the effective channel at the UEs. In other words, one additional phase is considered and $\tau_{d,p}$ samples per coherence interval are spent for DL channel estimation. By leveraging the capacity-bounding technique for fading channels with non-Gaussian noise and side information, a capacity lower bound for the $k$th UE was derived in~\cite[Eq. (25)]{Interdonato:TWC:2019}. 

\textbf{Non-coherent decoding:}
This approach does not rely on the channel hardening and explicit estimate of $a_{kk}$. To this end, UE $k$ can estimate the gain of its instantaneous effective channel $a_{kk}$ from the collection of $\tau_d$ DL signals in the current coherence block~\cite{Chen:TCOM:2018}. With non-coherent decoding, only a scalar must be deduced from the received signal, and thus there is no need for  DL pilot transmission.  The corresponding SE lower bound with no instantaneous CSI at the UE was given in~\cite{Chen:TCOM:2018}. While this result provides an accurate lower bound, especially when the channels change slowly, a practical decoding method has yet to be developed, representing an  interesting avenue for future research. Another open research direction is the application of blind algorithms to estimate the effective channel gain $a_{kk}$ for DL decoding. This idea has already been reported for cellular mMIMO in~\cite{Ngo:TWC:2017}.    
\subsubsection{Uplink Transmission} 
All UEs simultaneously send their data to the APs. UE $k$ weights its information symbol $s_k\sim\mathcal{CN}(0,1)$ by a power control coefficient $\sqrt{\varsigma_k}$, with $0\leq \varsigma_k \leq 1$. The received signal at AP $\ell$ can be written as
\begin{align}~\label{eq:y_ell_ul}
   \qy_{\ell}^{\ul} = 
  \sqrt{ \rho_{u}} \sum\nolimits_{k=1}^K \sqrt{\varsigma_k} \qh_{k\ell} s_k + \qn_{\ell}, 
\end{align}
where $\rho_{u} \triangleq p_u/\Sn$ is the normalized UL SNR, with $p_u$ being the UL power, and  $\qn_{\ell}\sim\mathcal{CN}(\boldsymbol{0}, \qI_N)$ is the additive noise vector at the $\ell$th AP. 

The $\ell$th AP forms the receive combining vector $\qv_{k\ell}\in\mathbb{C}^{N\times 1}$ and locally computes $\qv_{k\ell}^H\qy_{\ell}^{\ul}$. Then, data detection can be performed locally at each AP, or can be delegated to the CPU. Data detection at the CPU offers the opportunity to combine multiple copies of the received signal from all APs, which can improve the performance of UL data detection. This, in turn, requires a high-capacity fronthaul link, and thus a trade-off between complexity and performance must be carefully considered in the system level design. In Section~\ref{sec:UL comb}, different levels of cooperation between the APs and CPU for data detection process are discussed, and some corresponding results are provided.    

Given the broader application of TDD, our focus in the following sections will be on TDD-based CF-mMIMO systems.

\begin{figure}[!t]\centering \vspace{0em}
    \def\svgwidth{240pt} 
    \fontsize{8}{7}\selectfont 
\begingroup%
  \makeatletter%
  \providecommand\color[2][]{%
    \errmessage{(Inkscape) Color is used for the text in Inkscape, but the package 'color.sty' is not loaded}%
    \renewcommand\color[2][]{}%
  }%
  \providecommand\transparent[1]{%
    \errmessage{(Inkscape) Transparency is used (non-zero) for the text in Inkscape, but the package 'transparent.sty' is not loaded}%
    \renewcommand\transparent[1]{}%
  }%
  \providecommand\rotatebox[2]{#2}%
  \newcommand*\fsize{\dimexpr\f@size pt\relax}%
  \newcommand*\lineheight[1]{\fontsize{\fsize}{#1\fsize}\selectfont}%
  \ifx\svgwidth\undefined%
    \setlength{\unitlength}{1029.97503662bp}%
    \ifx\svgscale\undefined%
      \relax%
    \else%
      \setlength{\unitlength}{\unitlength * \real{\svgscale}}%
    \fi%
  \else%
    \setlength{\unitlength}{\svgwidth}%
  \fi%
  \global\let\svgwidth\undefined%
  \global\let\svgscale\undefined%
  \makeatother%
  \begin{picture}(1,0.652443)%
    \lineheight{1}%
    \setlength\tabcolsep{0pt}%
    \put(0,0){\includegraphics[width=\unitlength]{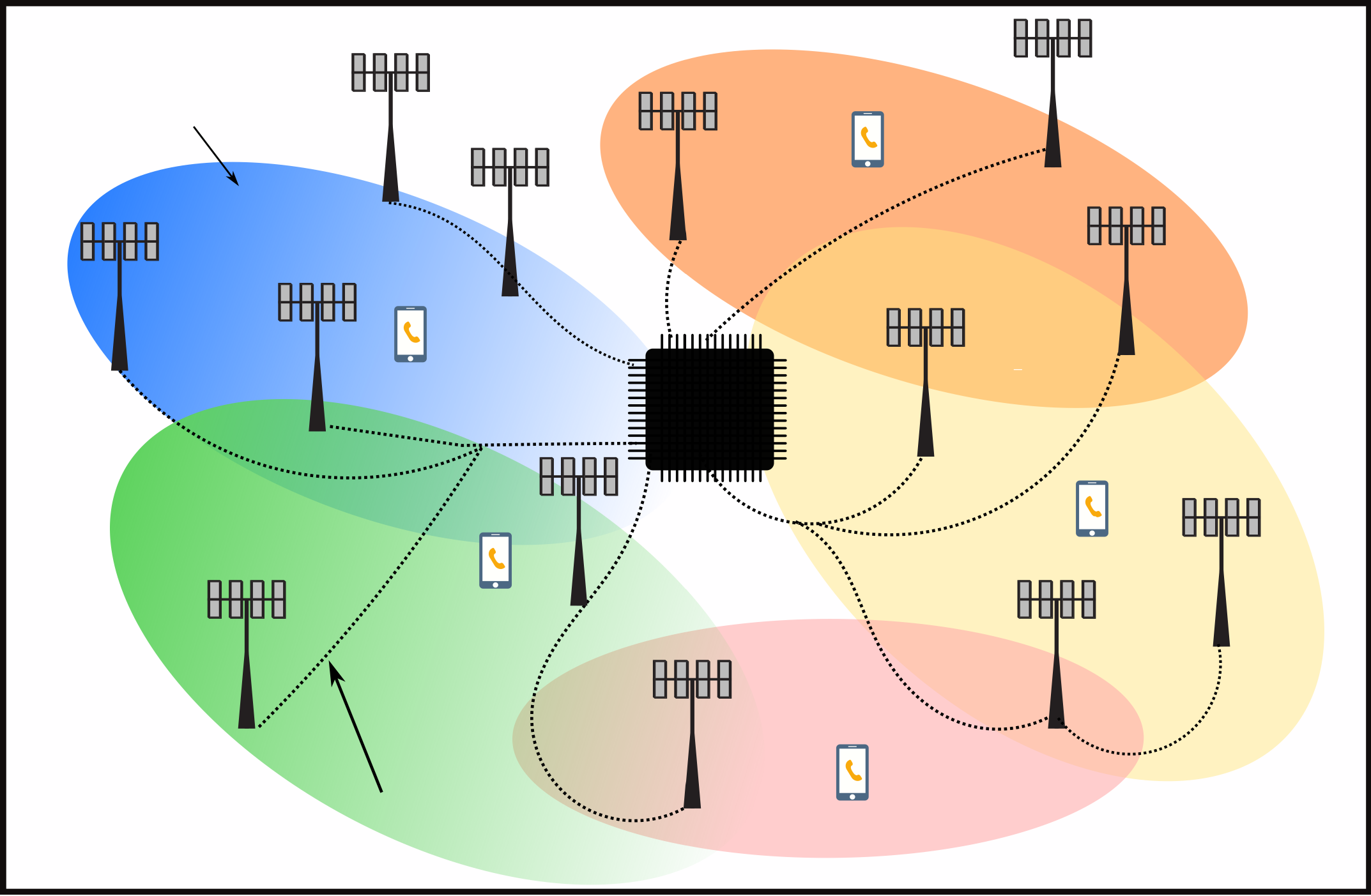}}%
    \put(0.24400246,0.04294796){\color[rgb]{0,0,0}\makebox(0,0)[lt]{\lineheight{1.25}\smash{\begin{tabular}[t]{l}Fronthaul\end{tabular}}}}%
    \put(0.31920532,0.38105262){\color[rgb]{0,0,0}\makebox(0,0)[lt]{\lineheight{1.25}\smash{\begin{tabular}[t]{l}UE $k$\end{tabular}}}}%
    \put(0.09605558,0.56864674){\color[rgb]{0,0,0}\makebox(0,0)[lt]{\lineheight{1.25}\smash{\begin{tabular}[t]{l}$\mathcal{M}_k$\end{tabular}}}}%
    \put(0.47902453,0.34180209){\color[rgb]{1,1,1}\makebox(0,0)[lt]{\lineheight{1.25}\smash{\begin{tabular}[t]{l}\textbf{CPU}\end{tabular}}}}%
  \end{picture}%
\endgroup%
 \vspace{0mm}
    \caption{Illustration of UC CF-mMIMO system, where each UE is
served by a subset of APs.}\vspace{0mm} \label{fig:NAFD_CFmMIMO}
    \vspace{1em}
\end{figure}
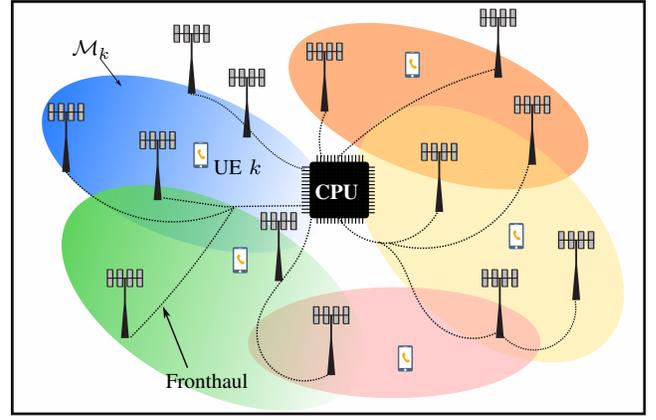

\subsection{User-Centric Cell-free Massive MIMO}
In the canonical form of CF-mMIMO networks, all UEs are served by all APs over the same time-frequency resources. Therefore, each AP must process the UL and DL information of all users and exchange the UL and DL data-payload and power control coefficients with the CPU through the fronthaul link. The computational complexity at the APs, related to channel estimation and data processing as well as the fronthaul load required for data and CSI sharing, grow linearly with the number of UEs, leading to the scalability problem in the CF-mMIMO networks~\cite{Emil:TCOM:2020}. To have a scalable implementation of a CF-mMIMO network, it is required to guarantee the complexity and resource requirement of signal processing to be finite for each AP as $K \rightarrow \infty$. On the other hand, due to the large path-loss variations, each UE receives most of its power from a small subset of APs during DL transmission. Likewise, not all APs are required for decoding a specific UE's data during UL data reception. 

To accommodate the above challenges, an alternative is to consider the UC approach (also referred to as dynamic cooperating clustering~\cite{Emil:TSP:2011}), where each UE is served by a subset of APs. A \textit{UC CF-mMIMO}  architecture is a special case of scalable CF-mMIMO, which creates a serving cluster for each UE, consisting of the APs that contribute a useful signal to the UE. These serving clusters are constructed based on different criteria, such as the serving distance~\cite{Buzzi:WCL:2017,Ammar:GLOBSIP:2019,Buzzi:TWC:2020} and two-stage process, where the serving cluster is first created based on the LSF and then is refined and optimized using some UE scheduling and resource allocation algorithms~\cite{Ammar:TWC:2022}. The recent studies in~\cite{Ammar:Tuts:2022,Shuaifei:DCN:2022} delve into resource allocation and signal processing issues in UC CF-mMIMO, surveying the latest schemes and algorithms.

Let each AP be responsible for a limited number of UEs denoted by $\mathcal{D}_{\ell}\subset\{1,\ldots,K\}$, $\ell=1,\ldots M$, i.e., AP $\ell$ receives/sends data related to UE subset $\mathcal{D}_{\ell}$ instead of all UEs. Therefore, the signal processing procedure of the UC CF-mMIMO shares the same terminology and mathematical expressions as canonical CF-mMIMO, given in~\eqref{eq:rk} and~\eqref{eq:y_ell_ul}, where $\qw_{\ell k}=\boldsymbol{0}_N$ and $\qv_{k\ell}=\boldsymbol{0}_N$ for $k \notin \mathcal{D}_{\ell}$, $\ell=1,\ldots M$. From the UE perspective, $\Mk\subset\{1,\ldots M\}$, $k=1,\ldots K$, denotes the subset of APs serving UE $k$.

\vspace{-1em}
\subsection{Preliminary Mathematical Results}
In analyzing the SE of CF-mMIMO systems, leveraging random matrix theory is essential. In the following, some important results from the literature are summarized.
\begin{Lemma}~\label{lemma:tracelemma}
(Trace Lemma~\cite[Lemma 4]{Wagner:IT:2012}) Let $\qx$, $\qw\sim\mathcal{CN}(\boldsymbol{0}, \frac{1}{M}\qI_M)$ be mutually independent vectors of length $M$ and also independent of $\qA\in\mathbb{C}^{M\times M}$ , which has a uniformly bounded spectral norm for all $M$. Then,
\begin{align}
    &\qx^H\qA\qx -\frac{1}{M}\trace(\qA)\xrightarrow{M\to\infty} 0,\\
    &\qx^H\qA\qw \xrightarrow{M\to\infty} 0,\\
    &\Ex\bigg\{\Big\vert  (\qx^H\qA\qx)^2 - \Big(\frac{1}{M}\trace(\qA)\Big)^2\Big\vert^2\bigg\}
   \xrightarrow {M\to\infty} 0\\
    &\Ex\bigg\{\Big\vert  \qx^H\qA\qx - \frac{1}{M}\trace(\qA)\Big\vert^p\bigg\}
    =\mathcal{O}\bigg(\frac{1}{N^{\frac{p}{2}}}\bigg).
\end{align}
\end{Lemma}
\begin{Lemma}~\label{lemma:Tchebyshev}(Tchebyshev's Theorem~\cite{cramer2004random})
    Let $X_1,\ldots,X_n$ be independent random variables such that $\Ex\{X_i\} = \bar{x}_i$ and $\mathtt{Var}(X_i)\leq c_i <\infty$. Then, Tchebyshev’s theorem states
    \begin{align}
        \frac{1}{n}\sum\nolimits_{n'=1}^n X_{n'} \xrightarrow[n\rightarrow\infty]{P}  \frac{1}{n}\sum\nolimits_{n'=1}^n \bar{x}_{n'}.
    \end{align}
\end{Lemma}
\begin{Lemma}~\label{lemma:Wishart}~\cite[Lemma 2.10]{tulino04}:
For a $K\times K$ central Wishart matrix with $\qX$ with $N$ degrees of freedom and covariance matrix $\qI_K$
\begin{align}
    \Ex\{\trace(\qX^{-1})\} = \frac{K}{N-K}.
\end{align}
\end{Lemma}
\begin{Lemma}~\label{lemma:expxhAx}~\cite[Eq. (15.14)]{Kay}: 
Let ${\qu}$ be a complex $n \times 1$ random vector with mean $\boldsymbol{\mu}$ and covariance matrix $\boldsymbol{\Sigma}$ and let $\qB$ be an $n\times n$ positive definite Hermitian matrix. Then, we have
\begin{align}
 \Ex\{{\qu}^H \qB  {\qu} \} = \boldsymbol{\mu}^H \qB \boldsymbol{\mu} + \trace (\qB \boldsymbol{\Sigma}).   
\end{align}
\end{Lemma}
\begin{Lemma}~\label{lemma:expxhAxsq}~\cite[Lemma 2]{Matthaiou:TCOM:2015}: 
Let ${\qu}$ be a complex $n \times 1$ random vector with mean $\boldsymbol{\mu}$ and covariance matrix $\boldsymbol{\Sigma}$ and let $\qB$ be an $n\times n$ positive definite Hermitian matrix. Then, we have
\begin{align}
  \Ex\{\lvert\qu^H \qB \qu\rvert^2\}=\lvert \trace(\pmb{\Sigma}\qB )\rvert^2+\trace(\pmb{\Sigma}\qB\pmb{\Sigma}\qB^H).  
\end{align}
\end{Lemma}
\begin{Lemma}~\label{Lemma:B}
~\cite[Lemma 1]{Mohamed_projection_2024} For the projection matrix
$\qB = \mathbf {I}_{M}-\qR^{H}\left (\qR\qR^{H}\right)^{-1}\qR$, with $\qR \in \mathbb{C}^{N \times M}$, we have
\vspace{-0.3em}
\begin{align}
    \Ex\{\qB\} = \frac{M-N}{M} \qI_M, \quad M>N.
\end{align}
\end{Lemma}
\begin{Lemma}~\label{Lemma:xMxxNx}
~\cite[Lemma 5]{van2021reconfigurable}
    For a random vector $\qx\in\mathbb{C}^N$ distributed as $\qx\sim\mathcal{CN}(\boldsymbol{0}, \bar{\qR})$ with $\bar{\qR}\in\mathbb{C}^{N\times N}$ and two deterministic matrices $\qM, \qN\in\mathbb{C}^{N\times N}$, it holds that
    \begin{align} 
    \mathbb {E} \{ \mathbf {x}^{H} \mathbf {M} \mathbf {x} \mathbf {x}^{H} \mathbf {N} \mathbf {x} \} = \mathrm {tr}(\bar {\mathbf {R}}\mathbf {M}\bar {\mathbf {R}}\mathbf {N}) + \mathrm {tr}(\bar {\mathbf {R}}\mathbf {M}) \mathrm {tr}(\bar {\mathbf {R}}\mathbf {N}).
    \end{align}
\end{Lemma}

\section{Signal Processing in CF-mMIMO systems}~\label{sec:SignalProcess}
\vspace{-2em}
\subsection{Uplink Training and Channel Estimation}~\label{sec:CHE}
During the UL training phase, all UEs send their pilot sequences to the APs. Assume that $\tauup$ mutually orthogonal pilot sequences are available as $\Bpsi_{1}, \ldots, \Bpsi_{\tauup}$ with $\Vert \Bpsi_{k}\Vert^2=\tauup$. These pilot sequences are assigned to the UEs in a deterministic way. The case of practical interest is a large-scale network with $K>\tauup$, so that more than one UE is assigned to one pilot sequence. We denote by $i_k\in\{1,\ldots,\tauup\}$ the index of the pilot sequence used by UE $k$, and define $\mathcal{P}_k \subset \{1,\ldots,K\}$ as the set of indices of UEs transmitting the same pilot as UE $k$. Therefore,  we have 
\begin{align}
    \Bpsi_{i_t}^H \Bpsi_{i_k} =
    \begin{cases}
        \tauup  & t \in \mathcal{P}_k, \\
        0  & t \notin \mathcal{P}_k.
    \end{cases}
\end{align}
When the UEs transmit their pilot sequences, the received pilot signal by AP $\ell$, $\qY_\ell\in\mathbb{C}^{N\times \tauup}$, is given by $\qY_\ell = \sum_{k=1}^{K} \qh_{\ell k} \sqrt{\varsigma_k p_p}\Bpsi_{i_k}^H + \qN_{\ell}$, where $\qN_{\ell}\in\mathbb{C}^{N\times \tauup}$ is the Gaussian noise matrix with i.i.d. $\mathcal{CN}(0,\Sn)$ elements; $0\leq \varsigma_k \leq 1$ is the pilot power control coefficient at UE $k$; $p_p$ is the maximum transmit power at each UE per each pilot symbol.  To compute an estimate of $\qh_{\ell k}$, AP $\ell$ first computes the projection of $ \qY_\ell$ onto $\frac{\Bpsi_{i_k}}{\sqrt{\tauup}}$ as 
\begin{align}
 \qy_{\ell k}=\frac{1}{\sqrt{\tauup}}\qY_\ell\Bpsi_{i_k} = \sum\nolimits_{i\in\mathcal{P}_k} \sqrt{\tauup \varsigma_i p_p} \qh_{\ell i} + \qn_{\ell k},  
\end{align}
with $\qn_{\ell k}\triangleq\frac{1}{\sqrt{\tauup}}\qN_{\ell}\Bpsi_{i_k} \sim\mathcal{CN}(\boldsymbol{0}, \Sn\qI_N)$. Note that, since the pilots are chosen from a set of mutually orthogonal sequences, $\qy_{\ell k}$ is a sufficient statistic, i.e. estimates based on $\qy_{\ell k}$ and $\qY_\ell$ are the same. Then, the minimum mean square error (MMSE) estimate of the channel between UE $k$ and AP $\ell$ is given by
\begin{align}
   \hat{\qh}_{\ell,k} =\sqrt{\tauup \varsigma_k p_p} \qR_{\ell k}
   \boldsymbol{\Phi}_{\ell k}^{-1}
   \qy_{\ell k},
\end{align}
where $\boldsymbol{\Phi}_{\ell k} =  \sum_{i\in\mathcal{P}_k} {\tauup \varsigma_i p_p} \qR_{\ell i} + \Sn\qI_N$. The channel estimate $\hat{\qh}_{\ell k}$ and estimation error $\tilde{\qh}_{\ell k}={\qh}_{\ell k}-\hat{\qh}_{\ell k}$ are independent vectors distributed as $\hat{\qh}_{\ell k}\sim\mathcal{CN}(\boldsymbol{0}, \tauup \varsigma_k p_p \qR_{\ell k} \boldsymbol{\Phi}_{\ell k}^{-1} \qR_{\ell k} )$ and $\tilde{\qh}_{\ell k}\sim\mathcal{CN}(\boldsymbol{0},  \qR_{\ell k}-\tauup \varsigma_k p_p \qR_{\ell k} \boldsymbol{\Phi}_{\ell k}^{-1} \qR_{\ell k} )$. Let $\hat{\qH}_\ell = [\hat{\qh}_{\ell 1},\ldots\hat{\qh}_{\ell K}] \in\mathbb{C}^{N\times K}$ denote the channel estimate matrix from AP $\ell$ to all UEs.

As the number of UEs is high in CF-mMIMO networks, a significant number of orthogonal pilots are required. However, due to the limited slots of the coherence interval in wireless fading channels, pilot resources are constrained. Consequently, pilots must be reused among UEs in CF-mMIMO systems, resulting in the phenomenon known as pilot contamination. Mathematically speaking, when $\tauup < K$, some of the channel estimates become parallel, leading to the rank deficiency of $\hat{\qH}_\ell$.

\begin{Definition}
\textit{(Pilot contamination):} For any pair of UEs $k$ and $t$ sharing the same pilot sequence, the respective channel estimates are linearly dependent as  $\hat{\qh}_{\ell k} = \frac{\varsigma_k \beta_{\ell k}}{\varsigma_t \beta_{\ell t}} \hat{\qh}_{\ell t}$. Therefore, the AP is unable to spatially differentiate between linearly correlated channels, which forms the core concept of pilot contamination.
\end{Definition}

\subsubsection{Pilot Assignment}
Pilot contamination has the potential to degrade the performance of CF-mMIMO systems, often becoming a significant bottleneck. Ensuring optimal system performance necessitates efficient management of both the available radio resources and pilot sequences, particularly in scenarios involving massive access where the number of UEs is roughly equivalent to the number of  APs. Hence, there is significant importance in devising pilot assignment schemes to alleviate pilot contamination. 

In the literature, various pilot assignment schemes have been proposed, each based on different principles. These schemes are summarized as follows:
\begin{itemize}
    \item Greedy algorithm~\cite{Hien:cellfree}: It iteratively updates the pilot sequence of the UE with lowest DL SE (or UL SE) so that its pilot contamination effect is minimized. This pilot assignment is recomputed on a large-scale fading
    time scale, which significantly simplifies the signal processing at the CPU.
    \item Tabu-search algorithm~\cite{Liu:TVT:2020}: This is a greedy-based algorithm founded on the neighborhood principle. For a given pilot assignment vector of length $K$, there are $K$ neighboring vectors, each differing from the original by just one pilot sequence. The best pilot assignment vector is then chosen to maximize the UL sum SE. To avoid being trapped in a local optimal, a Tabu list is maintained to record the previous solutions, ensuring an efficient search of the solution space.     
    \item K-means clustering algorithm~\cite{Chen:JSAC:2021}: The distances between all UEs and APs are considered as a metric for pilot assignment. The main idea is to separate $K$ UEs into $\lceil K/\tauup\rceil$ disjoint clusters, each centered around one of the $\lceil K/\tauup\rceil$ centroids,  with the centroids placed as far apart as possible. Every such cluster comprises at most $\tauup$ UEs and UEs in the same cluster are assigned mutually orthogonal pilots.
    \item Graph coloring-baased algorithm~\cite{Heng:TVT:2020}: The idea is to generate an interference graph based on AP selection for each UE. AP selection is conducted using large-scale fading coefficients between the APs and UEs to identify the most significant APs providing strong signals to each UE. The optimal pilot assignment is then achieved by updating the interference graph using a graph coloring algorithm, which aims to color the vertices of the graph with the minimum number of different colors.    
    \item Hungarian-based algorithm~\cite{Buzzi:WCL:2021}: This is an iterative procedure involving at each step the definition of a proper bipartite graph such that the Hungarian algorithm\footnote{The Hungarian algorithm, is a popular combinatorial algorithm used to solve weighted matching problems in a bipartite graph with polynomial complexity~\cite{munkres1957algorithms}.} can be used to perform matching. The algorithm parameters can be tuned so as to maximize either the sum-rate or the fairness across UEs.
\end{itemize}

\subsection{Downlink Precoding}~\label{sec:DL precod}
Channel hardening at the UE side can be affected by the adopted precoding design at the APs, as the effective DL channel gain is given by the inner product of the channel vector and precoding vector. Therefore, the channel hardening can be artificially boosted through the precoding design at the APs. Moreover, inter-user interference can be efficiently managed by deploying multi-antenna APs and appropriate precoder design. We notice that the more antennas, the more the channel hardens. DL precoding in CF-mMIMO can be classified into centralized~\cite{Nayebi:TWC:2017,Emil:TCOM:2020} and distributed (local) precoding design~\cite{Interdonato:TWC:2020,Interdonato:TCOM:2021}. 

\subsubsection{Centralized Precoding} When centralized precoding is designed, the DL data-payload in~\eqref{eq:sl:dl} is generated at  the CPU and then sent to the AP $\ell$. Therefore, the CPU requires all estimates of the channels between all APs and UEs to perform precoding design. A centralized zero-forcing (ZF) precoding design was exploited in~\cite{Nayebi:TWC:2017}, which can completely cancel the inter-user interference in CF-mMIMO networks with perfect CSI. With single-antenna AP and UEs, the centralized ZF precoding is given by $\qW^\CZF = \hat{\qG}^*(\hat{\qG}^T\hat{\qG}^*)^{-1}$, where $[\hat{\qG}]_{mk} = \hat{g}_{mk}$ and $\hat{\qG}^T\qW^\CZF =\qI_K$ for  $MN \geq K$. Nevertheless, the centralized ZF precoding matrix $\qW^\CZF$ is suboptimal for the power control problem subject to the per-antenna power constraints~\cite{Wiesel:TSP:2008}. To establish an orthogonal channel between the APs and UEs following ZF precoding, it is imperative that $\eta_{1k}=\ldots=\eta_{Mk}$ for all values of $k$. In other words, all APs must apply the same power control coefficient to a specific UE, as discussed in~\cite{Nayebi:TWC:2017}.

A joint maximum-ratio and ZF precoding scheme, termed as JMRZF, was proposed in~\cite{Liutong:TCOM:2021}, where part of the APs are combined to perform centralized ZF, while the other APs apply simple maximum-ratio transmission (MRT). The proposed precoder offers an adaptable trade-off between the SE and fronthaul signaling overhead. The findings presented in~\cite{Liutong:TCOM:2021} offered a generalized result encompassing both the fully distributed MRT and fully centralized ZF cases with multi-antenna APs.

\subsubsection{Distributed Precoding with Maximum Ratio (MR) Processing} While centralized ZF precoding can efficiently cancel all inter-user interference terms at UE $k$, it is not scalable when the number of APs and UEs is large. To preserve the system scalability, distributed precoder design has gained more attention in the literature. Two outstanding aspects of the CF-mMIMO, i.e., the large macro-diversity and favorable propagation, motivate the use of traditional MR processing, also known as conjugate beamforming (CB). The early work~\cite{Hien:cellfree} investigated the CB designs subject to long-term average power constraints, i.e,  $\qw_{\ell k}^{\CB} = \hat{\qh}_{\ell k}^*$. Later, a DL precoding design that satisfies a short-term average power constraint\footnote{When power allocation is used in CF-mMIMO systems, we need to take into account that power is constrained by either long-term average power constraints or short-term average power constraints~\cite{khoshnevisan2012power}. For long-term   power constraints, the average is taken over codewords and channel fading coefficients. By contrast, for short-term power
constraints, the average is just taken over the codewords} at the APs, named as normalized CB (NCB), was proposed in~\cite{Interdonato:CAMD:2016}. The authors derived an approximate closed-form expression for the per-user achievable DL SE by using  the hardening bounding technique. With NCB, the precoding factor consists of the conjugate of the channel estimate normalized by its magnitude, i.e., $\qw_{\ell k}^{\NCB} = \frac{\hat{\qh}_{\ell k}^*}{\Vert \hat{\qh}_{\ell k}\Vert}$. Therefore, compared to the conventional CB scheme~\cite{Hien:cellfree}, the NCB just performs a phase shift of the data signal. Nevertheless, it enables a reduction of the beamforming gain uncertainty due to the lack of CSI knowledge at the UE side, thereby resulting in a large advantage in terms of hardening with respect to CB. Therefore, NBC outperforms CB when the number of APs is moderate, thanks to the improved channel hardening. The authors in~\cite{polegre2020new} extended the results in~\cite{Interdonato:CAMD:2016} into multi-antenna APs and presented a closed-form expression for an achievable DL SE by using the hardening bound. A variant of the NCB precoding scheme described in~\cite{Interdonato:CAMD:2016,polegre2020new}, dubbed enhanced
NCB (ECB), was proposed in~\cite{Interdonato:TCOM:2021}, where the vector of the channel estimates between a multi-antenna APs and a given UE is normalized by its squared norm, i.e, $\qw_{\ell k}^{\ECB} = \frac{\hat{\qh}_{\ell k}}{ \Vert \hat{\qh}_{\ell k}\Vert^2}$. An exact closed-form expression for an achievable DL SE was derived by using the popular hardening bound, which accounts for channel estimation errors at the AP, pilot contamination due to pilot reuse, and lack of CSI at the UE side.

Moreover, an approximate closed-form expression for an
achievable DL SE for CB assuming DL training and multi-antenna APs, termed as CB-DT, was derived in~\cite{Interdonato:TCOM:2021}.

\subsubsection{Distributed Precoding with ZF Processing} The performance of the distributed CB designs is degraded by the unavoidable inter-user interference. A distributed ZF precoding design, named as \textbf{full-pilot ZF (FZF)} has been developed in~\cite{Interdonato:TWC:2020} for the CF-mMIMO with multi-antenna APs to mitigate the inter-user interference. With FZF precoding, each AP uses its respective local channel estimates to construct a ZF precoder as $\qw_{\ell k}^{\FZF} = \gamma_{\ell k} \sqrt{N-\tauup} \bar{\qH}_{\ell} \big( \bar{\qH}_{\ell}^H \bar{\qH}_{\ell}\big)^{-1} \qe_{i_k}$, where $\bar{\qH}_{\ell} = \qY_{\ell} \BBPsi \in \mathbb{C}^{N\times \tauup}$ with $\BBPsi=[\Bpsi_{1},\ldots,\Bpsi_{\tauup}]\in\mathbb{C}^{\tauup\times \tauup}$ and $\qe_{i_k}$ is the $i_k$-th column of the identity matrix $\qI_{\tauup}$. We note here that $\hat{\qh}_{\ell k} = \bar{\qH}_{\ell} \qe_{i_k}\sim\mathcal{CN}(\boldsymbol{0},\gamma_{\ell k}\qI_N)$ and $\bar{\qH}_{\ell}$ contains the $\tauup$ independent columns of rank-deficient matrix $\hat{\qH}_\ell$.

The FZF precoder can mitigate the inter-user interference within each AP's coverage, but not the interference from other APs. Nevertheless, the performance of the FZF is limited by the available spatial degrees of freedom, i.e., the number of AP antennas $N$ and the number of orthogonal spatial directions we wish to cancel the interference towards, i.e., the number of pair-wise orthogonal pilot sequences $\tauup$. More specifically, to implement FZF, the condition $N > \tauup$ must be satisfied. However, reducing $\tauup$ degrades the quality of the channel estimation and increasing $N$ imposes higher hardware and computational complexity at the AP. This motivates two fully distributed ZF-based precoding schemes, referred to as local partial zero-forcing (PZF) and local protective partial zero-forcing (PPZF)~\cite{Interdonato:TWC:2020}:

\textbf{PZF design:} The principle behind the PZF is that for each AP $\ell$, the set of active UEs is first divided in two groups 1) strong UEs, denoted by $\mathcal{S}_{\ell}\subset\{1,\ldots,K\}$, and 2) weak UEs, denoted by $\mathcal{W}_{\ell}\subset\{1,\ldots,K\}$, where $\mathcal{S}_{\ell}\cap\mathcal{W}_{\ell} = \emptyset$. Now, AP $\ell$ uses the FZF precoder to completely suppress the intra-group interference between the UEs inside $\mathcal{S}_{\ell}$, while the MRT precoder is used for weak UEs. Therefore, the UEs in $\mathcal{S}_{\ell} $ experience non-coherent inter-group interference from the signal transmitted towards the UEs in $\mathcal{S}_{\ell}$. 

\textbf{PPZF design:} To protect the strong UEs against the inter-group interference caused by the transmission for weak UEs, the PPZF design forces the MRT precoder at AP $\ell$ to place in the orthogonal complement of the all strong UEs channel space. To this end, the MRT precoder is multiplied by the projection matrix onto the orthogonal complement channel space.  
The UE grouping for PZF and PPZF can be performed based on the different criteria, including the LSF-based criterion defined in~\cite{Hien:TGCN:2018}.

In future wireless networks, many user devices of moderate physical size (e.g., laptops, tablets, and smart vehicles) will be equipped with multiple antennas to enhance multiplexing gain and improve system reliability through diversity gain. Therefore, it is crucial to evaluate the performance of CF-mMIMO systems with multiple antennas at the user end. Nevertheless, only a few works in the literature have investigated the DL performance of CF-mMIMO systems with multi-antenna UEs in conventional~\cite{Mai:TCOM:2020}, UC~\cite{Buzzi:TWC:2020,Kunnath:JOPCS:2024}, and multicast~\cite{Zhou:WCL:2021} scenarios.

\subsection{Uplink Processing and Detection}~\label{sec:UL comb}
UL data detection schemes can be performed locally at the APs or can be fully/partially delegated to the CPU. Local data detection at the APs relying on the local channel estimates is the simplest implementation level of UL data detection for CF-mMIMO systems, which is termed as ``Level 1: Small-Cell Network" design~\cite{Emil:TWC:2020}. This name stems from the observation that when data detection is  performed locally at the APs, CF-mMIMO is turned into conventional small-cell networks. AP selection or UC architectures can improve the performance of the local data detection, making CF-mMIMO system with Level 1 processing comparable with conventional small-cell networks~\cite{Emil:TWC:2020}. To improve the UL SE, the CPU can fully/partially take care of data detection process. In the following, the corresponding data processing designs are discussed.   

\subsubsection{Fully Centralized Detection} 
To implement a fully centralized detection design, which  is called ``Level 4: Fully Centralized Processing" in the literature~\cite{Emil:TWC:2020}, all APs (or the serving subset of APs in UC CF-mMIMO systems) transmit their received pilots and received signals given in~\eqref{eq:y_ell_ul} to the CPU via fronthaul links. Then, the CPU performs channel estimation and data detection. To this end, an arbitrary combining vector $\qv_k\in\mathbb{C}^{NM\times 1}$ is designed at the CPU based on the collective channel estimates, $\hat{\qh}_k$, $k=1,\ldots, K$, to create an estimate of the transmit signal by the $k$th UE, $s_k$, as 
\begin{align}
  \hat{s}_k \!= \!\sqrt{ \rho_{u} \varsigma_k} \qv_k^H\qh_k s_k \!+ \!\sqrt{ \rho_{u}}\qv_k^H\sum\nolimits_{i\neq k}^K \sqrt{ \varsigma_i}\qh_i s_i \!+\! \qv_k^H\qn, 
\end{align}
where $\qn = [\qn_{1}^T, \ldots,\qn_{M}^T]^T \in\mathbb{C}^{NM\times 1}$, with  $\qn_{k}\sim\mathcal{CN}(\boldsymbol{0}, \qI_N)$ collects all noise vectors. Heuristic combiners such as MR, ZF, and regularized ZF (RZF) can be used to design $\qv_k$. In~\cite{Emil:TWC:2020},  the centralized MMSE (C-MMSE) combiner, which minimizes the conditional mean square error, $\Ex\{ \vert s_k - \hat{s}_k \vert \big\vert \{\hat{\qh}_i, i\in\mathcal{D}_{\ell}\} \}$ was developed as
\begin{align}
\qv_{k}^{\CMMSE} \!= \!\varsigma_k \rho_u\Big(\sum\nolimits_{i\in\mathcal{D}_{\ell}} \varsigma_i \rho_u(\hat{\qh}_{i} \hat{\qh}_{i}^H \!+\! \qC_{i}) \!+\! \qI_N\Big)^{-1}\hat{\qh}_{k},
\end{align}
where $\qC_i \triangleq \Ex\big\{ (\tilde{\qh}_i - \Ex\{\tilde{\qh}_i\})(\tilde{\qh}_i - \Ex\{\tilde{\qh}_i\})^H \big\}$. By using C-MMSE at the CPU, the following capacity lower bound is achieved
    \begin{align}~\label{eq:SE:ul:L4}
      &\SE_{k,\ul}^{(4)}\!=\!
      \frac{\tau_u}{\tau_c} 
      \Ex\left\{\log_2\Bigg( 1 +\rho_u \varsigma_k \hat{\qh}_k^H 
      \right.\nonumber\\
      &\hspace{0em}\left.{
      \times\Bigg( \rho_u\sum_{i\neq k}^K  \varsigma_i \hat{\qh}_i \hat{\qh}_i^H 
      \!+\! \rho_u\sum_{i=1}^K\!  \varsigma_i \qC_i  \!+\!\Sn \qI_{MN} \Bigg)^{-1} \hat{\qh}_k}
      \Bigg)\right\}\!.  
    \end{align}
For single-antenna APs, a centralized MMSE combiner was conceived in~\cite{Nayebi:ICC:2016}, where the impact of noise was neglected. The computational complexity as well as feedback overhead required for the centralized detection scheme is prohibitively high, especially when the network density is increased. If the computational complexity is a concern, ZF and RZF designs can be used at the expense of SE reduction in the low SNR regime. In~\cite{Liu:TWC:2020}, the achievable UL rate performance of the CF-mMIMO systems with single-antenna APs and centralized ZF detector was derived.

\subsubsection{Partially Centralized Detection}
To reduce the fronthaul load, the APs can construct the receive combining vectors locally using the estimated CSI and coherently process the received signals. Then, the resulting local estimates of the transmit signals at each AP, can be sent to the CPU to perform a postcoding procedure. The local estimate of the received signal $s_k$ at $\ell$th AP can be expressed as
\begin{align}~\label{eq:hat_skl}
   \hat{s}_{k\ell}& =  \sqrt{ \rho_{u} \varsigma_k}\qv_{k\ell}^H \qh_{k\ell} s_k 
   \nonumber\\
   &+ \sqrt{ \rho_{u}}\qv_{k\ell}^H \sum\nolimits_{i\in\mathcal{D}_{\ell}\setminus k}  \sqrt{\varsigma_i} \qh_{i\ell} s_i + \qv_{k\ell}^H \qn_k,~\ell\in \mathcal{M}_k,
\end{align}
where $\qv_{k\ell}\in\mathbb{C}^{N\times 1}$ is the local combining vector at AP $\ell$ for UE $k$. During the local processing stage, various local combining vectors, $\qv_{k\ell}$, can be utilized at the APs, each offering different balance between complexity and performance. For example, a low-complexity MR combining vector $\qv_{k\ell}^{\MR} = \hat{\qh}_{k\ell}$, or a high-performance local MMSE (L-MMSE) combining vector with
\begin{align}
\qv_{k\ell}^{\LMMSE} \!= \!\varsigma_k \rho_u\Big(\sum_{i=1}^K\varsigma_i \rho_u(\hat{\qh}_{i\ell} \hat{\qh}_{i\ell}^H \!+\! \qC_{i\ell}) \!+\! \qI_N\Big)^{-1}\hat{\qh}_{k\ell},
\end{align}
can be designed at each AP based on the trade-off between complexity and performance. However, to compute $\qv_{k\ell}^{\LMMSE}$, we need to compute all $K$ channel estimates $\hat{\qh}_{i\ell}$, for $i=1,\ldots,K$, at any AP $\ell$ that serves UE $k$. Therefore, the complexity of L-MMSE combining grows with $K$.  To tackle this, the alternative local partial MMSE was proposed in~\cite{Emil:TCOM:2020}, which only suppress the interference from the UEs that are served by partially the same APs as UE $k$, i.e., 
\begin{align}
\qv_{k\ell}^{\LPMMSE} \!= \!\varsigma_k \rho_u\Big(\sum_{i\in\mathcal{D}_{\ell}} \varsigma_i \rho_u(\hat{\qh}_{i\ell} \hat{\qh}_{i\ell}^H \!+\! \qC_{i\ell}) \!+\! \qI_N\Big)^{-1}\hat{\qh}_{k\ell}. 
\end{align}

Then, the local estimates in~\eqref{eq:hat_skl} are sent to the CPU, where they are linearly combined via weight coefficients $\{a_{k\ell},\ell\in \mathcal{M}_k\}$ to obtain $\hat{s}_k = \sum_{\ell\in \mathcal{M}_k} a_{k\ell}^*\hat{s}_{k\ell}$. By using~\eqref{eq:hat_skl}, we have
\begin{align}~\label{eq:hat_sk:final}
   \hat{s}_{k} 
   &\!=\!\sqrt{ \rho_{u} \varsigma_k}\qa_k^H\qg_{k k} s_k +\sqrt{ \rho_{u}} \!\!\sum_{i\in\mathcal{D}_{\ell}\setminus k} \!\!\sqrt{\varsigma_i}\qa_k^H\qg_{k i} s_i +  \tilde{\qn}_k,
\end{align}
where $\qa_k =[a_{k,\ell}, \ell\in \mathcal{M}_k]$ is the collective weighting coefficient vector for UE $k$ and $\qg_{k i} = [ \qv_{k\ell}^H\qh_{i\ell}, \ell\in \mathcal{M}_k]$, $i\in\mathcal{D}_{\ell}$ represents the effective channel. Since only the knowledge of the statistic of CSI is available at the CPU, the achievable UL SE for UE $k$ can be obtained by using the use-then-forget capacity-bounding technique as
    \begin{align}~\label{eq:SE:ul:L3}
      &\SE_{k,\ul}^{(3)}\!=\!
      \frac{\tau_u}{\tau_c} 
      \log_2\Bigg( 1 +
      \nonumber\\
      &\hspace{0.2em}
      \frac
      {{ \rho_{u} \varsigma_k}\vert\qa_k^H\Ex\{\qg_{k k}\}\vert^2}
      {\sum_{i\in\mathcal{D}_{\ell}} { \rho_{u} \varsigma_i}\Ex\{\vert\qa_k^H\qg_{k i}\vert^2\} \!-\! { \rho_{u} \varsigma_k}\vert\qa_k^H\Ex\{\qg_{k k}\}\vert^2 \!+\! \qa_k^H\qD_k\qa_k }     \Bigg),  
    \end{align}
where $\qD_k \triangleq \diag(\Ex\{\Vert \qv_{ki} \Vert^2\},i\in\mathcal{D}_{\ell})$.  Noticing that the signal-to-interference-plus-noise ratio (SINR) expression in~\eqref{eq:SE:ul:L3} is a generalized Rayleigh quotient, the optimal weighting vector at the CPU that maximizes $\SE_{k,\ul}^{(3)}$ can be obtained as, 
\begin{align}~\label{eq:qk:LSFD}
    \qa_k = \bigg(\sum\nolimits_{i\in\mathcal{D}_{\ell}} { \rho_{u} \varsigma_i}\Ex\{\qg_{k i} \qg_{k i}^H\} + \qD_k\bigg)^{-1}
    \Ex\{\qg_{k k}\},
\end{align}
which totally depends on the channel statistics. This approach is known as LSF decoding (LSFD) and $\qa_k$ is termed as the LSFD receiver. The authors in~\cite{Emil:TWC:2020}, coined "Level 3: Local Processing \& LSF Decoding" for this combining/decoding design in CF-mMIMO systems. 

The optimal LSFD receiver in~\eqref{eq:qk:LSFD}  requires knowledge of the channel statistics, which must be transferred to the CPU, all via a fronthaul link.  When the statistics vary with time, the acquisition and transmission process might not be feasible. To tackle this issue, an alternative is to create an estimate of $s_k$ by simply taking the average of the local estimates as $s_k = \frac{1}{\vert \mathcal{M}_k\vert} \sum_{\ell\in \mathcal{M}_k} \hat{s}_{k\ell}$. This design process is called ``Level 2: Local Processing \& Simple Centralized Decoding" in the literature~\cite{Emil:TWC:2020} and yields the UL SE expression for UE $k$, given in~\eqref{eq:SE:ul:L2} at the top of the next page. The authors in~\cite{Fan:ICC:2019} analyzed the UL performance of the CF-mMIMO with Level-2 processing, taking into account the effects of spatial channel correlation when each AP is equipped with multiple antennas and applies MR precoding to the received signal.

\begin{figure*}
    \begin{align}~\label{eq:SE:ul:L2}
      &\SE_{k,\ul}^{(2)}\!=\!
      \frac{\tau_u}{\tau_c} 
      \log_2\left( 1 +
      \frac
      { \rho_{u} \varsigma_k\Big\vert \sum_{\ell\in \mathcal{M}_k}\Ex\{\qv_{k\ell}^H\qh_{k \ell}\}\Big\vert^2}
      { \sum_{i\in\mathcal{D}_{\ell}}
      \rho_{u} \varsigma_i\Ex\Big\{\big\vert \sum_{\ell\in \mathcal{M}_k}\qv_{k\ell}^H\qh_{i \ell}\big\vert^2\Big\}
      -\rho_{u} \varsigma_k\Big\vert \sum_{\ell\in \mathcal{M}_k}\Ex\{\qv_{k\ell}^H\qh_{k \ell}\}\Big\vert^2\!+\! \sum_{\ell\in \mathcal{M}_k}\Ex\{\Vert \qv_{k\ell}\Vert^2\}   }     \right),  
    \end{align} 
    \hrulefill
\vspace{-2mm}
\end{figure*}

Optimal LSFD decoding (Level-3 processing) was first introduced to reduce the interference in cellular mMIMO networks~\cite{Adhikary:TCOM:2017}. This framework has been then applied into CF-mMIMO systems with single-antenna ~\cite{Nayebi:ICC:2016,Hien:Asilomar:2018,bashar19TWC} and multi-antenna APs~\cite{Zhang:TCOM:2021,Wang:CLET:2021}. Nayebi~\ettall~\cite{Nayebi:ICC:2016} developed an LSFD receiver with MR precoding  and conducted a performance comparison with Level-4 processing with MMSE combining vectors, revealing a considerable performance gap between the two approaches. Ngo~\ettall~\cite{Hien:Asilomar:2018} examined the performance of LSFD receiver with MR combining and under Ricean channels. They proposed an AP selection algorithm based on the LSFD weighting parameters. More specifically, this algorithm enables certain APs to transmit their signals to the CPU, resulting in reducing fronthaul requirements. Bashar \textit{et al.}~\cite{bashar19TWC} studied the UL SE gain achieved by applying joint LSFD receiver and power control design. Wang~\ettall~\cite{Wang:CLET:2021} considered a practical CF-mMIMO system with multi-antenna APs and spatially correlated Ricean fading channels, where the phase-shift of the line-of-sight (LoS) induced by the UE movement is modeled randomly. They investigated the UL SE achieved by Level-3 processing with MR and L-MMSE combining based on the phase-aware MMSE, phase-aware element-wise MMSE and linear MMSE channel estimators.

Zhang \textit{et al.}~\cite{Zhang:TCOM:2021} studied the UL SE provided by Level 3 processing with distributed ZF precoding schemes, e.g., FZF, partial FZF, protective weak partial FZF, and local RZF (LRZF) combining.  All these ZF-based designs can suppress inter-user interference fully distributively or coordinately in a scalable fashion and outperform MR, while provide comparable performance to L-MMSE, but have the added benefit of leading to closed-form expressions. With FZF, all degrees of freedom are spent for inter-user interference cancellation. However, in practice, the inter-user interference that affects UE $k$ is mainly generated by a small subset of other UEs, termed as strong UEs. Therefore, from the $\ell$th AP perspective, the UEs are divided into two groups: $\mathcal{S}_{\ell}\subset\{1,\ldots,K\}$ gathers strong UEs and  $\mathcal{W}_{\ell}\subset\{1,\ldots,K\}$ gathers weak UEs. Accordingly, partial FZF combining is designed to suppress the interference generated by the strong UEs, while the interference from the weak UEs is tolerated. Nevertheless, weak UEs experience intra-group interference from strong UEs. To protect the weak UEs against this interference, protective weak partial FZF combining  for weak UEs was proposed to significantly reduce the intra-group interference. 

In a few recent studies, the UL SE for multi-antenna UEs with different processing schemes has been examined over Weichselberger Ricean~\cite{Wang:TWC:2022} and Rayleigh fading~\cite{Li:TVT:2022} channels, respectively.

Assuming that each UE is equipped with $L$ antennas, transmit signal $\qs_k =[s_{k,1},\ldots,s_{k,L}]^T\in\mathbb{C}^L$ is constructed as $\qs_k =\qP_k\qx_k$, while $\qx_k\sim\mathcal{CN}(\boldsymbol{0}, \qI_L)$ is the data symbol vector transmitted from UE $k$ and $\qP_k\in\mathbb{C}^{L\times L}$ denotes the precoding matrix, satisfying the power constraint of UE $k$, given by $\trace(\qP_k\qP_k^H)\leq p_k$, with $p_k$ being the maximum transmitted power of UE $k$. Let $\qH_{mk}\in\mathbb{C}^{N\times L}$ denote the channel between the UE $k$ and AP $m$. The MMSE estimate of $\qH_{mk}$ follows the same process as that in Section~\ref{sec:CHE}. Let $\hat{\qH}_{mk}$ and $\tilde{\qH}_{mk}$ denote the MMSE estimate and estimation error of ${\qH}_{mk}$, respectively.  

In order to apply ``Level $4$" processing for data detection, the received signal at the CPU is obtained as 
\begin{align} \qy=\sum\nolimits_{k=1}^{K}{\mathbf {H}_{k}\mathbf {P}_{k}\mathbf {x}_{k}}+\mathbf {n},
\end{align}
where $\qy= {[\mathbf {y}_{1}^{T},\cdots,\mathbf {y}_{M}^{T}]^{T}}\in\mathbb{C}^{MN\times 1}$ denotes the received signal at $M$ APs and $\qH_k={[\mathbf {H}_{1k}^{T},\cdots,\mathbf {H}_{Mk}^{T}]^{T}}\in\mathbb{C}^{MN\times L}$ is the overall channel matrix between all APs and UE $k$. Next, the CPU selects an arbitrary receive combining matrix $\qV_k\in\mathbb{C}^{MN\times L}$ according to collective channel estimates for UE $k$ to detect $\qx_k$ as $\hat{\qx}_k =\qV_k^H\qy $. An achievable SE for UE $k$ with MMSE-SIC  (successive interference cancellation)
detector is given by~\cite{Wang:TWC:2022}
\begin{align} 
\SE_{k,\ul}^{(4)}=\frac {\tau _{u}}{\tau _{c}}  \mathbb {E}\left \{{ \log _{2}\left |{ \mathbf {I}_{N}+\qD_{k,({4}}^{H}\boldsymbol{\Sigma }_{k,({4 })}^{-1}\mathbf {D}_{k,\left ({4 }\right)} }\right | }\right \}\!, 
\end{align}
where $\qD_{k,({4})} = \qV_k^H \hat{\qH}_k\qP_k$ and $\boldsymbol{\Sigma }_{k,({4 })} \triangleq \qV_k^H\big(\sum_{l=1}^K \hat{\qH}_{l}\bar{\qP}_l \hat{\qH}_{l}^H - \hat{\qH}_{k}\bar{\qP}_k \hat{\qH}_{k}^H + \sum_{l=1}^K \qC_{l} + \Sn\qI_{MN} \big)\qV_k$, with $\hat{\qH}_{k}={[\hat{\qH}_{1k}^{T},\cdots,\hat{\qH}_{Mk}^{T}]^{T}}\in\mathbb{C}^{MN\times L}$, $\bar{\qP}_k \triangleq\qP_k\qP_k^H $, and $\qC_l = \diag(\qC_{1l},\ldots,\qC_{Ml})$, while  $\qC_{ml} = \Ex\{\tilde{\qH}_{ml}\bar{\qP}_l \tilde{\qH}_{ml}^H\}$. For $\qV_k$, any combining receive matrix can be utilized. For example, MR combining $\qV_k = \hat{\qH}_k$ and MMSE combining which minimizes the mean-squared error
\begin{align*} 
\mathbf {V}_{k}^\CMMSE=\Big ({\sum\nolimits _{l=1}^{K}{\left ({\mathbf {\hat {H}}_{l}\mathbf {\bar {P}}_{l}\mathbf {\hat {H}}_{l}^{H}+\mathbf {C}_{l} }\right)}+\sigma ^{2}\mathbf {I}_{ML} }\Big) ^{-1}\mathbf {\hat {H}}_{k}\mathbf {P}_{k}.
\end{align*}

When ``Level $3$" processing is applied, AP $m$ utilizes the combining matrix $\qV_{mk}\in\mathbb{C}^{N\times L}$ for UE $k$ to obtain a local estimate of $\qs_k$ at AP $m$, denoted as $\hat{\qs}_{mk}$. Possible choices for $\qV_{mk}$ are the MR combining with $\qV_{mk} = \hat{\qH}_{mk}$ and Local MMSE  combiner~\cite{Wang:TWC:2022} 
\begin{align*} \mathbf {V}_{mk}^\LMMSE=\left ({\sum _{l=1}^{K}{\left ({\mathbf {\hat {H}}_{ml}\mathbf {\bar {P}}_{l}\mathbf {\hat {H}}_{ml}^{H}+\mathbf {C}_{ml}' }\right)}\!+\!\sigma ^{2}\mathbf {I}_{N} }\right) ^{\!-1}\mathbf {\hat {H}}_{mk}\mathbf {P}_{k}.
\end{align*}
Accordingly, the second layer decoding, called LSFD, is performed at the CPU, by weightening the received local estimates. The optimal  LSFD coefficient matrix, $\mathbf {A}_{k}\in\mathbb{C}^{ML\times L}$, to maximize the achievable SE of UE $k$ for ``Level $3$" is given by~\cite{Wang:TWC:2022} 
\begin{align*} \mathbf {A}_{k}=\left ({\sum\nolimits _{l=1}^{K}{\mathbb {E}\left \{{ \mathbf {G}_{kl}\mathbf {\bar {P}}_{l}\mathbf {G}_{kl}^{H} }\right \}}+\sigma ^{2}\mathbf {S}_{k} }\right) ^{-1}\mathbb {E}\left \{{ \mathbf {G}_{kk} }\right \} \mathbf {P}_{k}, 
\end{align*}
where $\qG_{kl}= [\qV_{1k}^H\hat{\qH}_{1l};\ldots;\qV_{Mk}^H\hat{\qH}_{Ml}]\in\mathbb{C}^{ML\times L}$ and $\qS_k\triangleq\diag(\Ex\{\qV_{1k}^H\qV_{1k}\}, \ldots,\Ex\{\qV_{Mk}^H\qV_{Mk}\})\in\mathbb{C}^{ML\times ML}$. Then, the achievable SE for UE $k$ in ``Level $3$" with
MMSE-SIC detectors is given by
\begin{equation} \mathrm {SE}_{k}^{\left ({3 }\right)}=\frac {\tau _{u}}{\tau _{c}}  \log _{2}\left |{ \mathbf {I}_{N}+\qD_{k, ({3 })}^{H}\boldsymbol{\Sigma }_{k,({3 })}^{-1}\mathbf {D}_{k,\left ({3 }\right)} }\right |\!,
\end{equation}
where $\qD_{k, ({3 })} \triangleq \qA_k^H \Ex\{\qG_{kk}\}\qP_k$ and $\boldsymbol{\Sigma }_{k,({3 })}\triangleq\sum_{l=1}^{K} \qA_k^H \Ex\{\qG_{kl}\bar{\qP}_{l}\qG_{kl}^H\}\qA_k-\qD_{k, ({3 })}\qD_{k, ({3 })}^H + \Sn\qA_k^H\qS_k\qA_k$. 

Wang \textit{et al.}~\cite{Zhe:TCOM:2023} developed efficient UL precoding schemes based on an iteratively weighted sum-minimum mean square error (I-WMMSE) algorithm to maximize the weighted sum SE for ``Level 3" and ``Level 4" processing schemes over Weichselberger Rayleigh fading channels.  The proposed I-WMMSE precoding schemes in~\cite{Zhe:TCOM:2023} are more efficient with a larger number of UE antennas.

 \begin{flushleft} 
\begin{table*}
\centering
\small
\caption {Summary of DL precoding and UL detection designs for CF-mMIMO}\label{tab:PLYS}
\vspace{-0.4em}

\begin{tabular}{|>{\centering\arraybackslash}m{1.4cm}|>{\centering\arraybackslash}m{2.5cm}|>{\centering\arraybackslash}m{2.1cm}|>{\centering\arraybackslash}m{4.4cm}|>{\centering\arraybackslash}m{4.4cm}|}
\hline
\textbf{Category} 
&\centering \textbf{} 
& \centering \textbf{Processing} 
& \centering\textbf{Pros} 
&\textbf{Cons} \\ \hline
\centering

\multirow{4}{*}{\vspace{-16em} Downlink} 
&\hspace{1em} \multirow{2}{*}{\vspace{-2em}Centralized} 
&\vspace{-0.1em}\centering CZF~\cite{Nayebi:TWC:2017} 
& Capable of fully eliminating inter-user interference 
& Not scalable and requires extensive backhaul signaling \\ 
\cline{3-5}                      
&                              
&\vspace{-0.1em}\centering JMRZF~\cite{Liutong:TCOM:2021}
& Generalized form of distributed MRT and fully centralized ZF 
& Not scalable and requires extensive backhaul signaling \\ 
\cline{2-5}                       
&\hspace{1em} \multirow{6}{*}{\vspace{-7em}Distributed}  
&\vspace{0.3em}\centering MR (CB)~\cite{Hien:cellfree} 
& Scalable, computationally simple and less complex to implement compared to other schemes
& Ineffective at managing inter-user interference compared to the ZF-based designs \\ 
\cline{3-5}
&                              
&\vspace{0.01em}\centering NCB~\cite{Interdonato:CAMD:2016} 
& Reduction of uncertainty caused by UE's lack of CSI knowledge 
& less effective at managing inter-user interference compared to the ZF-based designs \\ 
\cline{3-5}
&                              
&\vspace{0.2em}\centering  ECB~\cite{Interdonato:TCOM:2021}
& Boost channel hardening and reducing beamforming gain uncertainty compared to NCB and CB
& less effective at managing inter-user interference compared to the ZF-based designs \\ 
\cline{3-5}
&                             
&\vspace{0.01em}\centering FZF~\cite{Interdonato:TWC:2020} 
&  Mitigation of inter-user interference within each AP’s coverage
&  Performance is constrained by spatial degrees of freedom at APs \\ 
\cline{3-5}
 &                              
 &\vspace{0.1em}\centering PZF~\cite{Interdonato:TWC:2020}  
 & Protect strong UEs against the intra-group interference
 & Non-coherent inter-/intra-group interference among the UEs  \\ 
 \cline{3-5}
 &                              
 &\vspace{0.01em}\centering PPZF~\cite{Interdonato:TWC:2020} 
 & Protect strong UEs against the inter-/intra-group interference 
 & Non-coherent intra-group interference among the weak UEs \\ \hline
\multirow{4}{*}{\vspace{-8em}Uplink} 
&\centering Fully centralized detection (Level $4$) 
&\vspace{-0.1em}\centering  C-MMSE~\cite{Emil:TWC:2020}
& Achieves maximum mutual information of channels
& High computational complexity as well as fronthaul load \\ 
\cline{2-5}
&\vspace{-0.1em}\centering Local processing \& LSFD (Level $3$)                            
&\vspace{-0.1em}\centering L-MMSE, LP-MMSE~\cite{Emil:TWC:2020}  
& Lower overhead and complexity compared to Level $4$ due to local preprocessing at APs 
& High fronthaul load to transfer all channel statistics to CPU for optimal LSFD  design\\ 
\cline{2-5}
&\centering Local processing \& simple centralized decoding (Level $2$)  
&\centering  L-MMSE,~\cite{Emil:TWC:2020} MR~\cite{Fan:ICC:2019} 
& Much lower fronthaul load and signal processing complexity compared to the Level $3$
& Performance degradation with respect to Level $3$ design  \\ 
\cline{2-5}
&\vspace{-0.1em}\centering  Small-cell network (Level $1$)                              
&\centering L-MMSE~\cite{Emil:TWC:2020}, MR  
& Local decoding at AP via local channel estimates without exchange anything with the CPU
& Provides the worst SE performance among all designs \\ 
\hline
\end{tabular}

\label{table2}
\vspace{-1em}
\end{table*}
 \end{flushleft} 

\vspace{-2em}
\subsection{Power Control and Performance Optimization}
Power control can profoundly impact the performance of CF-mMIMO network through the inter-user interference control and network-wide performance optimization. A range of objective functions have been utilized for power control. We discuss here some of the important and related studies in the literature.    

\subsubsection{Downlink Power Control} There exist three main power control formulations in the literature as: 

\textbf{Fairness-oriented, max-min optimization:} This is an egalitarian policy that ensures maximized identical SE throughout the network. This policy suits CF-mMIMO which, by nature, guarantees a more uniform quality of service than co-located mMIMO~\cite{interdonato2019ubiquitous}. For DL transmission in CF-mMIMO systems, the max-min optimization problem is formulated as
\begin{align}~\label{eq:maxmin}
\max_{\Bmu} \Big\{\min_{\ 1\leq k \leq K} \SE_{k,\dl}^\ps (\Bmu^\ps) \vert \Bmu^\ps \in \mathcal{S} \Big\}, 
\end{align}
where the set $\mathcal{S} = \big\{\Bmu^\ps\vert \Bmu^\ps \geq 0; \Vert \Bmu_m^\ps\Vert^2 \leq \zeta^\ps, m=1,\ldots,M \big\}$ defines the feasibility set of the optimization problem,  and $\SE_{k,\dl}^{\ps}$ is the DL SE for a given precoding design $\ps = \{\CB, \NCB, \ECB, \FZF, \PZF, \PPZF\}$. Moreover, $\Bmu^\ps=[\Bmu_1^\ps;\ldots;\Bmu_M^\ps]\in \mathbb{R}_{+}^{MK\times 1}$ includes the power control coefficients with $\Bmu_m^\ps=[\mu_{mk}^\ps,\ldots,\mu_{mK}^\ps]$  denoting the vector of all power control coefficients associated with the $m$th AP. 

The max-min fairness problem~\eqref{eq:maxmin} requires a centralized approach. In other words, the APs need to send long-term channel statistics to the CPU, where the power control coefficients are computed and subsequently fed back. On the other hand, solving this problem for large scale networks with a large number of APs and UEs might be prohibitive since the computation time is polynomial in the number of optimization variables, $MK$. These challenges might undermine the system scalability, increase the fronthaul load, and cause latency. Nevertheless, since the computation of power control coefficients depends on the LSF coefficients, they need to be updated whenever there are macroscopic network variations. Therefore, the update frequency of the power control coefficients is relatively low and the optimal power control is practical. Moreover, through the implementation of UC approach, this computation can be confined within few APs. As a result, the network can efficiently manage computational complexity, fronthaul load, and latency. 

The globally optimal solution to~\eqref{eq:maxmin} can be obtained in polynomial time via the bisection method by solving a sequence of convex (more specifically, second-order cone) feasibility problems. Ngo~\ettall~\cite{Hien:cellfree} provided the solution for the CB scheme with single antenna APs. Interdonato~\ettall~\cite{Interdonato:TCOM:2021} presented a detailed description of the bisection search algorithm for the NCB and ECB schemes, as well as the CB scheme with DL training and for multi-antenna APs. The required computational complexity of the second-order interior-point methods used in~\cite{Hien:cellfree,Interdonato:TCOM:2021} and memory requirement scale quickly with the problem size. Farooq~\ettall~\cite{Hien:Utility:TCOM:2021} developed an accelerated projected gradient method to solve~\eqref{eq:maxmin}. Particularly, each iteration of the proposed iterative power control algorithms is given in closed-form and can be done in parallel. 

The max-min fairness power control problem for ZF-based precoding designs, including FZF, PZF, and PPZF was studied in~\cite{Hien:cellfree}. Driven by the structural similarity to the CB-based problem in~\eqref{eq:maxmin}, the problem was reformulated as a second-order cone feasibility problem and solved through the bisection method. Du~\ettall~\cite{Liutong:TCOM:2021} addressed the max-min power control problem for joint MRT and ZF precoding design. They employed both second-order cone and first-order methods to tackle the problem. The former approach can yield the globally optimal solution but comes with a significantly high computational complexity. On the other hand, the latter technique offers a suboptimal solution with very low computational complexity.

\textbf{Sum-SE maximization:} In distributed wireless systems, providing fairness to UEs having ``bad" channels, may seriously compromise the overall network performance~\cite{interdonato2019ubiquitous}. Conversely, sum-SE maximization prioritize UEs with ``good" channel conditions by maximizing the throughput and without any fairness guarantees. The problem of sum-SE maximization is given by
\vspace{-1em}
\begin{align}~\label{eq:maxsum}
\max_{\Bmu} \bigg\{\sum\nolimits_{ k=1}^{ K} \SE_{k,\dl}^\ps (\Bmu^\ps) \vert \Bmu^\ps \in \mathcal{S} \bigg\}. 
\end{align}
An accelerated project gradient method was proposed in~\cite{Hien:Utility:TCOM:2021}, to solve~\eqref{eq:maxsum}, where CB has been adopted at APs. A combination of sequential and alternating optimization was used in~\cite{Buzzi:PIMRC:2017} to derive the solution of~\eqref{eq:maxsum}. Chakraborty~\ettall~\cite{Chakraborty:JOPC:2021} developed two  low-complexity algorithms for sum-SE power optimization inspired by weighted minimum mean square error (WMMSE) minimization and fractional programming. The authors used  the alternating direction method of multipliers to solve specific convex subproblems in the proposed algorithms.

\textbf{Proportional fairness maximization:} This optimization provides excellent performance balance between max-min fairness and sum-SE maximization. It is also equivalent to the geometric mean  maximization, and is given by~\cite{Hien:Utility:TCOM:2021}
\begin{align}~\label{eq:geometricmin}
\max_{\Bmu} \bigg\{\sum\nolimits_{ k=1}^{ K} \log_2\big(\SE_{k,\dl}^\ps (\Bmu^\ps)\big) \vert \Bmu^\ps \in \mathcal{S} \bigg\}. 
\end{align}
Proportional fairness maximization gives rise to rates for UEs suffering from poor channel conditions without enforcing QoS constraints, thus maintaining a good sum-rate in the network. Farooq~\ettall~\cite{Hien:Utility:TCOM:2021} proposed an accelerated projected gradient method to solve a geometric mean  maximization problem for DL CF-mMIMO systems with CB at the APs. The proposed method exhibits significantly reduced runtime compared to the existing solutions, which often rely on successive convex approximation and use off-the-shelf convex solvers, and implement an interior-point algorithm to solve the derived convex problems. Tuan~\ettall~\cite{Tuan:TCOM:2022} developed a scalable algorithm to solve~\eqref{eq:geometricmin}, where a CB scheme is applied at the APs. This algorithm iterates across linear-complex closed-form expressions, making it practical and applicable regardless of the network's scale.

\textbf{Average energy efficiency (EE) maximization:}
An important issue in deploying CF-mMIMO systems is the high power consumption, which is proportional to the number of APs.  In addition to the transmit power and hardware dissipation, the fronthaul links in CF-mMIMO systems can potentially increase the total power consumption to a level that can undermine the achieved SE gains. This issue has raised the question around the suitability of CF-mMIMO for green communications in terms of the total EE, which is defined as how many bits can be transmitted by one Joule.
Specifically, the total EE (bit/Joule) can be calculated as
\begin{align}~\label{eq:EE:SR}
    \mathtt{EE} = \frac{B_w \times \SE }{P_{\mathtt{total}}},
\end{align}
where $B_w$ is the system bandwidth, $\SE$ denotes the overall SE of the system, and $P_{\mathtt{total}}$ denotes the total power consumption. 

A general power consumption model for CF-mMIMO systems was provided in~\cite{Mohammad:JSAC:2023}, capturing both UL and DL power requirements in the network. The total power consumption $P_{\mathtt{total}}$ consists of four elements: 1) power consumption for transmitting signals and the required power consumption to run circuit
components for the UL transmission at UL UEs, 2) power consumption to run circuit components for the DL transmission at DL UEs, 3) power consumption at APs that includes the power consumption of the transceiver chains (e.g., converters, mixers, and filters) and the power consumed for the DL or UL transmission, 4) power consumption of the fronthaul signal load to each AP $m$, which is proportional to the fronthaul rate through the traffic-dependent fronthaul power as $P_{\mathtt{bh},m} = P_{0,m} + B_{w} \SE P_{\mathtt{bt},m}$, where $P_{0,m}$ is the fixed power consumption of each fronthaul (traffic-independent power) which may depend on the distances between the APs and the CPU and the system topology and $P_{\mathtt{bt},m}$ is the traffic-dependent power (in Watt per bit/s)~\cite{Hien:TGCN:2018}. 

Regarding the energy consumption, there are several works that have studied the EE optimization problem of CF-mMIMO systems~\cite{Hien:TGCN:2018,Kamel:CLET:2017,Papazafeiropoulos:TGCN:2021,Tuan:TCOM:2022,Chien:TWC:2020,mai2022energy}. Nguyen~\ettall~\cite{Kamel:CLET:2017} proposed a low-complexity power control technique with centralized ZF precoding design to maximize the EE of CF-mMIMO, taking into account the fronthaul power consumption and the imperfect channel state information. Papazafeiropoulos~\ettall~\cite{Papazafeiropoulos:TGCN:2021} formulated a DL EE problem, assuming that the multi-antenna APs follow a Poisson point process. Analytical expressions formulated based on the optimal pilot reuse factor, the AP density, and the number of AP antennas and UEs that maximize the EE, were derived. The results in~\cite{Papazafeiropoulos:TGCN:2021} showed that an optimal pilot reuse factor and AP density exist, while larger values result in an increase of the interference, and subsequently, lower EE. Mai~\ettall~\cite{mai2022energy}  studied the DL EE optimization under a sum power constraint at each AP and a QoS constraint at each UE. They developed an iterative power control algorithm based on the framework of accelerated projected gradient method. This method serves as an alternative solution to the sequential second-order cone programs based method for addressing the high computational complexity in large-scale CF-mMIMO. Tuan~\ettall~\cite{Tuan:TCOM:2022} introduced the geometric mean EE concept, as the ratio of the geometric mean rate to the total power consumption, as a meaningful index for quantifying the EE. Through simulation results, the authors showed that maximization of the geometric mean EE still maintains the QoS to all UEs. This finding is in contrast to the conventional EE index in~\eqref{eq:EE:SR}, which is a meaningful index only under additional QoS constraints.

To enhance the EE even further, considering that only a subset of the APs is likely required to meet the UE’ performance demands, implementing AP selection can significantly reduce the power consumption attributed to the fronthaul. Inspired by this fact, Ngo~\ettall~\cite{Hien:TGCN:2018} proposed two AP selection schemes, namely received-power-based selection and LSF-based selection scheme. They developed an optimal power allocation algorithm at the selected multi-antenna APs, aiming at maximizing the total EE, subject to a per-UE SE constraint and a per-AP power constraint. The effects of channel estimation, power control, non-orthogonality of pilot sequences, and fronthaul power consumption were taken into consideration. Chien~\ettall~\cite{Chien:TWC:2020} minimized the total power consumption optimization in CF-mMIMO networks by jointly optimizing the DL transmit powers and the number of active APs, while satisfying the SEs requested by all the UEs. They found a globally optimal solution by formulating the considered problem as a mixed-integer second-order cone program by utilizing the branch-and-bound approach.

\subsubsection{Uplink Power Control} In UL CF-mMIMO networks, different power control (and receive filtering)  designs have been investigated with different utility functions, including 1) \textbf{max-min fairness optimization}, 2) \textbf{sum-SE maximization}, 3)\textbf{ proportional fairness maximization},  and 4) \textbf{total power minimization}. More specifically, Bashar \ettall~\cite{Bashar:TWC:2019} solved a max-min fairness problem by alternatively solving two subproblems: (i) the receiver coefficient design, and (ii) the power control problem, which was solved using geometric programming (GP). A similar alternating optimization 
based scheme was used in~\cite{Bashar:WCL:2018} to address a mixed QoS problem, including the max-min fairness for a set of UEs and a fixed QoS for the remaining UEs. Again, the power control subproblem was solved using GP. The authors in~\cite{Nguyen:ACCESS:2018} examined the impact of AP selection along with three power control design problems, total power minimization with predefined QoS requirements at the UEs, max-min fairness optimization, and sum-SE maximization. A successive approach was deployed to approximate the constraints into the form of GP, and then off-the-shelf convex
solvers were utilized.  Mai~\ettall~\cite{Trang:TVT:2018} proposed a pilot power coefficients design to improve the channel estimation accuracy during the training phase and manage pilot contamination. To reduce the fronthaul load, AP selection is first applied, using a received-power-based selection or LSF-based selection criterion. Then, a min-max optimization problem which minimizes the largest of all UE normalized mean-squared errors was formulated. This optimization problem was solved via sequential convex approximation method. The results indicate that pilot power control is preferable for high mobility environments.

A common feature of all the aforementioned studies is the use of GP by means of off-the-shelf convex solvers, which renders them suitable for small-scale scenarios. To address this challenge, a method based on mirror prox was proposed in~\cite{Farooq:CLET:2022}. This approach involves reformulating the problem as a convex-concave one, allowing the application of the mirror prox algorithm to find a saddle point. The proposed scheme offers an opportunity to assess the performance of CF-mMIMO in large-scale scenarios. 

On the other side, cross-layer flow control and rate allocation in UL co-located and CF-mMIMO systems with randomly arriving data traffic was studied in~\cite{Chen:TGCN:2020}. The authors proposed a dynamic scheduling algorithm that maximizes predefined utility functions including max-min fairness, proportional fairness, and maximum sum-rate. This algorithm is based on Lyapunov optimization theory and determines, at each time slot, the amount of data to admit to the transmission queues and the transmission rates over the wireless channel. The proposed power control algorithm substantially reduces the average delay experienced by the UEs as their locations change over time. It also ensures finite delay in scenarios where the conventional schemes fall short.

Learning-based approaches can be exploited to design the optimal power allocation, potentially lowering the computational complexity, and thereby making the power allocation strategies appealing for online implementation.The state-of-the-art in this paper will be discussed in subsection~\ref{sec:cell-free learning}.

\subsection{Practical Issues/Challenges}
The aforementioned (sophisticated) UL/DL processing designs and optimization frameworks are based on the simplifying, but impractical, assumption of unlimited fronthaul capacity and perfect hardware. On the other hand, the usage of large number of multi-antenna APs with high-quality hardware and high-resolution analog-to-digital converters (ADCs)/ digital-to-analog converters (DACs) entail significant financial cost and energy consumption. Therefore, in this subsection, we delve into these challenges and provide an overview of the existing solutions to address them.   
\subsubsection{Fronthaul-Limited}
In practical scenarios, the communication between the APs and CPU is carried out over finite capacity fronthaul links. As the number of APs and the number of antennas per AP increase, the fronthaul links require significant capacity to ensure precise signal transfer between the APs and the CPU. To address this challenge, one option is to rely on local processing and decoding schemes, though these provide inferior performance compared to centralized processing and decoding schemes. Alternatively, fronthaul data compression has emerged as a viable solution, where only a quantized version of the data and/or channel estimates/pilots is sent to the CPU through the fronthaul link. By using low-resolution ADCs at the APs, the transmitted signal over the fronthaul link is quantized using fewer bits, leading to a lower fronthaul load, reduced power consumption, and decreased hardware cost.

In the literature, four quantization protocols have been proposed for UL transmission:
\begin{enumerate}
    \item \textbf{Quantize-and-estimate}: The APs quantize the received pilot and signal and send them to the CPU. From these quantized received pilots, the CPU performs channel estimation and then constructs the combining vectors for data detection~\cite{Maryopi:TVT:2019, Masoumi:TWC:2020,Bashar:TCOM:2021}.
    \item \textbf{Quantize-and-forward}: This refers to the case where both the channel estimates and the received signals are quantized at the AP and forwarded to a CPU. Signal combining and data detection are then performed at the CPU via combining vectors~\cite{Bashar:TCOM:2019,Masoumi:TWC:2020,Bashar:TCOM:2021}.
    \item \textbf{Combine-quantize-and-forward }: The APs combine the received signals by multiplying them with the combining vectors, and the quantized versions of these combined signals are sent to the CPU for signal detection~\cite{Bashar:TCOM:2019,Bashar:TCOM:2021}.
    \item \textbf{LSF-based quantize-and-forward/combine-quantize-and-forward}: To enhance the performance, the forwarded signal from the APs is multiplied by the receiver filter coefficients at the CPU before data detection~\cite{Bashar:TGCN:2019,Masoumi:TWC:2020}. 
\end{enumerate}

\begin{figure}[t]
	\centering
	\vspace{-0.8em}
	\includegraphics[width=0.50\textwidth]{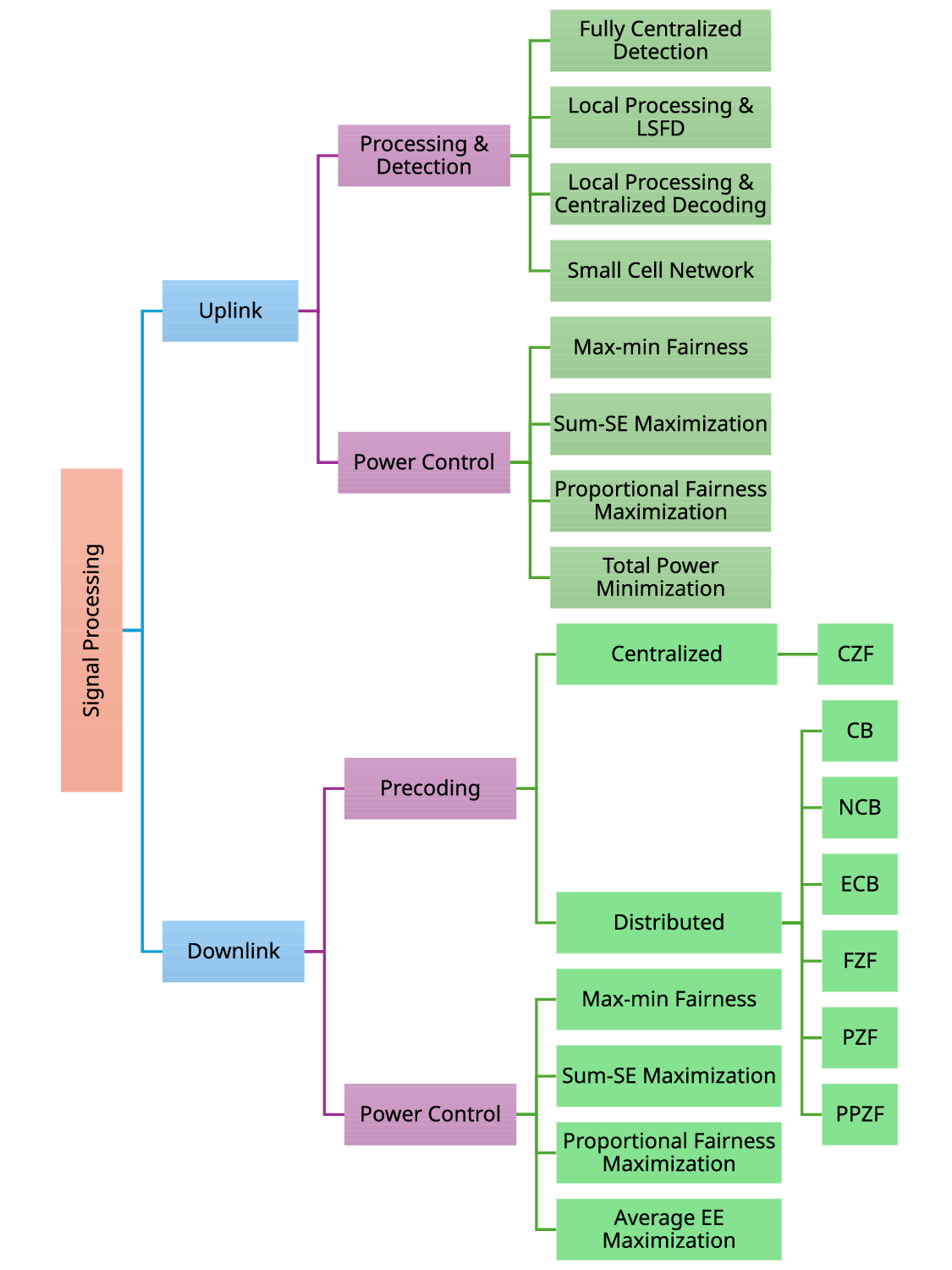}
	\vspace{-0.7em}
	\caption{Summary of the signal processing techniques in CF-mMIMO systems.}
	\vspace{0.4em}
	\label{fig:chart}
\end{figure}

\section{Network-assisted full-duplex CF-mMIMO}~\label{sec:NAFD}
Traditional TDD-based CF-mMIMO systems with half-duplex (HD) APs offer very simple control of UL and DL traffic at UEs.  More specifically, a UE with UL (DL) data demand has to wait for a slot where all APs are operating in the UL (DL) mode in order to complete its transmission. However, in widely-used delay-sensitive services such as cloud storage, video chat, and innovative IoT application, both UL and DL transmissions share equal priority, and, hence, must be carried out simultaneously. To meet this requirement and also to address the SE loss associated with HD transceivers, FD radios can be employed at the APs to enable UL and DL transmissions over the same time-frequency resources~\cite{Sabharwal:JSAC:2014}. This approach holds the potential for achieving nearly a two-fold enhancement in SE compared to the HD counterparts~\cite{tung19ICC,Nguyen:JSAC:2020}.

Despite the aforementioned benefits, FD CF-mMIMO systems suffer from two major challenges: 1) residual self-interference (SI) at the FD APs, i.e., in-band leakage from the transmitter to the receiver, and 2) a large number of APs and legacy UEs gives rise to significant interference termed as cross-link interference (CLI), i.e., the interference received by the receiving antennas of one AP from the transmitting antennas of another AP, as well as the interference received by a DL UE from other UL UEs. Although experimental and practical breakthroughs in SI cancellation techniques~\cite{Duarte:TWC:2012,Hong:CMG:2014}, have rendered the implementation of FD transceivers possible, SI suppression requires power-hungry hardware at the FD APs. Furthermore, CLI results in increased power consumption to achieve the UEs’ SE requirements, consequently leading to a significant reduction in EE. In other words, a critical question is how to support UL and DL communications simultaneously, while maintaining the EE at the optimal level.

Recent studies in~\cite{Nguyen:JSAC:2020,Anokye:TVT:2021,Dey:TWC:2022,Datta:JOP:2022} have delved into addressing the trade-off between the SE-EE through the design of UL and DL power control, as well as AP-UE association and AP selection. However, from the energy consumption perspective, CLI and SI are still the major bottlenecks, rendering the application of FD CF-mMIMO networks impractical for future wireless networks. Fortunately, \textit{network-assisted full duplexing} (NAFD) has recently emerged as a new physical-layer paradigm for CF-mMIMO networks to tackle this issue~\cite{Wang:TCOM:2020}. In NAFD CF-mMIMO, the APs can operate in FD (all APs perform DL and UL at the same time over the same frequency), hybrid-duplex (both FD and HD APs exist in the network), or flexible-duplex (APs operate in HD mode, termed as DL/UL APs), thus, NAFD unifies all duplex modes in the network, cf. Fig~\ref{fig:NAFD_CFmMIMO}. As a result, with a flexible-duplex (hybrid-duplex) architecture, the requirement for FD APs is eliminated (reduced), while both the UL and DL traffic within the network can be managed. Moreover, there is a notable decrease in inter-AP interference, when compared to the FD CF-mMIMO networks. This reduction occurs because only a subset of APs operates in DL, leading to a decrease in interference with the remaining APs operating in UL. Finally, by implementing a dynamic assignment of UL/DL transmission modes at the APs and power control at the DL APs and UL UEs, along with applying LSFD, the management of CLI becomes more effective.
\begin{figure}[!t]\centering \vspace{0em}
    \def\svgwidth{240pt} 
    \fontsize{8}{7}\selectfont 
\begingroup%
  \makeatletter%
  \providecommand\color[2][]{%
    \errmessage{(Inkscape) Color is used for the text in Inkscape, but the package 'color.sty' is not loaded}%
    \renewcommand\color[2][]{}%
  }%
  \providecommand\transparent[1]{%
    \errmessage{(Inkscape) Transparency is used (non-zero) for the text in Inkscape, but the package 'transparent.sty' is not loaded}%
    \renewcommand\transparent[1]{}%
  }%
  \providecommand\rotatebox[2]{#2}%
  \newcommand*\fsize{\dimexpr\f@size pt\relax}%
  \newcommand*\lineheight[1]{\fontsize{\fsize}{#1\fsize}\selectfont}%
  \ifx\svgwidth\undefined%
    \setlength{\unitlength}{1029.97503662bp}%
    \ifx\svgscale\undefined%
      \relax%
    \else%
      \setlength{\unitlength}{\unitlength * \real{\svgscale}}%
    \fi%
  \else%
    \setlength{\unitlength}{\svgwidth}%
  \fi%
  \global\let\svgwidth\undefined%
  \global\let\svgscale\undefined%
  \makeatother%
  \begin{picture}(1,0.652443)%
    \lineheight{1}%
    \setlength\tabcolsep{0pt}%
    \put(0,0){\includegraphics[width=\unitlength]{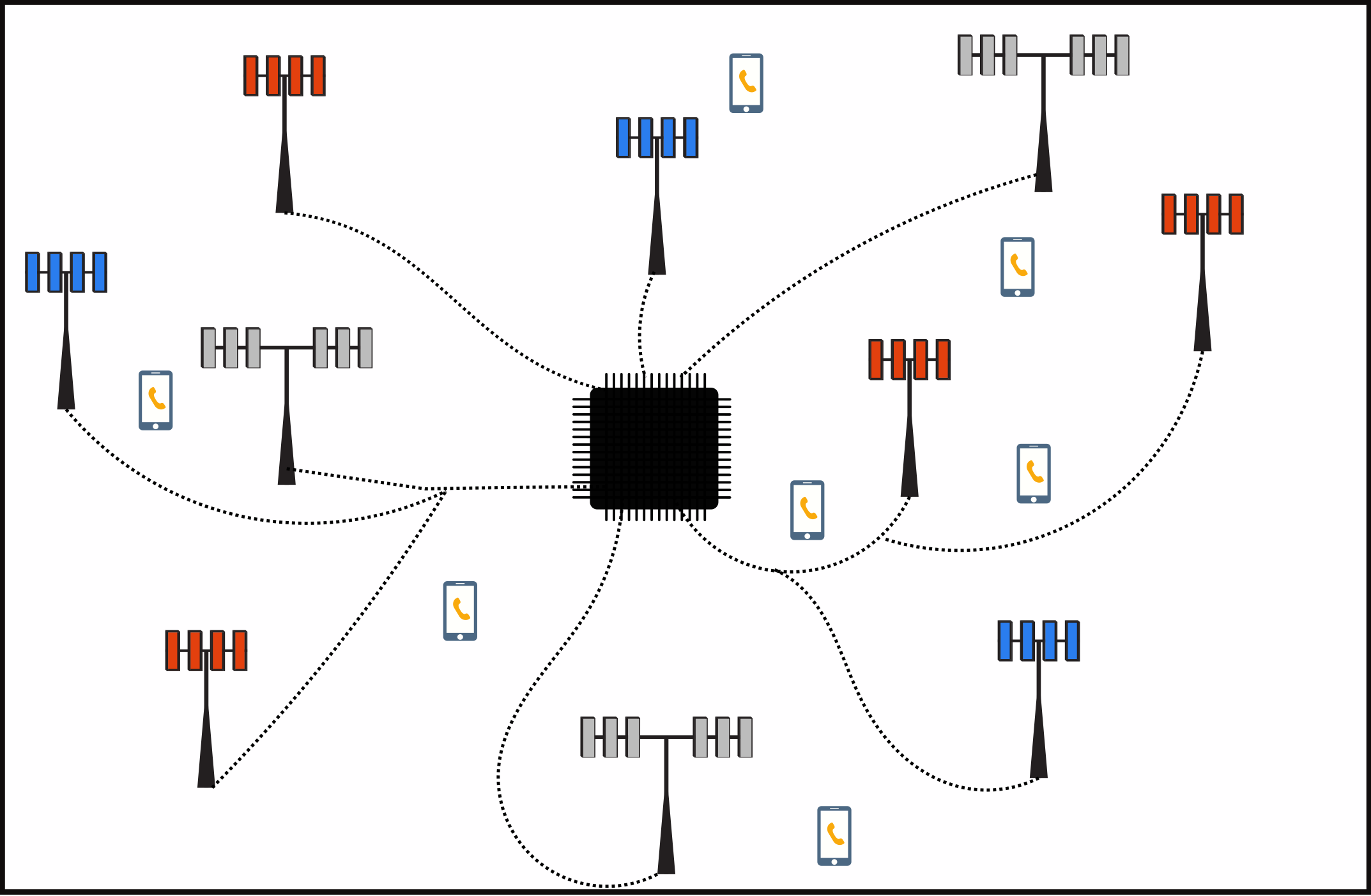}}%
    \put(0.77114382,0.53048798){\color[rgb]{0,0,0}\makebox(0,0)[lt]{\lineheight{1.25}\smash{\begin{tabular}[t]{l}FD AP\end{tabular}}}}%
    \put(0.62555873,0.02537312){\color[rgb]{0,0,0}\makebox(0,0)[lt]{\lineheight{1.25}\smash{\begin{tabular}[t]{l}UE\end{tabular}}}}%
    \put(0.21727642,0.51301003){\color[rgb]{0,0,0}\makebox(0,0)[lt]{\lineheight{1.25}\smash{\begin{tabular}[t]{l}HD UL AP\end{tabular}}}}%
    \put(0.77126088,0.11330517){\color[rgb]{0,0,0}\makebox(0,0)[lt]{\lineheight{1.25}\smash{\begin{tabular}[t]{l}HD DL AP\end{tabular}}}}%
    \put(0.4384769,0.31329983){\color[rgb]{1,1,1}\makebox(0,0)[lt]{\lineheight{1.25}\smash{\begin{tabular}[t]{l}\textbf{CPU}\end{tabular}}}}%
  \end{picture}%
\endgroup%
 \vspace{0mm}
    \caption{Illustration of a NAFD CF-mMIMO system with distributed HD and FD APs.}\vspace{0mm} \label{fig:NAFD_CFmMIMO}
    \vspace{2em}
\end{figure}
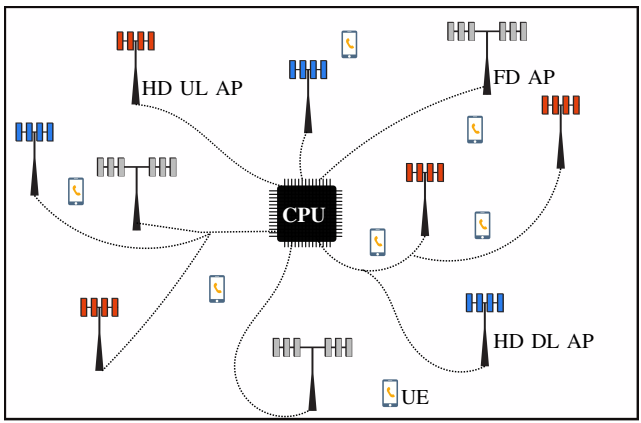
\vspace{-1em}
\subsection{Literature Review}
The concept of NAFD share similarities with the existing nomenclature in the literature, such as dynamic-TDD~\cite{Haas:JSAC:2001,Liu:TCOm:2017,Jiamin:TWC:2018}, dynamic UL-DL configuration in time-division Long Term Evolution~\cite{Shen:MCOM:2012}, coordinated multipoint for In-band wireless FD (CoMPflex)~\cite{Thomsen:WCL:2016}, and bidirectional dynamic networks~\cite{Xin:ACCESS:2017},  which were proposed to accommodate the asymmetric UL-DL data traffics in wireless (cellular) networks. In dynamic-TDD, based on the UL-DL traffic demands within each cell, a specific time slot can be dynamically allocated for either reception or transmission. Nevertheless, this scheme does not fully cater to heterogeneous data demands within the cells, as UL (DL) UEs within a cell assigned a DL (UL) time slot must wait for the UL (DL) time slot to transmit their data. Moreover, the effectiveness of the dynamic-TDD is constrained by inter-BS interference and the need for strict synchronization among cells, which necessitates cell-cooperation. This results in significant overhead and complexity. The rationale behind the dynamic UL-DL configuration in time-division Long Term Evolution is to establish opposite transmission directions across various small cells in heterogeneous cellular networks~\cite{Shen:MCOM:2012}. CoMPflex, which is motivated by the coordinated multipoint concept in cellular networks, involves emulating an FD BS and using two spatially separated and coordinated HD BSs~\cite{Thomsen:WCL:2016}. In bidirectional dynamic networks, the number of UL and DL remote radio heads is adjusted flexibly to facilitate simultaneous UL and DL communications~\cite{Xin:ACCESS:2017}.

\begin{table*}
	\centering
	\caption{\label{tabel:NAFD}Summary of NAFD CF-mMIMO Literature}
	\vspace{-0.6em}
	\small
\begin{tabular}{|m{0.7cm}|m{0.9cm}|m{0.9cm}|>{\centering\arraybackslash}m{1cm}|>{\centering\arraybackslash}m{1cm}|>{\centering\arraybackslash}m{1cm}|>{\centering\arraybackslash}m{8cm}|}
	\hline
	\multirow{2}{*}{\textbf{Ref.}} & \multicolumn{2}{c|}{\textbf{CSI}} & 
	\multicolumn{3}{c|}{\textbf{Architecture}} & \multirow{2}{*}{\textbf{Technical Contribution}}\\

	\cline{2-6}
	\centering
	  &\textbf{Stat.}  &\textbf{Instant.} &\textbf{Flexible-duplex} &\textbf{Hybrid-duplex} &\textbf{FD}  & \\
\hline
	\hline
	\centering

~\cite{tung19ICC} &\centering\checkmark   &\centering- &\centering- &\centering- &\centering\checkmark &SE performance with MR combining and precoding at the receive and transmit side of the FD APs \\
	\hline

~\cite{Nguyen:JSAC:2020} &\centering\checkmark   &\centering- &\centering- &\centering- &\centering\checkmark &Joint optimization of power control, AP-UE association and AP selection to reduce the network interference\\
	\hline

~\cite{Anokye:TVT:2021} &\centering\checkmark   &\centering- &\centering- &\centering- &\centering\checkmark &Characterizing joint impact of the residual SI, CLI, pilot contamination and quantization noise by deriving closed-form solutions for the UL/DL SE\\
	\hline
 
~\cite{Dey:TWC:2022} &\centering\checkmark   &\centering- &\centering- &\centering- &\centering\checkmark &SE/EE analysis and optimization in the presence of implications of low-resolution ADCs at the APs and UEs, with DL pilot transmission\\
	\hline
 
~\cite{Wang:TCOM:2020} &\centering-   &\centering\checkmark &\centering\checkmark &\centering- &\centering- & DL/UL SE analysis with centralized UL and DL signal processing with inter-AP interference mitigation at the CPU\\
	\hline	

~\cite{Jiamin:TWC:2021} &\centering\checkmark  &\centering- &\centering\checkmark &\centering- &\centering- &Coherent data detection at DL UEs and centralized inter-AP interference cancellation at the CPU using estimated CSI  \\
	\hline

~\cite{Xia:China:2021} &   &\centering\checkmark & &\centering\checkmark & &Duplex mode selection and transceiver design to maximize the aggregated SE of DL and UL \\
	\hline

~\cite{Zhu:COML:2021} &\centering-   &\centering\checkmark &\centering\checkmark &\centering- &\centering- & Duplex mode optimization at the APs to enhance the aggregated SE of DL and UL \\
	\hline

~\cite{Chowdhury:TCOM:2022} &\centering\checkmark   &\centering- &\centering\checkmark &\centering- &\centering- &Suboptimal greedy algorithm for dynamic AP-scheduling to maximize the aggregated SE of DL and UL\\
	\hline
 
~\cite{Mohammad:JSAC:2023} &\centering\centering\checkmark   &\centering- &\centering\checkmark &\centering- &\centering- &SE/EE enhancement via AP mode assignment, UL and DL power control, and LSFD design, considering the QoS requirements of the UEs \\
	\hline

	\hline

~\cite{Song:SYST:2023} &\centering-   &\centering\checkmark &\centering\checkmark &\centering- &\centering- &Design bit allocation algorithm for low-resolution ADCs at the APs to maximize the total SE/EE \\
	\hline
 
\end{tabular}
\vspace{0em}
\end{table*}

\subsubsection{Performance Analysis and Channel Estimation} Inspired by the aforementioned concepts that aimed at supporting simultaneous UL-DL traffic in wireless networks, the proposal for NAFD CF-mMIMO systems was introduced in~\cite{Wang:TCOM:2020}. The authors assumed that the transmission mode of the APs is fixed and derived the sum-rate for UL with MMSE receiver as well as DL with RZF and ZF precoders. They also proposed a genetic algorithm based user scheduling strategy to alleviate the inter-AP interference. Li~\emph{et al.}~\cite{Jiamin:TWC:2021}  proposed to utilize a beamforming training scheme, which estimates inner products of beamforming and channel vectors, to perform centralized interference cancellation among the APs and coherent decoding at DL UEs. The CPU performs
interference cancellation by subtracting the reconstructed interference signal from the received signal from UL APs. Moreover, closed-form expressions for the DL ergodic achievable rates with MRT and ZF beamforming under different CSI were derived.

\subsubsection{AP Mode Assignment} The main idea of AP mode assignment is to facilitate simultaneous UL and DL transmissions. If APs can dynamically select the slots for operating in either UL or DL modes, any UE with a specific data demand can locate some nearby APs operating in the corresponding mode in the same slot. Xia \emph{et al.}~\cite{Xia:China:2021}  concentrated their attention on the duplex mode selection and transceiver design in NAFD CF-mMIMO systems, where each antenna at the APs can operate in three modes, i.e., UL reception, DL transmission, and sleep. They formulated a mixed-integer optimization problem to maximize the aggregated SE of DL and UL, where the QoS constraints and power budget constraints are considered. Zhu~\emph{et al.}~\cite{Zhu:COML:2021} proposed two algorithms based on parallel successive convex approximation and a reinforcement learning (RL) algorithm based on enhanced Q-learning to dynamically optimize the duplex mode of APs with the objective of SE enhancement in NAFD CF-mMIMO systems. Chowdhury~\emph{et al.}~\cite{Chowdhury:TCOM:2022} formulated the AP-scheduling problem with the objective of  maximizing the sum of UL and DL SEs, where the UL and DL SEs were derived with the use-and-then-forget capacity bounding technique. They developed a  greedy algorithm for dynamic AP-scheduling where, at each step, the transmission mode of the AP that maximizes the incremental SE is added to the already scheduled AP-subset. A framework for joint AP mode assignment (UL or DL mode for HD APs), power control, and LSFD design  was developed for NAFD CF-mMIMO systems in~\cite{Mohammad:JSAC:2023}. This framework aims to maximize the total SE/EE of the system while ensuring the QoS requirements for all UEs in the network.  
\subsubsection{Hardware Impairments} Song~\emph{et al.}~\cite{Song:SYST:2023} extended the results in~\cite{Jiamin:TWC:2021} and investigated the impact of low-resolution ADC on the SE and EE performance of the system. In order to maximize the total SE/EE, the authors in~\cite{Song:SYST:2023} proposed a bit allocation algorithm for low-resolution ADCs based on a deep Q network. Table~\ref{tabel:NAFD} shows a summary of the major related works on NAFD CF-mMIMO systems.

\subsection{Case Study and Discussion}
Consider a NAFD CF-mMIMO system  under TDD operation, where $M$ APs serve $K_u$ UL UEs and $K_d$ DL UEs, c.f. Figure~\ref{fig:NAFD_CFmMIMO}. Each UE is equipped with one single antenna, while each AP is equipped with $N$ antennas. All APs and UEs are HD devices. Local CB is applied at the APs for UL reception and DL transmission. Assume that the binary element UL and DL mode assignment vectors of size $M\times 1$ are denoted by $\qa$ and $\qb$, where $a_m=1$ ($b_m=1$) indicates that AP $m$ operates in the DL (UL) mode and otherwise if $a_m=0$ ($b_m=0$). Moreover, let $\THeta=\{\theta_{mk}\}$ denote the power control coefficients at AP $m\in\mathcal{M}=\{1,\ldots,M\}$ for DL UE $k\in\mathcal{K}_d=\{1,\ldots,K_d\}$, $\VARSIGMA=\{\varsigma_k\}$ present the power control coefficients at UL UE $k\in\mathcal{K}_u=\{1,\ldots,K_u\}$, while $\boldsymbol \alpha=\{\alpha_{mk}\} $ denotes the LSFD weight at AP $m$ for UL UE $k$. We consider the sum-SE maximization problem, subject to the power constraint at the APs and per-UE QoS constraint, with the optimization variables $\qx=\{\qa, \qb, \THeta, \VARSIGMA, \boldsymbol \alpha\}$ as
\begin{subequations}\label{P:SE}
	\begin{align}
		\underset{\qx}{\max}\,\, &
         \!\sum\nolimits_{\ell\in\mathcal{K}_u}\!\! \mathcal{S}_{\ul,\ell}^{\vFD} (\qb, \boldsymbol \varsigma, \boldsymbol \theta, \boldsymbol \alpha)  \! + \! \sum\nolimits_{k\in\mathcal{K}_d}\!\!\mathcal{S}_{\dl,k} ^{\vFD}(\qa, \boldsymbol \theta, {\boldsymbol{\varsigma}})
		\\
		\mathrm{s.t.} \,\,
		&\sum\nolimits_{k\in\mathcal{K}_{d}} \gamdmk \theta_{mk}^2 \leq \frac{1}{\Ntx}, \forall m\label{eq:per-APpower}
		\\
        &|\alphml|^2\leq 1,  \quad \forall \ell, m
        \\
        &
        0\leq {\varsigma}_{\ell} \leq 1, \forall \ell,
        \\
        & a_m + b_m = 1, \forall m,\label{eq:ambm}
        \\
        &\theta_{mk} = 0, \forall k,\,\, \text{if}\,\, a_m = 0,  \forall m,\label{eq:AP:DL}
        \\
		& \mathcal{S}_{\ul,\ell}^{\vFD} (\qb, \boldsymbol \varsigma, {\boldsymbol{\theta}},\boldsymbol \alpha) \geq \mathcal{S}_\ul^o,~\forall \ell
		\label{UL:QoS:cons}
		\\
		&\mathcal{S}_{\dl,k}^{\vFD} (\qa, \boldsymbol \theta, {\boldsymbol{\varsigma}}) \geq  \mathcal{S}_\dl^o,~\forall k, 
		\label{DL:QoS:cons}
	\end{align}
\end{subequations}
where  $\mathcal{S}_\ul^o$ and $\mathcal{S}_\dl^o$ are the minimum SE required by the $\ell$-th UL UE and $k$-th DL UE, respectively, to guarantee the QoS in the network. Constraint~\eqref{eq:per-APpower} represents the power constraint at the APs,~\eqref{eq:ambm} guarantees that AP $m$ only operates in either the DL or UL mode,~\eqref{eq:AP:DL} ensures that if AP $m$ does not operate in the DL mode, its transmit power is zero. Moreover, the UL and DL SE can be expressed as~\eqref{eq:SINRulell} at the top of the next page,  where $\gamdmk \triangleq
\frac{{\tauup\rho_p}(\betamkd)^2} {\tauup\rho_p\betamkd+1}, \gamuml \triangleq \frac{{\tauup\rho_p}(\betamlu)^2}{\tauup\rho_p
\betamlu+1}$, with $\rho_p = p_p/\Sn$ being the normalized transmit power of each pilot symbol, $\betamkd$ ($\betamlu$) denotes the LSF coefficient between AP $m$ and DL UE $k$ (UL UE $\ell$), and $\beta_{mi}$ presents the LSF coefficient between AP $m$ and AP $i$, $m\neq i$.

\begin{figure}[t]
	\centering
	\vspace{-0.25em}
	\includegraphics[width=0.47\textwidth]{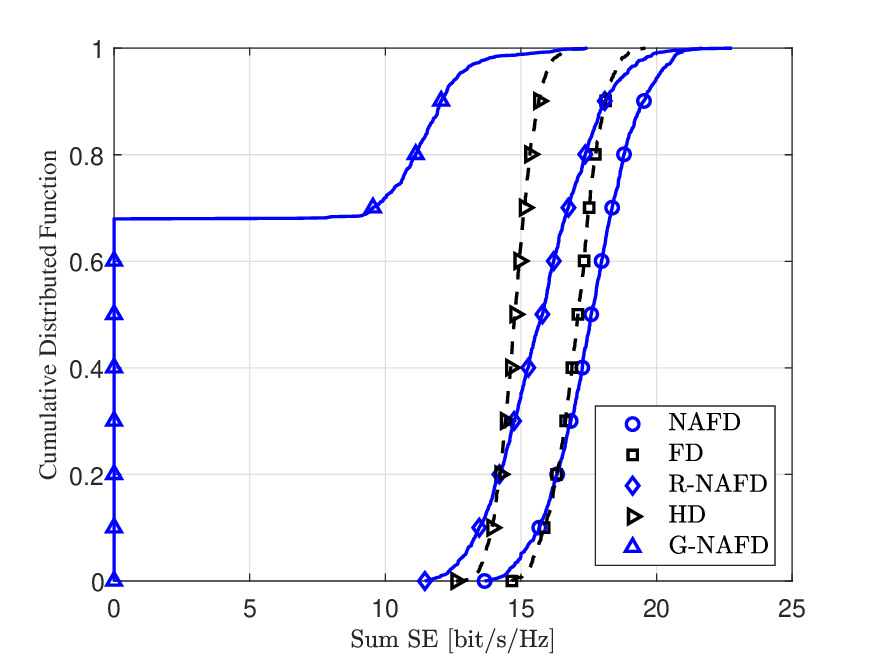}
	\vspace{-0.7em}
	\caption{CDF of the sum SE over different network structures with $M=40$, $K_d=K_u=5$, $N=2$, $\tau_c=200$, $\tauup =10$, $p_u=p_p=0.1$ W, $p_d=1$ W, $\mathcal{S}_\dl^o=\mathcal{S}_\ul^o =0.2$ bit/s/Hz, $N_r=N_t=1$, and $\sigma_{\mathtt{SI}}^2/\Sn=50$ dB.}
	\vspace{1em}
	\label{fig:NAFD}
\end{figure}

\begin{figure*}
\begin{align}~\label{eq:SINRulell}
\mathcal{S}_{\ul,\ell}^{\vFD} (\qb, \boldsymbol \varsigma, {\boldsymbol{\theta}},\boldsymbol \alpha)&\!=\! \frac{\tau_c\!-\!\tauup}{\tau_c}
\log_2 \!\left(\!1\!+\!
\frac{
	\Nrx \rho_{u} \left(\sum\limits_{\substack{m\in\mathcal{M}}} \sqrt{b_m \varsigma_{\ell}} \alphml \gamuml \right)^2
	}
	{\rho_{u}\!\!\!\!
		\sum\limits_{\substack{m\in\mathcal{M}}}
		 \sum\limits_{q\in\mathcal{K}_u}\!\!
		\!\!
		b_m \varsigma_{q}
		\alphml^2
		\betamlu
		\gamma_{m\ell}^{\ul}
		\!+\!
		\rho_{d}\Ntx \!\!\!
		\!\sum\limits_{\substack{m\in\mathcal{M}}}
		\sum\limits_{\substack{i\in\mathcal{M}}}
		\sum\limits_{k\in\mathcal{K}_d}\!\!\!\!
		b_m \theta_{ik}^2 \alphml^2  \gamuml \beta_{mi} \gamma_{ik}^{\dl}
		\!+\!\!\!\!\!
		\sum\limits_{\substack{m\in\mathcal{M}}}\!\!\!\!		
		b_m\alphml^2\gamuml}\!\!\right)\!,
  \nonumber\\
  \mathcal{S}_{\dl,k}^{\vFD} (\qa, \boldsymbol \theta, {\boldsymbol{\varsigma}}) &=  \frac{\tau_c-\tauup}{\tau_c}
\log_2 \left(1  
    + \frac{ \big(\Ntx \sqrt{\rho_{d}}
                           \sum_{{m\in\mathcal{M}}}
                            \theta_{mk} \gamdmk\big)^2}
                            {\rho_{d}\Ntx
    \!\sum_{k'\in\mathcal{K}_{d}}\!\sum_{m\in\mathcal{M}}\!
    \theta_{mk'}^2 \betamkd\gamdmkp + \rho_u\!\sum_{\ell\in\mathcal{K}_u} \! {\vsl} \betakldu \!+\! 1}
	\right),
  \end{align}
  	\hrulefill
	\vspace{-4mm}
  \end{figure*}

Note that the optimization problem~\eqref{P:SE} encompasses the FD CF-mMIMO network as a special case. By setting $a_m=b_m=1$, $\forall m\in\mathcal{M}$ and $\beta_{mm} =\sigma_{\mathtt{SI}}^2$ into~\eqref{eq:SINRulell}, the UL and DL SEs, termed as $\mathcal{S}_{\ul,\ell}^{\FD} (\boldsymbol \varsigma, {\boldsymbol{\theta}},\boldsymbol \alpha)$ and, $ \mathcal{S}_{\dl,k}^{\FD} (\boldsymbol \theta, {\boldsymbol{\varsigma}})$ are obtained. By substituting $\mathcal{S}_{\ul,\ell}^{\vFD} (\qb, \boldsymbol \varsigma, {\boldsymbol{\theta}},\boldsymbol \alpha)$ with $\mathcal{S}_{\ul,\ell}^{\FD} (\boldsymbol \varsigma, {\boldsymbol{\theta}},\boldsymbol \alpha)$ and $ \mathcal{S}_{\dl,k}^{\vFD} (\qa, \boldsymbol \theta, {\boldsymbol{\varsigma}})$ with $ \mathcal{S}_{\dl,k}^{\FD} (\boldsymbol \theta, {\boldsymbol{\varsigma}})$ in equation~\eqref{P:SE}, and then removing the constraints~\eqref{eq:ambm} and~\eqref{eq:AP:DL}, we obtain the sum-SE maximization problem for FD CF-mMIMO systems.  

When HD APs are deployed in the network, the DL-and-UL payload data transmission phase is divided into two equal time fractions of length $(\tau_c-\tauup)/2$. Each UL or DL data transmission is performed in one time fraction and each AP uses $N_t$ transmit antennas and $N_r$ receive antennas to support UL and DL communications simulatanously.  In each time fraction, all UL
or DL UEs are served by all the APs, i.e., $a_m = b_m = 1, \forall m$. There is no interference from the UL UEs to DL UEs, and from the DL APs to UL APs. Furthermore, there is an additional factor of $\frac{1}{2}$ applied in the SE expression. Therefore, the sum-SE maximization problem is reduced to 
\begin{subequations}\label{P:SE:HD}
	\begin{align}
		\underset{\{\THeta, \VARSIGMA, \boldsymbol \alpha\} }{\max}\,\, &
         \!\sum\nolimits_{\ell\in\mathcal{K}_u}\!\! \mathcal{S}_{\ul,\ell}^{\HD} (\boldsymbol \varsigma, \boldsymbol \theta, \boldsymbol \alpha)  \! + \! \sum\nolimits_{k\in\mathcal{K}_d}\!\!\mathcal{S}_{\dl,k} ^{\HD}(\boldsymbol \theta, {\boldsymbol{\varsigma}})
		\\
		\mathrm{s.t.} \,\,
		&\sum\nolimits_{k\in\mathcal{K}_{d}} \gamdmk \theta_{mk}^2 \leq \frac{1}{\Ntx}, \forall m\label{eq:per-APpower}
		\\
        &|\alphml|^2\leq 1,  \quad \forall \ell, m
        \\
        &
        0\leq {\varsigma}_{\ell} \leq 1, \forall \ell,
        \\
		& \mathcal{S}_{\ul,\ell}^{\HD} (\boldsymbol \varsigma, {\boldsymbol{\theta}},\boldsymbol \alpha) \geq \mathcal{S}_\ul^o,~\forall \ell
		\label{UL:QoS:cons}
		\\
		&\mathcal{S}_{\dl,k}^{\HD} (\boldsymbol \theta, {\boldsymbol{\varsigma}}) \geq  \mathcal{S}_\dl^o,~\forall k, 
		\label{DL:QoS:cons}
	\end{align}
\end{subequations}
where 
\begin{align*}
\mathcal{S}_{\dl,k}^{\HD} (\boldsymbol \theta, {\boldsymbol{\varsigma}}) &=  \frac{\tau_c-\tauup}{2\tau_c} \log_2 \Big(1 + \text{SINR}_{\dl,k}^{\HD} (\boldsymbol \theta, {\boldsymbol{\varsigma}}) \Big)\\
\mathcal{S}_{\ul,\ell}^{\HD} (\boldsymbol{\varsigma}, \boldsymbol{\theta}, \boldsymbol{\alpha} )
	&=\! \frac{\tau_c\!-\!\tauup}{2\tau_c}\log_2
	\Big(1+ \text{SINR}_{\ul,\ell}^{\HD}(\boldsymbol{\varsigma}, \boldsymbol{\theta}, \boldsymbol{\alpha} ) \Big),      
\end{align*}
with the effective SINR expressions
\begin{subequations}
  \begin{align}
  &\text{SINR}_{\dl,k}^{\HD} (\boldsymbol \theta, {\boldsymbol{\varsigma}}) \triangleq \frac{N_t^2 \rho_{d} \left(
                           \sum_{{m\in\mathcal{M}}}
                            \theta_{mk} \gamdmk \right)^2 }{\rho_{d} N_t
    \sum_{k'\in\mathcal{K}_{d}}\sum_{m\in\mathcal{M}}
    \theta_{mk'}^2 \betamkd\gamdmkp +
    1}\vspace{0em},\\
    &\text{SINR}_{\ul,\ell}^{\HD} (\boldsymbol \theta, {\boldsymbol{\varsigma}},\boldsymbol{\alpha}) 
\triangleq
\nonumber\\
&\hspace{1em}\frac{
	N_r \rho_{u} \left(\sum_{\substack{m\in\mathcal{M}}} \sqrt{ \varsigma_{\ell}} \alphml \gamuml \right)^2
	}
	{\!\rho_{u}
		\! \sum_{\substack{m\in\mathcal{M}}}
		\! \sum_{q\in\mathcal{K}_u}\!
		\!
		\varsigma_{q}
		\alphml^2
		\betamlu
		\gamma_{m\ell}^{\ul}
		\!+\!
		\!\sum_{\substack{m\in\mathcal{M}}}
		\!
		\alphml^2 \!\gamuml}.
\end{align}  
\end{subequations}

Figure~\ref{fig:NAFD} illustrates the CDF of the total SE across various network structures. The results obtained from solving the proposed scheme in~\eqref{P:SE} are denoted as 'NAFD.' In the 'R-NAFD' scheme, the AP mode assignment vectors ($\qa, \qb$) are randomly assigned, while we optimize the power control coefficients ($\boldsymbol \theta, {\boldsymbol{\varsigma}}$) and LSFD weights $\boldsymbol \alpha$. On the other hand, the 'G-NAFD' scheme employs a greedy algorithm introduced in~\cite{Chowdhury:TCOM:2022} for AP mode assignment, with fixed power control coefficients and LSFD weights, i.e., $\theta_{mk} = \frac{a_{m}}{\sqrt{N K_d \gamdmk}}$, $\varsigma_{\ell} = 1$, $\alpha_{m\ell}=1$ for all $m$, $k$, and $\ell$.  For a fair comparison, FD scheme deploys the same number of antennas as the other schemes, i.e.,
$N=N_r=N_t$, which is called a ``antenna-number-preserved" condition~\cite{himal14TWC,mohammad18TWC}. It is observed that NAFD, with both optimal and random AP mode assignments, can significantly enhance the sum SE, when compared to the HD scenario. Furthermore, NAFD with optimal AP mode assignment outperforms the FD scenario in terms of total SE. The superiority of NAFD over FD becomes even more evident when we consider the EE. More specifically, FD transceivers consume significantly more power than HD ones, resulting in inferior EE compared to NAFD~\cite{Mohammad:JSAC:2023}.

\subsection{Future Research Directions}
NAFD CF-mMIMO can efficiently handle both asymmetric UL and DL traffic over the same time-frequency resources. Inter-AP interference continues to be a significant bottleneck in such networks, resulting in a severe degradation of the UL performance. Consequently, exploring methods to mitigate or suppress this inter-AP interference represents a promising avenue for future research. Moreover, the EE of NAFD CF-mMIMO systems can be further enhanced through enabling UC clustering. In other words, each AP can select a subset of UEs to serve. For instance, by selecting $N$ UEs based on the largest Frobenius norms of their channel vectors (matrices) or by applying an LSF-based  selection algorithm to choose $N$ UEs for each AP, we can then apply more advanced processing such as local ZF or MMSE to effectively manage the interference in the network, leading to SE and EE improvement. Enhancing the system performance can be further achieved by considering joint AP clustering, AP mode assignments, and power control strategies. Exploring resource allocation and clustering techniques with low complexity, including ML-based algorithms, to achieve satisfactory SE and EE performance in NAFD CF-mMIMO represents an interesting field for future research.

\section{Cell-free Massive MIMO and Non-Orthogonal Transmission}~\label{sec:NOMA}
Traditional OMA techniques, such as FDMA in first generation, TDMA in second generation, code division multiple access (CDMA) in third generation, and  OFDMA in fourth generation seem to reach their fundamental limits, and therefore are no longer suitable to meet the formidable challenges of massive connectivity and resource scarcity~\cite{Dai:MCOM:2015,Zhang:MVT:2019}. The advent of MIMO networks  has ushered in the opportunity to utilize space-division multiple access, which avails of multi-antenna processing at the transmitter to serve multiple UEs in the same time-frequency resources. To further enhance the SE, multiple-access techniques have been progressed toward the direction of NOMA~\cite{Ding:JSAC:2017}, where UEs are superposed in the same time-frequency resources via the power domain, e.g., power-domain NOMA~\cite{Islam:COMST:2017},  or code domain, e.g, code-domain NOMA~\cite{Dai:CSTO:2018}. Moreover, RSMA has been also recognized as another promising paradigm for non-orthogonal transmission~\cite{Clerckx:COMMG:2016}. In this section, non-orthogonal transmission in CF-mMIMO systems is reviewed.

\vspace{-0.5em}
\subsection{Power-Domain Non-Orthogonal Multiple Access}
In power-domain NOMA, multiple UEs are allowed to share the same time/frequency resource block by using different power levels. The key benefit of power-domain NOMA in DL systems is attributed to the fact that UEs with better channel conditions (termed as strong UEs) utilize SIC to cancel the interference caused by UEs with poorer channel conditions (termed as weak UEs). On the other hand, by allocating more portion of the available power to the weak UEs, to guarantee the feasibility of SIC at strong UEs, user fairness can be achieved.       

The key idea of applying NOMA in CF-mMIMO networks is to group the UEs into clusters, in which UEs in the same cluster use NOMA, i.e., one beam is sent to one cluster. As a result, the same pilot sequence is assigned to all UEs in a given cluster. In each cluster, SIC is carried out to improve the achievable SE. Therefore, the performance of NOMA-aided CF-mMIMO systems is hindered by intra-cluster pilot contamination and error propagation due to imperfect SIC. In the presence of these two challenges, NOMA-aided CF-mMIMO systems become inferior to the OMA counterpart in scenarios with a low number of users (relative to the length of the coherence interval). However, in ultra-dense scenarios where numerous UEs are supported by many distributed co-coverage APs and pilot reuse is inevitable, applying NOMA with strategic UE clustering allows the NOMA-aided CF-mMIMO system to outperform the OMA counterpart. Therefore, combining NOMA with CF-mMIMO networks offers the potential to achieve significant benefits in supporting massive connectivity~\cite{Yikai:WCL:2018,Bashar:TCOM:2020}.  Another use case for NOMA-aided CF-mMIMO networks is in scenarios with a short coherence interval (e.g., due to high UE mobility), where the available data transmission interval is limited whenever orthogonal pilots are used. The pilot overhead can be significantly reduced by clustering UEs and then applying NOMA, allowing UEs within each cluster to be supported over the same time-frequency resources.

\begin{figure}[!t]\centering \vspace{0em}
    \def\svgwidth{240pt} 
    \fontsize{8}{7}\selectfont 
\begingroup%
  \makeatletter%
  \providecommand\color[2][]{%
    \errmessage{(Inkscape) Color is used for the text in Inkscape, but the package 'color.sty' is not loaded}%
    \renewcommand\color[2][]{}%
  }%
  \providecommand\transparent[1]{%
    \errmessage{(Inkscape) Transparency is used (non-zero) for the text in Inkscape, but the package 'transparent.sty' is not loaded}%
    \renewcommand\transparent[1]{}%
  }%
  \providecommand\rotatebox[2]{#2}%
  \newcommand*\fsize{\dimexpr\f@size pt\relax}%
  \newcommand*\lineheight[1]{\fontsize{\fsize}{#1\fsize}\selectfont}%
  \ifx\svgwidth\undefined%
    \setlength{\unitlength}{1029.97503662bp}%
    \ifx\svgscale\undefined%
      \relax%
    \else%
      \setlength{\unitlength}{\unitlength * \real{\svgscale}}%
    \fi%
  \else%
    \setlength{\unitlength}{\svgwidth}%
  \fi%
  \global\let\svgwidth\undefined%
  \global\let\svgscale\undefined%
  \makeatother%
  \begin{picture}(1,0.652443)%
    \lineheight{1}%
    \setlength\tabcolsep{0pt}%
    \put(0,0){\includegraphics[width=\unitlength]{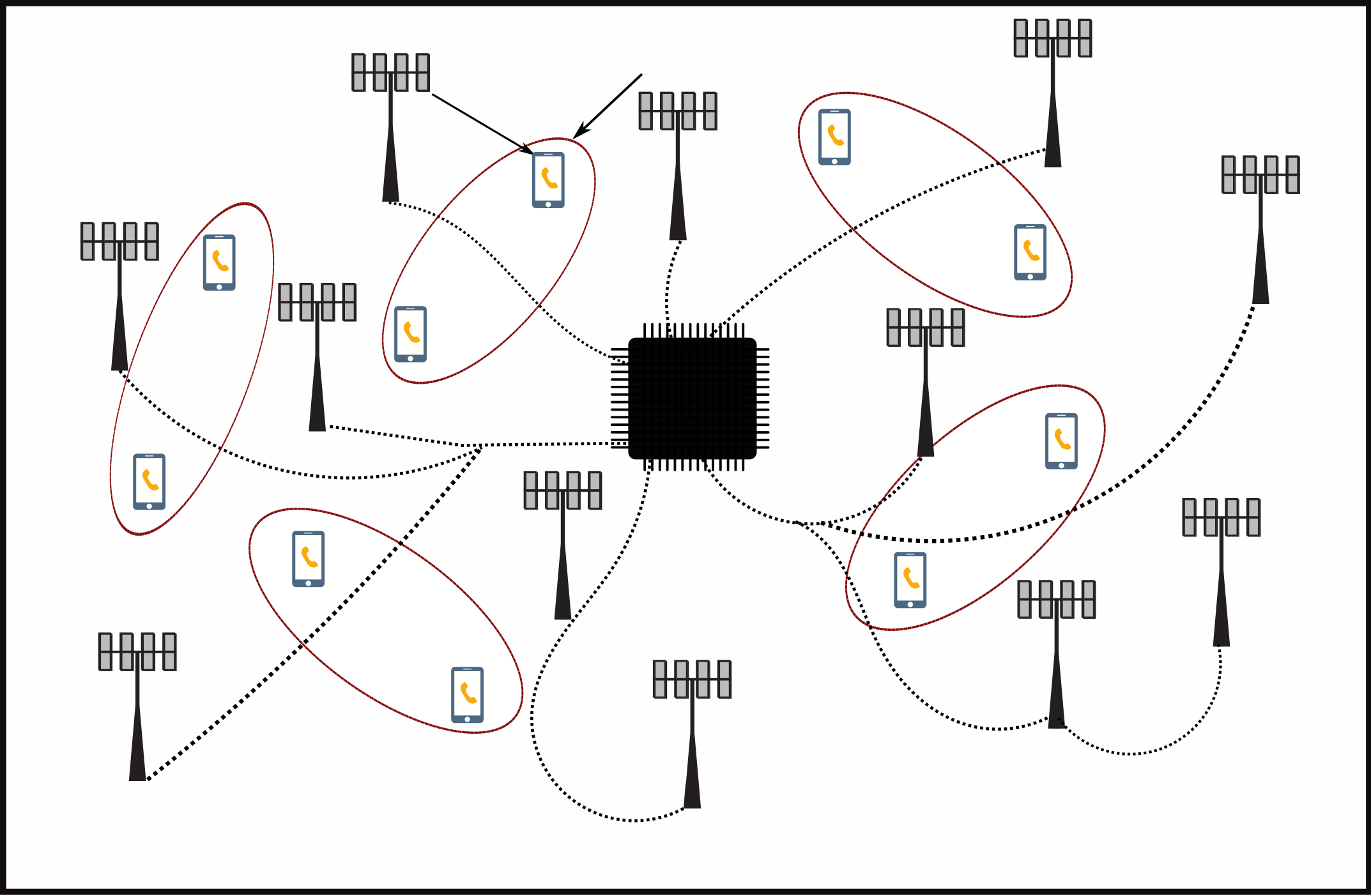}}%
    \put(0.43917436,0.60414508){\color[rgb]{0,0,0}\makebox(0,0)[lt]{\lineheight{1.25}\smash{\begin{tabular}[t]{l}Cluster $\ell$\end{tabular}}}}%
    \put(0.46643249,0.34977031){\color[rgb]{1,1,1}\makebox(0,0)[lt]{\lineheight{1.25}\smash{\begin{tabular}[t]{l}\textbf{CPU}\end{tabular}}}}%
    \put(0.180247,0.52689405){\color[rgb]{0,0,0}\makebox(0,0)[lt]{\lineheight{1.25}\smash{\begin{tabular}[t]{l}AP $m$\end{tabular}}}}%
    \put(0.37490633,0.47613041){\color[rgb]{0,0,0}\makebox(0,0)[lt]{\lineheight{1.25}\smash{\begin{tabular}[t]{l}UE $\ell k$\end{tabular}}}}%
    \put(0.33464198,0.57647359){\color[rgb]{0,0,0}\makebox(0,0)[lt]{\lineheight{1.25}\smash{\begin{tabular}[t]{l}$g_{m\ell k}$\end{tabular}}}}%
  \end{picture}%
\endgroup%
 \vspace{0mm}
    \caption{Illustration of a NOMA CF-mMIMO system with clustered UEs.}\vspace{0mm} \label{fig:NOMA_CFmMIMO}
    \vspace{1em}
\end{figure}
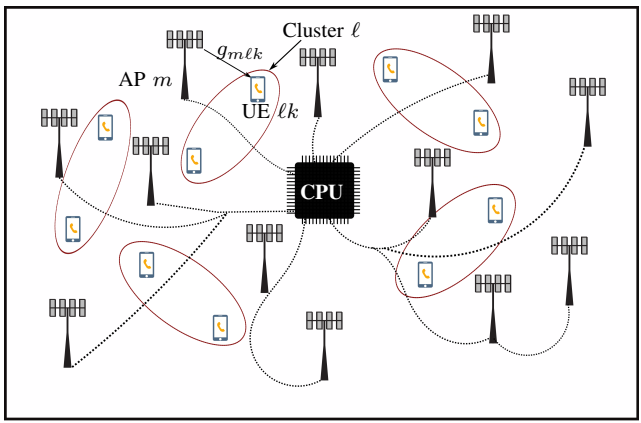

\subsection{Rate Splitting Multiple Access}
RSMA has emerged as a  potent strategy for non-orthogonal transmission and robust interference management technique for multi-antenna wireless networks. RSMA is viewed as a technique that subsumes space-division multiple access and power-domain NOMA as special cases~\cite{Mao:Tut:2022}. In RSMA, the transmitted signal is split into a common and private signal, where the common signal is required to be first decoded by all the receivers using SIC; then, each receiver decodes its intended private signal by treating the unintended private signals as noise. Clerckx \emph{et al.}~\cite{Mao:Tut:2022} highlighted the benefits of RSMA in terms of SE, EE, and reduction of CSI feedback overhead. Furthermore, the authors in~\cite{Clerckx:JOP:2021}, showed that  RSMA can fully exploit the multiplexing gain and the benefits of SIC to boost the rate and the number of UEs in multi-antenna settings, thus outperforms NOMA. In other words, RSMA is more attractive for practical implementation in multi-antenna scenarios, owning to the stronger robustness to user deployments, network loads and inaccurate CSI and lower receiver complexity than NOMA~\cite{Clerckx:JOP:2021}.

RSMA CF-mMIMO tends to outperform OMA CF-mMIMO in environments featuring dynamic network conditions, high interference levels, and dense networks like those found in massive machine type communications. RSMA's inherent capability to mitigate inter-user interference provides distinct advantages over conventional OMA setups in such conditions. In dynamic networks, the movement of UEs induces channel aging, causing the estimated CSI to become outdated. This challenge is particularly pronounced in CF-mMIMO systems, where each UE is served by multiple APs, leading to diverse angles of arrival~\cite{Zheng:TWC:2021}. In scenarios of high mobility with long frame durations, the impact of channel aging becomes even more pronounced in CF-mMIMO systems. Moreover, in dense networks where pilot reuse is inevitable, RSMA can efficiently mitigate the imperfect CSI caused by pilot contamination.

\subsection{Power-Domain NOMA/RSMA and CF-mMIMO Networks}
In this subsection, we summarize the state-of-the-art on the coexistence of power-domain NOMA/RSMA and CF-mMIMO networks and implications of performance optimization and AP/UE clustering. 

\subsubsection{Performance Analysis and Optimization} The pioneering work by Li and Baduge\cite{Yikai:WCL:2018} investigated the DL achievable sum rate of NOMA-based CF-mMIMO systems using the NCB, considering the joint detrimental effects of intra-cluster pilot contamination, inter-cluster interference, and imperfect SIC. Zhang~\ettall\cite{Zhang:ICC:2019} looked into a SE maximization problem for an UL NOMA-enabled CF-mMIMO network, where they showed that a better performance can be achieved by controlling the per-user transmission power. Rezaei~\ettall~\cite{Rezaei:TCOM:2020} applied three linear precoding schemes, namely MRT, FZF and modified regularized ZF, originally designed for mMIMO systems~\cite{Wagner:IT:2012}, into NOMA CF-mMIMO systems. These precoders are locally designed at each AP with the same front-hauling overhead. While MRT maximizes the signal gain at the intended cluster and ignores the inter-cluster interference, FZF sacrifices some of the array gain to cancel the inter-cluster interference and modified regularized ZF balances the inter-cluster interference mitigation and intra-cluster power enhancement.  Bashar~\ettall~\cite{Bashar:TCOM:2020} proposed an adaptive NOMA/OMA selection scheme to maximize the DL per-user transmission rate. It was found that NOMA outperforms OMA when the number of UEs is relatively high. Moreover, the switching point between the NOMA and OMA depends both on the channel's coherence time and on the total number of UEs.  Zhang~\ettall~\cite{Jiayi:IOT:2022} investigated the DL sum SE and EE of NOMA-based CF-mMIMO for IoT over spatially correlated Ricean fading channels. The analysis captures the joint effects of inter-cluster pilot contamination, inter-cluster interference, and imperfect SIC. Kusaladharma~\ettall~\cite{Kusaladharma:TCOM:2021} derived the asymptotic achievable rate of NOMA-based CF-mMIMO systems by employing the stochastic geometry approach, under the assumption of realistic SIC. Simulation results showed that with no prior DL CSI knowledge at the APs, NOMA is inferior to OMA in terms of overall rate and its advantage will merely be within the arena of reduced latency.

Gao~\ettall\cite{Gao:TCOM:2023} invoked NOMA as the MA technique in space-ground integrated networks to improve the SE of terrestrial FD CF-mMIMO system. The sum-rate maximization problem was studied by jointly optimizing the power allocation factors of the NOMA DL, the UL transmit powers, and the beamformers for satellite and APs. 

Motivated by the robustness of the RSMA to imperfect CSI, the authors in~\cite{Mishra:CLET:2022} investigated a RSMA-assisted CF-mMIMO for massive machine type communication with random access, assuming all active UEs utilize the same pilot for channel estimation. On the base of the DL SE of the common and private message, a heuristic common precoder design and a max-min power control strategy were proposed.  The potential of RSMA in enhancing the performance of the CF-mMIMO systems in the presence of asynchronous reception was studied in a recent work~\cite{Zheng:JSAC:2023}. Asynchronous reception destroys the orthogonality of pilots and the coherent transmission of data, leading to an  unsatisfactory SE performance of CF-mMIMO systems. This performance loss can be efficiently compensated via RSMA by splitting the messages into common and private parts. To alleviate the negative impact of asynchronous reception, optimal power allocating to the two messages was determined and robust precoding design for the common messages was developed that maximizes the minimum individual SE of the common message. Zheng~\ettall~\cite{Zheng:TVT:2023} obtained sum-SE of the rate-splitting-assisted CF-mMIMO system with channel aging. A bisection-based precoding scheme was designed for common messages, and it was shown to significantly outperform superposition-based and random precoding schemes, especially in complex mobile environments.

\subsubsection{UE/AP Clustering} User clustering has been proposed in the NOMA-based CF-mMIMO to mitigate/reduce the pilot contamination.
In~\cite{Bashar:TCOM:2020}, three distance-based pairing schemes, including near pairing, far pairing, and random pairing, were proposed to group UEs into disjoint clusters. It was shown that close pairing, where two UEs with the smallest distance between them are paired, provides the worst performance, which is also aligned with the NOMA principle~\cite{Islam:COMST:2017}.  Nguyen~\ettall~\cite{Khai:TVT:2020} studied the UL data transmission of a NOMA-assisted CF-mMIMO network and developed a user grouping algorithm, where two UEs with minimum LSF profiles are grouped. The authors in~\cite{Rezaei:WCL:2020} investigated the DL performance of a cognitive CF-mMIMO NOMA system underlaid a primary mMIMO system and proposed a low complexity suboptimal user pairing method based on the Jaccard distance coefficient to find and group the most dissimilar UEs.

A cluster (UC)-based CF-mMIMO system with rate-splitting was proposed in~\cite{Flores:TCOM:2023}, which groups the transmit antennas in several clusters to reduce the computational and signaling loads. Multiple common streams, each one associated with a different cluster, and one private stream per UE are sent. Common messages from other clusters are considered as noise. Then, at each cluster, the common message is decoded first. Once the common messages are decoded, each receiver decodes its private message. Analysis of the sum-rate, computational cost, and signaling load of linear CB, ZF, and MMSE precoders was provided.  A summary of the existing works on NOMA/RSMA-assisted CF-mMIMO networks is shown in Table~\ref{tab:NOMA}.
 \begin{flushleft} 
\begin{table*}
\caption {CF-mMIMO and non-orthogonal transmission} \label{tab:NOMA}
\vspace{-0.8em}
\small
 \begin{center}
\begin{tabular}
{
|
>{\centering\arraybackslash}m{1.7cm}|
>{\centering\arraybackslash}m{0.7cm}|
>{\centering\arraybackslash}m{0.4cm}|
>{\centering\arraybackslash}m{0.4cm}|
>{\centering\arraybackslash}m{0.6cm}|
>{\centering\arraybackslash}m{0.7cm}|
>{\centering\arraybackslash}m{9cm}|}
\hline
\textbf{Category}  
& \textbf{Ref.} 
&\multicolumn{4}{c|}{\textbf{System Setup}} 
&\vspace{0.5em} \textbf{Technical Contributions} 
\\
\cline{3-6}
	\centering
	&\textbf{}  &\textbf{UL}  &\textbf{DL}  &\textbf{SISO} &\textbf{MISO} & \\

     \hline\hline
    \multirow{2}{*}{\vspace{-9em}NOMA }
     &\vspace{-1em}\cite{Yikai:WCL:2018}   
      &\centering- 
      &\centering\checkmark  
      &\centering\checkmark 
      &\centering-               
     &Achievable sum rate analysis
     \\ \cline{2-7}

     &\vspace{-1em}\cite{Zhang:ICC:2019}             
     &\centering\checkmark 
     &\centering- 
     &\centering- 
     &\centering\checkmark               
     &SE optimization subject to per-user QoS constraint and SIC constraint
     \\ \cline{2-7}
     
     &\vspace{-1em}\cite{Rezaei:TCOM:2020}             
     &\centering- 
     &\centering\checkmark  
     &\centering-
     &\centering\checkmark               
     &Performance analysis for linear precoding scheme with distance-based pairing schemes
     \\ \cline{2-7}

     &\vspace{-1em}\cite{Jiayi:IOT:2022}                       
     &\centering- 
     &\centering\checkmark 
     &\centering- 
     &\centering\checkmark               
     &DL sum-SE and EE analysis over spatially correlated Ricean fading channels \\\cline{2-7}   

      &\vspace{-1em}\cite{Kusaladharma:TCOM:2021}   
     &\centering- 
     &\centering\checkmark  
     &\centering-
     &\centering\checkmark               
     &Achievable rate analysis using the stochastic geometry tools, under the assumption of realistic SIC
     \\\cline{2-7}
     
     &\vspace{-1em}\cite{Gao:TCOM:2023}   
     &\centering\checkmark 
     &\centering\checkmark  
     &\centering-
     &\centering\checkmark             
     &Sum-rate maximization in space-ground integrated networks
     \\\cline{2-7}
     
     &\vspace{-1em}\cite{Khai:TVT:2020}                       
     &\centering\checkmark 
     &\centering-  
     &\centering- 
     &\centering\checkmark               
     &User grouping based on minimum LSF profiles 
     \\\cline{2-7} 

     &\vspace{-1em}\cite{Rezaei:WCL:2020}                     
     &\centering- 
     &\centering\checkmark  
     &\centering- 
     &\centering\checkmark               
     &Low complexity user grouping based on Jaccard distance coefficients
     \\\hline 

          

     \multirow{2}{*}{\vspace{1em}NOMA/OMA }
     &\vspace{-1em}\cite{Bashar:TCOM:2020}            
     &\centering- 
     &\centering\checkmark  
     &\centering- 
     &\centering\checkmark            
     &Introducing three distance-based pairing schemes as well as mode switching algorithm    
     \\\hline    

    &\vspace{-1em}\cite{Mishra:CLET:2022}        
    &\centering- 
    &\centering\checkmark  
    &\centering- 
    &\centering\checkmark                  
    &Achievable SE analysis for the common and private streams, as well as heuristic common precoder and max-min power control design 
    \\ \cline{2-7}   
    
    \multirow{ 2}{*}{RSMA } 
    &\vspace{-1em}\cite{Zheng:JSAC:2023}        
    &\centering- 
    &\centering\checkmark  
    &\centering- 
    &\centering\checkmark                  
    &SE analysis and robust precoding design for the common messages
    \\ \cline{2-7}    
  
    &\vspace{-1em}\cite{Flores:TCOM:2023}        
    &\centering- 
    &\centering\checkmark  
    &\centering- 
    &\centering\checkmark                  
    &Sum-rate, computational cost, and signaling load analysis for linear CB, ZF, and MMSE precoders   
   \\\cline{2-7}

    &\vspace{-1em}\cite{Zheng:TVT:2023}        
    &\centering- 
    &\centering\checkmark  
    &\centering- 
    &\centering\checkmark                  
    &Sum-SE analysis with  bisection-based precoding design for common messages 
   \\\hline

\end{tabular}
 \end{center}
\label{table2}
\vspace{0em}
\end{table*}
 \end{flushleft} 

\vspace{-1.5em}
\subsection{Case Study and Discussion}
Consider the DL transmissions in a NOMA-based CF-mMIMO system, where the $K$ UEs are grouped into $L$ clusters with $K_l$ UEs in each cluster, resulting in $K=K_l L$, c.f. Fig.~\ref{fig:NOMA_CFmMIMO}. The channel vector between the AP $m$ and UE $k$ in cluster $\ell$ (UE-$\ell k$) is denoted by $\qg_{m\ell k} = \sqrt{\beta_{m\ell k}} \qh_{m\ell k}$, where $\beta_{m\ell k}$ and $\qh_{m\ell k}\sim\mathcal{CN}(\boldsymbol{0}, \qI_N)$ represent the LSF and the small-scale fading, respectively. Assume that the same pilot sequences are assigned to the UEs within the same cluster, whereas orthogonal pilots are assigned to different clusters~\cite{Bashar:TCOM:2020}. AP $m$ utilizes pilot sequences received from UEs within the cluster $\ell$ to provide an estimate of the linear combination of the UEs' channel, i.e, $\qf_{m\ell} = \sum_{k=1}^{K_l} \qg_{m\ell k}$, $\forall \ell$.  This channel estimation is inspired by the findings in~\cite{Cheng:TWC:2018}, which suggests that estimating $\qf_{m\ell}$ leads to improved performance  when compared to estimating the individual channels, i.e, $\qg_{m\ell k}$. The MMSE estimate of  $\qf_{m\ell}$ is given by 
\begin{align}
 \hat {\mathbf {f}}_{m\ell}=c_{m\ell}\big ({\sqrt {\tauup \rho _{p}}\sum\nolimits _{k= 1}^{K}\mathbf {g}_{m\ell k}+\mathbf {W}_{p,m}{\Bpsi_{\ell}}}\big),   
\end{align}
where $\mathbf {W}_{p,m}\in\mathbb{C}^{M\times K_i}$ denotes the noise sequence at the AP $m$ whose elements are i.i.d. $\mathcal{CN}(0,1)$; $\Bpsi_{\ell}$ is the pilot sequence for cluster $\ell$; and $c _{m\ell}\triangleq\frac {\sqrt {\tauup \rho _{p}}\sum _{k^\prime =1}^{K_l}\beta _{m\ell k^\prime }}{\tauup \rho _{p}\sum _{k^\prime =1}^{K_l}\beta _{mlk^\prime }+1}$. Moreover, the MMSE variance is given by $\gamma _{m\ell}=\mathbb {E}\big \{{\big \Vert\big [{{\hat {\mathbf {f}}_{m\ell}}}\big]_{n}\big \Vert^{2}}\big \} =\sqrt {\tauup \rho _{p}}\sum _{k^\prime =1}^{K_l}\beta _{m\ell k^\prime }c_{m\ell}$. With CB at AP $m$, the transmit signal is represented by 
$\mathbf {x}_{m}= \sqrt {\rho _{d}}\sum _{\ell=1}^{L}\sum _{k=1}^{K} {\sqrt {\eta _{m\ell k}}} \hat {\qf}_{m\ell}^\ast s_{\ell k}$, where $s_{\ell k}$ ($\Ex \{\vert s_{\ell k\}\vert2}=1$) and $\eta_{m\ell k}$ denotes the transmitted symbol and the power control coefficient at the AP $m$, respectively. The signal received at the UE $k$ in the cluster $\ell$ is $r_{\ell k}^{{\text {B}}}= \sum _{m=1}^{M} \mathbf {g}_{m\ell k}^{T}\mathbf {x}_{m}+n_{lk}$, while  $n_{lk}\sim\mathcal{CN}(0,1)$ is the noise at UE $k$ in the cluster $\ell$. 

NOMA is utilized exclusively within individual clusters and does not extend to communication between clusters. According to~\cite{Ding:CLET:2015}, UE pairing plays a crucial role in NOMA systems, simplifying the practical adoption of NOMA for numerous UEs by minimizing the complexity of SIC. Given the fact that CSI is inaccessible at both the CPU and UEs, it is infeasible to make use of the UE pairing strategies proposed in the literature~\cite{Ding:CLET:2015,Ali:ACCESS:2017}. In~\cite{Bashar:TCOM:2020}, the authors proposed three UE pairing schemes for NOMA-based CF-mMIMO systems, which are based on channel statistics. These schemes are as follows: 1) close pairing, the UEs who have the smallest distance from each other are paired. The closest UEs are paired until all the UEs are grouped into clusters;  2) far pairing, the UEs who have the largest distance from each other, are paired. The farthest UEs are paired until all the UEs are grouped into clusters; 3) random pairing, UEs are randomly paired into clusters. Within the cluster $\ell$, “UE-$\ell 1$” is the least-contaminated UE (strongest UE) whose signal is detected first. The signals of other UEs are then detected by exploiting SIC, while “UE-$\ell K_l$” is the weakest UE whose signal becomes automatically available after SIC. 

The max-min fairness problem, where the minimum DL SE of the UEs is maximized while satisfying per-AP power constraints, was considered in~\cite{Bashar:TCOM:2020}, and it is mathematically represented as follows:
\vspace{-0.5em}
\begin{subequations}~\label{eq:PALOC:NOMA}
 \begin{align}
&\max _{\eta _{m\ell k}}\quad ~ \min _{k=1\ldots K_l, \ell=1 \ldots L}\quad \SE_{\ell k}^{\ell k,\text {final}, \CB}\\ &\text {s.t. } \sum\nolimits _{\ell=1}^{L}\sum\nolimits _{k=1}^{K} \eta _{m\ell k}\gamma _{m \ell}\le \frac {1}{N},\forall m,~\label{eq:per:AP:NOMA}\\ & \qquad \eta _{m\ell k}\ge 0, ~~ \forall m,\forall \ell, \forall k,
\end{align}   
\end{subequations}
where $\SE_{\ell k}^{\ell k,\text {final,CB}}=\frac {\tau _{d}}{\tau _{c}}\log _{2}\big ({1+ \SINR_{\ell k}^{\ell k,\text {final}, \CB}}\big)$ denotes the SE of UE $k$ in the cluster $\ell$ with $\SINR_{\ell k}^{\ell k,\text {final}} = \min \big ({\SINR_{\ell j}^{\ell k, \CB},\SINR_{\ell k}^{\ell k, \CB}}\big)$, where $\SINR_{\ell j}^{\ell k, \CB}$ refers to the effective SINR of UE-$\ell j$ when UE-$\ell j$ is decoding the signal intended for UE-$\ell k$, given as~\eqref{eq:SINRNOMACB} at the top of the next page. 
Moreover,~\eqref{eq:per:AP:NOMA} presents the power constraint at the AP $m$. The max-min optimization problem~\eqref{eq:PALOC:NOMA} was solved in~\cite{Bashar:TCOM:2020} using second-order cone programming, and the optimal solution was found by employing a bisection search method.
\begin{figure}[t]
	\centering
	\vspace{-0.25em}
	\includegraphics[width=0.47\textwidth]{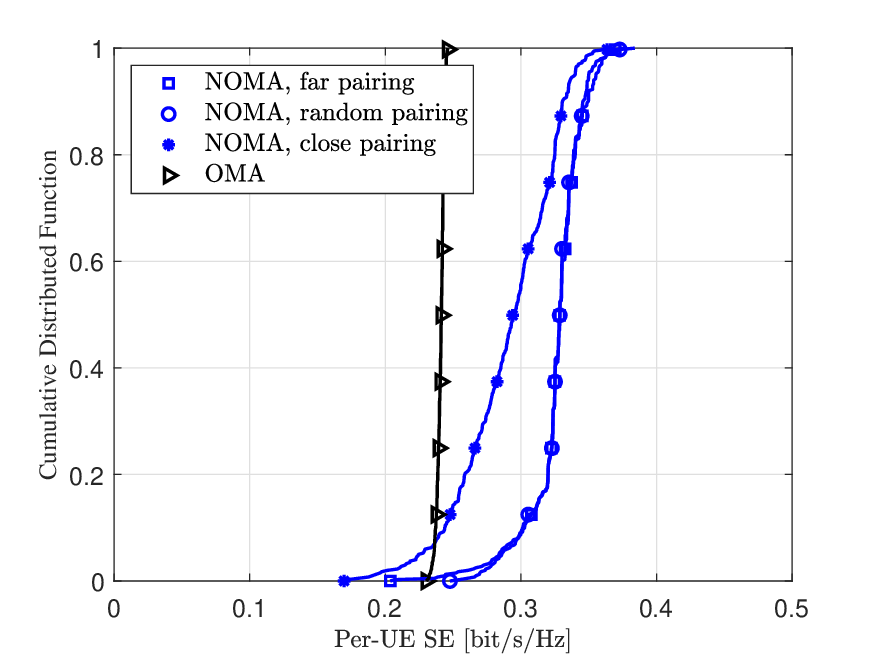}
	\vspace{-0.7em}
	\caption{CDF of the per-UE SE for NOMA and OMA CF-mMIMO systems with per-AP power constraints and for $M=20$, $L=50$, $K_l=2$, $N=15$, $\tau_c=110$, $p_p=0.1$ W, and $p_d=0.2$ W.}
	\vspace{1em}
	\label{fig:NOMA}
\end{figure}
\begin{figure*}
\begin{align}~\label{eq:SINRNOMACB}
\SINR_{\ell j}^{\ell k, \CB} = 
\frac{N^2\left ({\!\sum _{m=1}^{M}\!\sqrt {\eta_{m\ell k}}\frac {\gamma_{m\ell}\beta _{m\ell j}}{\sum_{i=1}^{K_l}\beta _{m\ell i}}}\right)^{2}}
{N^2\sum_{k'=1}^{k-1}\left ({\!\sum _{m=1}^{M}\!\sqrt {\eta_{m\ell k'}}\frac {\gamma_{m\ell}\beta _{m\ell j}}{\sum_{i=1}^{K_l}\beta _{m\ell i}}}\right)^{2} + N 
\sum_{\ell'=1}^{L} \sum_{k'=1}^{K_l} \sum_{m=1}^{M}
\eta_{m\ell'k'\beta_{m\ell j}\gamma_{m\ell'} + \frac{1}{\rho_d}}
}.
\end{align}
  	\hrulefill
	\vspace{-4mm}
  \end{figure*}

Figure~\ref{fig:NOMA} shows the CDF of the per-UE DL SE with CB at the APs, supporting $K=100$ UEs. Power control coefficients were determined by solving the optimization problem~\eqref{eq:PALOC:NOMA}.  In this figure, orthogonal pilots are assigned to the UEs for the OMA system to avoid pilot contamination, i.e., $\tauup= K$. With $K_l=2$ UEs per each cluster, we have $\tauup= K/2$ in the NOMA counterpart. Therefore, compared to OMA, there is more time available for payload data transmission. Nevertheless, in both OMA and NOMA systems, long pilot sequences leave less time for data transmission, thereby reducing the overall SE. From Fig.~\ref{fig:NOMA}, we observe that NOMA with far pairing outperforms other UE pairing schemes, while random-UE pairing achieves comparable performance to far-UE pairing. Results not shown in this figure confirm that, with an increase in the coherence time, OMA performs better than NOMA in CF-mMIMO systems, as OMA does not suffer from pilot contamination, while UEs in NOMA system still suffer from some residual interuser interference. Moreover, as the number of users ($N$) decreases, NOMA with close UE-pairing tends to become inferior to OMA.

\subsection{{Future Research Directions}}
To exploit the promising features of NOMA, it is crucial to group a sufficiently large number of UEs with distinct channel conditions that perform NOMA jointly~\cite{Islam:COMST:2017}. In cellular networks, the UEs with better channels (closer UEs to the BS) are grouped with the UEs with worse channels and SIC is performed inside each group~\cite{Islam:COMST:2017}. However, this approach cannot be readily utilized in CF-mMIMO systems. This is due to the fact that each UE is communicating with multiple APs, and determining the UE with the best channel is not straightforward.  Moreover, pilot reuse for UEs belonging to the same group (especially those in the close vicinity of each other) results in severe pilot contamination. Accordingly, perfect SIC becomes practically unattainable due to intra-group (cluster) pilot contamination and channel estimation errors. The design of UE grouping (clustering) algorithms in NOMA CF-mMIMO systems that alleviate the impact of pilot contamination is, therefore, of paramount importance.

Moreover, there has been very limited work on RSMA CF-mMIMO networks, and it is still unclear how this approach is compared with NOMA in terms of performance and complexity and under different setups. Investigating the RSMA integration in UC, FD, and NAFD CF-mMIMO networks with practical impairments deserves further studies in the future. Furthermore, it is interesting to investigate the impact of different power splitting factors, defined as proportion of power devoted to the transmission of common messages at each AP.  Additionally, exploring power control designs aimed at enhancing  the SE/EE in RSMA CF-mMIMO networks is of considerable interest. 

\section{Cell-free Massive MIMO and Security}~\label{sec:PLS}
Exploiting the physical-layer characteristics and impediments of the wireless channel, such as noise, fading, and interference, is a promising idea to complement the traditional cryptography schemes. This  technology, which has been coined as PLS, significantly improves the overall security of the wireless communication networks against potential external and/or internal eavesdroppers (Eves)~\cite{ZouIEEE:2016,Nguyen:MCOM:2018}.   
The Eves can be categorized into two general classes; i) passive Eves, who overhear legitimate messages silently, ii) active Eves, who can transmit malicious jamming signals while eavesdropping or spoofing pilot sequences~\cite{LI:TIFS:2017,Jiangbo:TCOM:2020}. The latter scenario is related to the so-called \emph{pilot spoofing attack}. From the eavesdropping point of view, active eavesdropping is more harmful than passive eavesdropping, since the channel estimation is severely vulnerable to pilot spoofing attacks. In CF-mMIMO networks, these channel estimation errors lead to improper beamforming design at the APs towards the Eve. Therefore, the confidential information leakage to active Eve is possibly higher~\cite{Zhou:TWC:2012}. From this perspective,  substantial research efforts have been invested towards the analysis and enhancement of the PLS in the CF-mMIMO systems with single and multiple active Eves.     
 
\subsection{Literature Review}
In this subsection, we summarize the key design insights, implications of practical transmission impairments, and concluding remarks for CF-mMIMO networks in the presence of pilot spoofing attacks.

\subsubsection{Pilot Spoofing Attack Mitigation}
To intercept the information intended to a specific UE, the active Eve sends an imitated version of the legitimate UE pilot sequence during the training phase, bewildering the APs to inadvertently steer the legitimate UE signals towards the Eve. Artificial noise (AN) transmission in conjunction with the information signal transmission has been developed in PLS context with the aim of confusing  Eve and protecting the confidential signals~\cite{Mukherjee:TUT:2014,Hamamreh:CSTO:2019}. Timilsina~\ettall~\cite{Timilsina:GC:2018} analyzed the achievable secrecy rate of the CF-mMIMO networks with AN transmission and compared with that of co-located mMIMO. 

To mitigate an active Eve, awareness of the pilot contamination attack is of crucial importance to reduce the information leakage in wireless networks~\cite{Yang:TPDS:2013,Xiong:TIFS:2015}. Hoang~\ettall~\cite{Hien:Secrecy:TCOM2018} proposed a simple training phase to recognize the presence of a spoofing attack to a particular UE  in CF-mMIMO networks. This method compares the asymmetry of received signal power levels to detect Eves. Then, relying on the spoofing attack detection outcome, the problem of maximizing the achievable rate of the attacked UE or its achievable secrecy rate has been studied to degrade the eavesdropping leakage during the DL data transmission phase. Zhang~\ettall~\cite{Zhang:Accesssec:2019}  developed a pilot spoofing assault detection scheme based on the minimum description length for a multigroup multicasting system with both UL and DL training. The minimum description length is an information theoretic criterion for determining the number of present signals and was initially utilized for active attack detection in multi-antenna systems~\cite{Haddadi:TSP:2010}.

\subsubsection{Practical Transmission Impairments}
The utilization of imperfect transceivers at both the APs and UEs as well as economical coarse ADCs/DACs is a feasible solution to tackle the bottleneck of cost and energy consumption~\cite{Bashar:TCOM:2019,Bashar:TGCN:2019}. Nevertheless, from the secure transmission point of view, the ADC/DAC distortion acts, in some sense, as a special type of AN sequence, since it can hamper the achievable rate of both the legitimate UE and Eve. Therefore, some research efforts have been dedicated to analyze the secrecy performance of the CF-mMIMO networks under hardware impairments. Zhang~\ettall~\cite{Zhang:JSYS:2020} concentrated on investigating the impact of the hardware impairments on the PLS of the CF-mMIMO networks in presence of pilot spoofing attack. Besides, Zhang~\ettall~\cite{Zhang:TVt:2021}  investigated secure transmission,  considering both the non-ideal RF chains and low-resolution ADCs/DACs at both the APs and legitimate UEs. Zhang~\ettall~\cite{Zhang:TCOM:2022} examined the secrecy performance of DL CF-mMIMO systems over Ricean fading channels, assuming that the APs are equipped with low-resolution DACs, in the presence of an active multi-antenna Eve. Through simulation results, the authors have shown that for extreme $1$-bit DACs, AN is not necessary since the severe quantization impairment serves as a special type of and helps to degrade the capacity of the Eve. 
\subsubsection{Secrecy Energy Efficiency}
To explore the system EE with information security, the secrecy EE has been defined as the sum secrecy rate per unit of energy and investigated for different PLS setups~\cite{Lu:TCOM:2020}. In a CF-mMIMO system with multiple active non-colluding Eves, Jiang~\ettall~\cite{Jiang:TVT:2023} formulated a power allocation optimization problem to maximize the average secrecy EE, provided that the QoS requirements of all legitimate UEs and per-AP power constraints are met.
\subsubsection{Precoding Design and Null-Space Artificial Noise}
Optimal precoding design to secure DL transmission of a UC-CF mMIMO system against multiple active collusive Eves was addressed by Gao~\ettall~\cite{Gao:WCOML:2023}. More specifically, the APs estimate the DL CSI by UL pilot training and then share the CSIs of serving UEs to CPU via the fronthaul link. Then, the CPU designs the precoding vectors for secure and reliable transmissions subject to the rate requirements of all UEs and power constraints of APs.  By leveraging tools from stochastic
geometry, Ma~\ettall~\cite{Ma:TIFS:2023}
analytically characterized the secrecy performance of scalable CF-mMIMO system in terms of both the outage-based secrecy transmission rate  and ergodic secrecy rate. In order to enhance the secrecy performance, they considered the use of null-space AN by the APs when transmitting confidential messages to the UEs. Atiya~\ettall~\cite{Zahra:VTC2023} studied a CF-mMIMO system with multi-antenna APs experiencing an active eavesdropping attack during UL training. They considered distributed PPZF precoding scheme and deduced closed-form expressions for the SE at the legitimate user and Eve.

Xia~\emph{et al.}~\cite{Xia:SYSJ:2022} examined the secrecy performance of the NAFD CF-mMIMO systems,  focusing on the optimization of local transmit and receiver beamforming vectors at the APs, AN covariance matrix, and  AP mode selection vectors. This optimization process was carried out using a two-layer algorithm. Table~\ref{tab:PLYS} shows a number of existing contributions to CF-mMIMO from the PLS perspective.

 \begin{flushleft} 
\begin{table*}
\centering
\caption {Summary of CF-mMIMO and PLS literature} \label{tab:PLYS}
\vspace{-0.4em}
\small
\begin{tabular}
{|>{\centering\arraybackslash}m{2.0cm}
|>{\centering\arraybackslash}m{0.7cm}
|>{\centering\arraybackslash}m{0.7cm}
|m{1.3cm}|m{0.7cm}|m{0.7cm}|>{\centering\arraybackslash}m{8cm}|}
\hline
\textbf{Category}  & \textbf{Ref.} 
&\multicolumn{4}{c|}{\textbf{System Setup}} 
&\textbf{Technical Contributions} \\
\cline{3-6}
	\centering
	&\textbf{}   &\textbf{UC} &\textbf{HD/NAFD} &\textbf{MISO} &\textbf{SISO} & \\

\hline\hline
\centering
 \multirow{2}{*}{ \shortstack{\\\\\\\\\\\\\\\\\\\\\\\\\\\\ Single active \\ Eve} }
     &\vspace{-2em}\cite{Timilsina:GC:2018}
     &\centering - 
     &\centering HD 
     &\centering - 
     &\centering \checkmark   
     &Achievable secrecy rate analysis  
     \\ \cline{2-7}
     
     &\vspace{-1em}\cite{Hien:Secrecy:TCOM2018}            
      &\centering - 
      &\centering HD 
      &\centering - 
      &\centering \checkmark            
     &Pilot spoofing attack detection and mitigation through the power control design at APs 
     \\ \cline{2-7}    

     &\vspace{-1em}\cite{Zhang:Accesssec:2019}            
      &\centering- 
      &\centering HD 
      &\centering - 
      &\centering \checkmark            
     &Pilot spoofing attack detection in multigroup multicasting CF-mMIMO 
     \\ \cline{2-7}  

     &\vspace{-1em}\cite{Zhang:JSYS:2020}            
      &\centering - 
      &\centering HD 
      &\centering- 
      &\centering\checkmark              
     &Investigating the impact of the hardware impairments on the PLS 
     \\ \cline{2-7}

     &\vspace{-1em}\cite{Zhang:TVt:2021}            
     &\centering- 
     &\centering HD 
     &\centering\checkmark 
     &\centering-             
     &Achievable secrecy rate analysis in presence of nonideal RF chains and low-resolution ADCs/DACs  
     \\ \cline{2-7}
     
     &\vspace{-1em}\cite{Zhang:TCOM:2022}            
      &\centering- 
      &\centering HD 
      &\centering\checkmark 
      &\centering-           
       &Secrecy performance analysis with low-resolution DACs and in the presence of an active multi-antenna Eve
       \\\hline      

    \multirow{ 2}{*}{\shortstack{\\\\\\\\\\\\\\\\\\\\\\ Multiple \\ Eves} }
      &\vspace{-1em}\cite{Jiang:TVT:2023}   
      &\centering - 
      &\centering HD 
      &\centering \checkmark 
      &\centering -               
     &Secrecy EE optimization in presence of non-colluding active Eves
     \\ \cline{2-7}
     
     &\vspace{-1em}\cite{Gao:WCOML:2023}                       
     &\centering\checkmark 
     &\centering HD 
     &\centering\checkmark 
     &\centering-                
     &Optimal precoding design to protect DL transmissions against multiple active collusive Eves  
     \\  \cline{2-7}   
     
    &\vspace{-1em}\cite{Ma:TIFS:2023}   
    &\centering\checkmark 
    &\centering HD 
    &\centering\checkmark 
    &\centering-               
    &Secrecy performance evaluation in terms of outage-based
      secrecy transmission rate and ergodic secrecy rate in presence of non-colluding passive Eves and AN injection
      \\ \cline{2-7}   

    &\vspace{-1em}\cite{Xia:SYSJ:2022}   
      &\centering- 
      &\centering NAFD 
      &\centering\checkmark 
      &\centering-               
     &Secrecy SE optimization via duplex mode optimization as well as AN covariance matrix and beamforming design at the APs in the presence of non-colluding/colluding active Eves
     \\\hline  
  
\end{tabular}
\label{table2}
\vspace{-1em}
\end{table*}
 \end{flushleft} 
\begin{figure}[!t]\centering \vspace{0em}
    \def\svgwidth{240pt} 
    \fontsize{8}{7}\selectfont 
    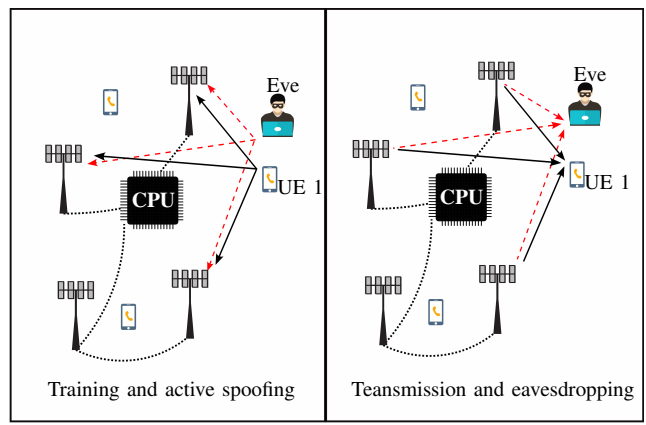 \vspace{0mm}
    \caption{CF-mMIMO with $K$ legitimate users under an active spoofing attack, where an Eve contaminates the UL channel estimation phase by sending an identical pilot sequence with the legitimate UE $1$.}\vspace{0mm} \label{fig:PLS_CFmMIMO}
    \vspace{1em}
\end{figure}
\vspace{-3em}
\subsection{Case Study and Discussion}
Consider a CF-mMIMO system with $M$  APs and $K$ single-antenna legal UEs under spoofing attack as shown in Fig.~\ref{fig:PLS_CFmMIMO} where  an Eve is actively involved in attacking the system in the UL training phase. Each AP is equipped with $N$ antennas. In the training phase, all UEs and Eve transmit the pilot sequences to the APs for requesting the messages. Suppose an Eve wants to overhear the confidential information destined to UE $1$ and designs its pilot sequence  $\Bpsi_E$ as $\Bpsi_E=\Bpsi_1$. After applying the MMSE channel estimation, the mean square of the estimate $\hhatlmk$ and $\hhatlmE$ can be obtained as 
\begin{align}~\label{eq:gamalmk}  
		\gamalmk= 
  \begin{cases}
  \frac{\tauup \rho_{u} \betalk^{2}}{\tauup \rho_{u} \betalk+1}, & k \neq 1, 
  \\ 
  \frac{\tauup \rho_{u} \betalone^{2}}{\tauup \rho_{u} \betalone+\tauup \rho_{E} \betalE+1}, 
  & k=1,
		\end{cases} 
\end{align}
and $\gamalmE\triangleq\alpha_{E 1} \gamma_{\ell,1}$, respectively, where  $\alpha_{E 1}=\big(\rho_{E} \betalE^{2}\big) /\big(\rho_{u} \betalone^{2}\big)$. 
In addition, the received signals at Eve and UEs during the DL transmission phase are given by
\begin{subequations}
	\begin{align}~\label{eq:z_k}   
		{y}_{k}&= \sum\nolimits_{\ell=1}^{M}\mathbf h^{\mathrm{H}}_{\ell k}\bigg(\sum\nolimits_{t=1}^{K} \sqrt {\eta _{\ell k}} \mathbf {w}_{\ell t} s_{t}\bigg)+{n}_{k},\\
		y_{E}&= \rho_d\sum\nolimits_{\ell=1}^{M}\mathbf h^{\mathrm{H}}_{\ell E}\bigg(\sum\nolimits_{t=1}^{K} \sqrt {\eta_{\ell k}} \mathbf {w}_{\ell t} s_{t}\bigg)+{n}_{E},
\end{align} 
\end{subequations}
where $n_{k}$ and $n_{E}$ are the received AWGN at UE $k$ and Eve, distributed as $\mathcal{C N}(0,1)$. Now, let us evaluate the secrecy performance of a CF-mMIMO system with PPZF precoding design and denote the set of indices of APs that transmit to the UE $k$ relying on PZF and protective MRT  by $\mathcal{Z}_{k}$ and $\mathcal{M}_{k}$, respectively, as  $\mathcal{Z}_{k} \triangleq\left\{\ell: k \in \mathcal{S}_{\ell}, \ell=1, \ldots, M\right\}$, and $\mathcal{M}_{k} \triangleq\left\{\ell: k \in \mathcal{W}_{\ell}, \ell=1, \ldots, M\right\}$, with $\mathcal{Z}_{k} \cap \mathcal{M}_{k}=\varnothing$, and $\left|\mathcal{Z}_{k}\right|+\left|\mathcal{M}_{k}\right|=M$.  Then,  by applying the use-and-then-forget capacity-bounding technique~\cite{Hien:cellfree}, the received SINR at UE $k$ and Eve for PPZF precoding scheme  can be expressed as~\cite{Zahra:VTC2023}
 \vspace{-1em}
 \begin{subequations}
   \begin{align} 
\SINRk&= \frac{\left(\!\!\!\!\quad\sum\nolimits_{\ell=1}^{M} \sqrt{\left(N-\tausl\right) \rholk \gamma_{\ell k}}\right)^{2}}{\sum\nolimits_{t=1}^{K} \sum\nolimits_{\ell=1}^{M} \rholt\left(\beta_{\ell k}-\delta _{\ell k}\gamma_{\ell k}\right)+1},~\label{eq:SINR_k2}\\
\SINRE&=\!\frac {\left(\sum\limits_{\ell=1}^{M} \!\!\sqrt {{\eta_{\ell1}}({N}\!\!-\!\!\tausl)\gamma_{\ell E}}\right)^{2}\!\!\!\!+\!\!\sum\limits_{\ell=1}^{{M}}\eta_{\ell  1}\beta_{\ell E}\!\!-\!\!\!\sum\limits_{\ell \in {\Zone}}\!\! \eta_{\ell 1}\gamma_{\ell E}}{{\sum\limits_{\substack{ t \neq 1}}^{K}} \sum\limits_{\ell=1}^{M}\eta_{\ell t}(\beta_{\ell E}\!-{\delta_{\ell 1}}\gamma_{\ell E})\!+\!1/\rho_d}, ~\label{eq:SINR_E2}
\end{align}   
 \end{subequations}
where $\delta_{\ell k} \triangleq 1$ if UE $k\in \mathcal{S}_{l}$ and $\delta_{\ell k} \triangleq 0$ if UE $k\in \mathcal{W}_{l}$.
Now, using the derived  SINR expressions for Eve in~\eqref{eq:SINR_E2}  and UEs in~\eqref{eq:SINR_k2}, we can  derive the lower bound for the   secrecy SE at UE $1$ as 
\begin{align}~\label{eq:R_sec} 
R_{\mathtt{sec}}=  \left[\log_{2}\left(\frac{1+\SINRone}{1+ \SINRE} \right)\right]^+,
\end{align}
where $[x]^+ = \max \{0, x\}$.

\begin{figure}[t]
	\centering
	\vspace{-0.8em}
	\includegraphics[width=0.47\textwidth]{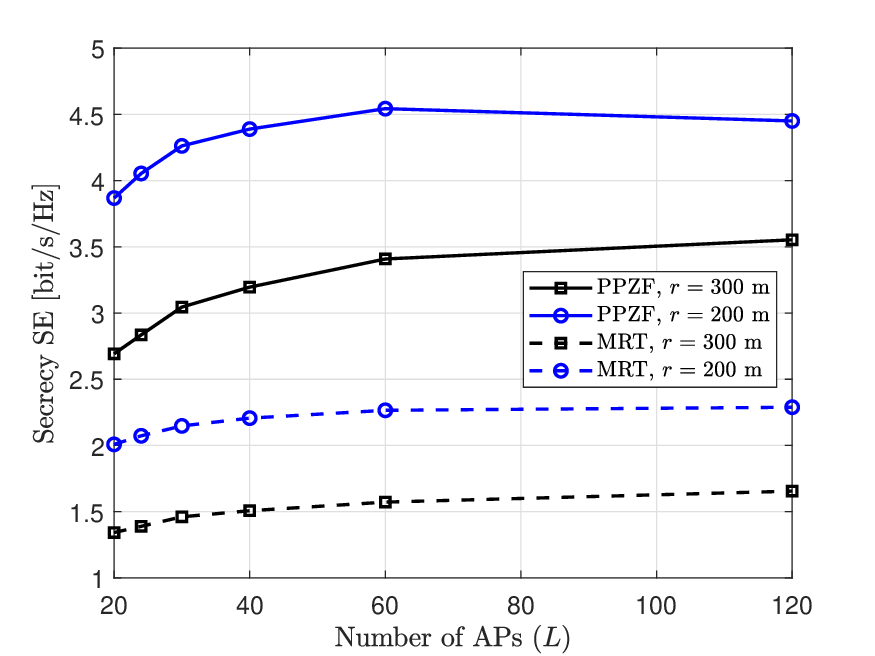}
	\vspace{-0.7em}
	\caption{Average secrecy SE versus the number of service antennas for different circle radii $r$ around UE $1$.}
	\vspace{1em}
	\label{fig:PLS}
\end{figure}

In Fig.~\ref{fig:PLS}, we examine the secrecy SE performance of the CF-mMIMO system under active eavesdropping for PPZF and MRT precoding designs as a function of the number of APs when the total number of service antennas is fixed, i.e., $MN=240$. Also, it is assumed that the Eve is randomly located in a circle with radius $r$ m around UE $1$. It is evident that, upon increasing the number of APs, the secrecy SE performance enhances due to the high degree of macro diversity and low path loss. We can also see that while PPZF achieves a promising performance, the MRT design fails to meet the secrecy performance specifications of the CF-mMIMO system, which demonstrates the importance of the precoding design.

\subsection{{Future Research Directions}}
The existing works in Table~\ref{tab:PLYS}  focus on analyzing the secrecy performance of the system relying on MRT precoding design without considering AP selection (clustering). Since multiple APs are available to multiple UEs, one possible area for future research  is the optimization of UE-AP association to enhance the secrecy of CF-mMIMO system. In this context, it is important to address the trade-off between secrecy and energy consumption when designing CF-mMIMO networks. Moreover, the interference generated by multiple APs may be properly exploited to degrade the performance of the Eve. Additionally, measures to prevent active Eve's intrusion during pilot training should also be considered.

Other concepts, including covert communication~\cite{Bloch:TIF:2016} and proactive eavesdropping~\cite{Xu:WCL:2016}, have remained untouched in CF-mMIMO literature. Covert communication aims to enable wireless transmission with negligible detection probability by unauthorized parties, ensuring the privacy of the transmitter. On the other hand, in proactive eavesdropping, legitimate monitors are exploited for wireless surveillance of suspicious UEs. In a recent work, Mobini~\ettall~\cite{mobini2023cell} proposed a new CF-mMIMO wireless surveillance system, where a large number of distributed multi-antenna aided legitimate monitoring nodes embark on either observing or jamming untrusted communication links. They derived closed-form expressions for the monitoring success probability of the legitimate monitoring nodes and proposed a greedy algorithm for the observing vs, jamming mode assignment of legitimate monitoring nodes. 

\section{Cell-free massive MIMO and Energy Harvesting}~\label{sec:EH}
With the explosive growth of telecommunication market resulting from the high data traffic demand with the launch of 5G and 6G networks as well as the penetration of IoT, the energy consumption of wireless networks is expected to increase dramatically. On the other hand, how to  efficiently power up massive wireless devices would be a challenging task. Therefore, EE and sustainable solutions are prime considerations for 5G and 6G networks. In this regard, wireless energy transfer (WPT) via RF and EH technologies have gained prominence in their ability not only to provide perpetual energy supply at low-power wireless devices but also to reduce the energy costs and environmental effects. To make the best use of RF spectrum and wireless network infrastructures (without any considerable modification requirements), dual use of RF signals for delivering energy and transporting information led to the emergence of unified design of wireless information and power transmission (WIPT) for future wireless networks~\cite{Clerckx:JSAC:2019,Lu:ToTs:2015,Clerckx:JSTSP:2021}.

 The two main categories of WIPT are simultaneous wireless information and power transmission (SWIPT) and wirelessly powered communication networks (WPCNs)~\cite{Clerckx:JSAC:2019}. With SWIPT, information could be jointly transmitted with energy from the same AP(s) using the same RF waveform(s) towards information receivers  and energy receivers.  Energy receivers and information receivers can be co-located over the same device that is simultaneously receiving information and harvesting energy, or separated, where energy receiver and information receiver are different devices.  WPT and wireless information transfer (WIT) exhibit a fundamental rate-energy trade-off, which needs to be taken into consideration, when designing related protocols. To exploit the RF signals for both energy harvesting and information reception, different SWIPT receiver structures, namely the power splitting, time-switching, and antenna-switching have been proposed~\cite{Ding:COMMAG:2015}.  In WPCN, wireless devices harvest energy from RF signal transmitted by AP(s) in DL, and utilize the harvested power to transmit their information back to the AP(s) in UL. Nevertheless, in WIPT systems, the low WPT efficiency due to the scattering and severe end-to-end path loss is usually the main limiting factor in practice. To fully unlock the WIPT potential, it must be combined with other advanced technologies. As effective countermeasures, MIMO, and especially mMIMO techniques (owning to their channel hardening and aggressive spatial multiplexing gains), have been adopted in WIPT networks. In this context, multi-antenna transmit energy beamforming techniques have been proposed to steer the RF signals towards desired locations for ERs~\cite{Ding:COMMAG:2015}. In general, the larger number of antennas installed at the APs, the sharper the energy beam that could be generated in a particular spatial direction~\cite{Bi:COMMG:2015,Yang:JSAC:2015}.

Although the efficiency of WPT can be improved through multi-antenna techniques, the EH opportunities of cell-boundary UEs is still compromised due to heavy path loss in the DL WPT phase and the consecutive UL data transmission phase. In contrast to co-located (cellular) mMIMO, a distributed topology that allows huge macro diversity, and coverage ratio can greatly improve the performance of cell-edge terminals by shortening the communication distances between the terminals and serving BSs (APs) as well as by reducing the probability of blockage. To this end, distributed antenna systems (DAS) and consequently CF-mMIMO have been deployed to improve the performance of WIPT. In the context of DAS, Fangchao \ettall ~\cite{Yuan:COMMG:2015} proposed the concept of WIPT in massive distributed antenna systems and discussed the key architectural designs in both the transmitter and receiver ends, and focused on technologies that  lead to the improvement of WPT  and WIT.  Optimal energy beamforming design in DAS  with and without coordination was investigated in~\cite{Lee:TSP:2017}. The authors in~\cite{Kim:ELLET:2016} studied the joint time allocation and energy beamforming problem to maximize the EE of WPCN with distributed antennas. In~\cite{Zhu:JSYS:2019}, the problem of joint beamforming and power splitting factor design for the multiuser DAS was addressed. CF-mMIMO systems not only leverage the advantages of the distributed systems to provide seamless energy harvesting opportunities for all UEs, but also circumvent the practical limitations and system scalability issues of the DAS, thanks to the UC architecture~\cite{interdonato2019ubiquitous}. Therefore, the integration of CF-mMIMO and WIPT has recently garnered significant research attention. 

\subsection{Literature Review}
The synergetic deployment of CF-mMIMO and WIPT has been the subject of increasing attention in recent years, for instance in~\cite{Shrestha:GC:2018, Femenias:TCOM:2021,Mohammadi:GC2023, Demir:TWC:2021,Diluka:TGCN:2021,Zhang:IoT:2022,Alageli:TIFS:2020, Wang:JIOT:2020,Wang:IoT:2021,Xinjiang:TWC:2021,Yang:SYSJ:2022,Xia:TVT:2023}. 

\subsubsection{SWIPT with Separated Energy and Information Receiver}
Shrestha and Amarasuriya~\cite{Shrestha:GC:2018} studied the performance of SWIPT-assisted CF-mMIMO networks, where two groups of UEs, i.e., energy UEs (EUs) and information UEs (IUs) are served at the same time. EUs harvest the energy using the time switching protocol in DL and then forward their information in the UL via the harvested energy. The achievable DL/UL rates for two groups of UEs and harvested energy at the EUs were derived and the impact of the time switching factor on the UL/DL energy-rate trade-off was quantified. Femenias \ettall~\cite{Femenias:TCOM:2021} considered a CF-mMIMO system over spatially correlated Ricean fading channels, capable of servicing separated ERs and IRs.  Based on the CSI estimates from the training phase, APs use ZF precoding to power up EUs during DL WPT phase. ERs use the harvested energy to transmit their UL data and, also, to transmit the pilot corresponding to the next UL training phase. A max-min power control problem to maximize the minimum of the weighted UL SINR of EUs and IUs was formulated with DL and UL transmit power constraint at both the APs (during DL energy transmit phase) and IRs (during UL data transmit phase). Finally, trade-offs among the SE, EE, and the amount of the harvested energy at the ERs were studied through simulations. Mohammadi~\ettall~\cite{Mohammadi:GC2023} proposed a joint operation mode selection and power control design in a CF-MIMO system, where certain APs are designated for energy transmission to EUs, while others are dedicated to information transmission to IUs. The problem of maximizing the total HE for EUs, subject to SE constraints for individual IUs and minimum HE requirements for individual EUs, was formulated and solved.

\subsubsection{SWIPT with Co-located Energy and Information Receiver}
Demir~\ettall~\cite{Demir:TWC:2021} developed a power control algorithm to maximize the minimum UL SE for a DL wireless-powered CF-mMIMO under the spatially uncorrelated Ricean fading. They considered the maximum ratio processing with LSFD based on either linear MMSE or least-squares channel estimation and for coherent and non-coherent energy transmission schemes. Their findings showed that coherent energy transmission increases the SE of each UE in comparison to its non-coherent counterpart, with an additional burden of DL synchronization among the APs. The performance improvement becomes more visible with the increase in the number of APs that further improves the SE.  Diluka \ettall~\cite{Diluka:TGCN:2021} investigated the feasibility of utilizing the harvested energy during DL WPT for UL data payload transmission in CF-mMIMO systems. Energy-rate trade-offs by considering the availability of statistical CSI and estimated DL CSI at the UEs were also quantified. Furthermore, with the aim to maximize the  minimum achievable UE rate and the harvested energy for both time switching and power splitting receiver structures, optimal transmit power control was derived, which enables near-far effects mitigation in SWIPT-enabled CF-mMIMO systems. Zhang \ettall~\cite{Zhang:IoT:2022} investigated the impact of normalizing the DL beamformer on the harvested energy in SWIPT-enabled CF-mMIMO systems. Accelerated projected gradient-based power control policy has been proposed to reduce the optimization run time. It has been concluded that normalized conjugate beamforming benefits from a better achievable rate, but performs unsatisfactorily in capturing RF energy compared to the conjugate beamforming. 

With the vision of providing secure communication against internal Eve, Alageli \ettall~\cite{Alageli:TIFS:2020} consider a SWIPT CF-mMIMO system, where the APs serve a large number of IUs and a single information-untrusted dual-antenna active energy-harvesting UE. The UE uses one antenna to legitimately harvest energy and the other antenna to eavesdrop information intended for a certain IU (i.e., separate SWIPT structure). Power control design at the APs was investigated to maximize the worst-case ESR, with a constraint on the minimum average harvested energy requirement of the legitimate energy harvester. 

\begin{figure}[!t]\centering \vspace{0em}
    \def\svgwidth{240pt} 
    \fontsize{8}{7}\selectfont 
\begingroup%
  \makeatletter%
  \providecommand\color[2][]{%
    \errmessage{(Inkscape) Color is used for the text in Inkscape, but the package 'color.sty' is not loaded}%
    \renewcommand\color[2][]{}%
  }%
  \providecommand\transparent[1]{%
    \errmessage{(Inkscape) Transparency is used (non-zero) for the text in Inkscape, but the package 'transparent.sty' is not loaded}%
    \renewcommand\transparent[1]{}%
  }%
  \providecommand\rotatebox[2]{#2}%
  \newcommand*\fsize{\dimexpr\f@size pt\relax}%
  \newcommand*\lineheight[1]{\fontsize{\fsize}{#1\fsize}\selectfont}%
  \ifx\svgwidth\undefined%
    \setlength{\unitlength}{1029.97503662bp}%
    \ifx\svgscale\undefined%
      \relax%
    \else%
      \setlength{\unitlength}{\unitlength * \real{\svgscale}}%
    \fi%
  \else%
    \setlength{\unitlength}{\svgwidth}%
  \fi%
  \global\let\svgwidth\undefined%
  \global\let\svgscale\undefined%
  \makeatother%
  \begin{picture}(1,0.652443)%
    \lineheight{1}%
    \setlength\tabcolsep{0pt}%
    \put(0,0){\includegraphics[width=\unitlength]{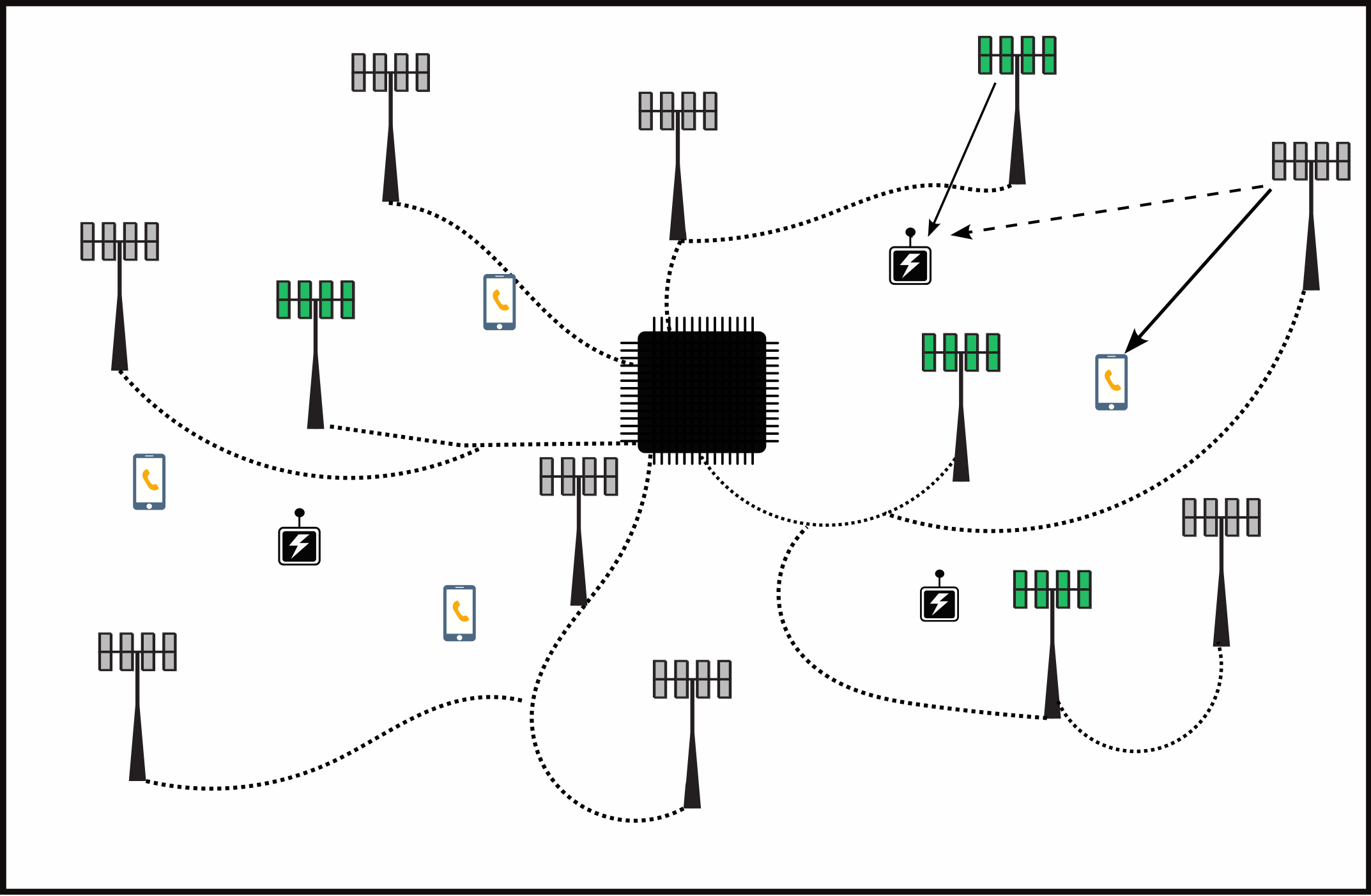}}%
    \put(0.70780467,0.30559449){\color[rgb]{0,0,0}\makebox(0,0)[lt]{\lineheight{1.25}\smash{\begin{tabular}[t]{l}E-AP\end{tabular}}}}%
    \put(0.76353847,0.51899767){\color[rgb]{0,0,0}\makebox(0,0)[lt]{\lineheight{1.25}\smash{\begin{tabular}[t]{l}Interference\end{tabular}}}}%
    \put(0.50977529,0.49125469){\color[rgb]{0,0,0}\makebox(0,0)[lt]{\lineheight{1.25}\smash{\begin{tabular}[t]{l}I-AP\end{tabular}}}}%
    \put(0.68419076,0.4390497){\color[rgb]{0,0,0}\makebox(0,0)[lt]{\lineheight{1.25}\smash{\begin{tabular}[t]{l}EU\end{tabular}}}}%
    \put(0.82547894,0.35724643){\color[rgb]{0,0,0}\makebox(0,0)[lt]{\lineheight{1.25}\smash{\begin{tabular}[t]{l}IU\end{tabular}}}}%
    \put(0.47334158,0.35437278){\color[rgb]{1,1,1}\makebox(0,0)[lt]{\lineheight{1.25}\smash{\begin{tabular}[t]{l}\textbf{CPU}\end{tabular}}}}%
  \end{picture}%
\endgroup%
 \vspace{0mm}
    \caption{CF-mMIMO system with I-APs and E-APS serving IUs and EUs at the same time.}\vspace{0mm} \label{fig:WPT_CFmMIMO}
    \vspace{2em}
\end{figure}
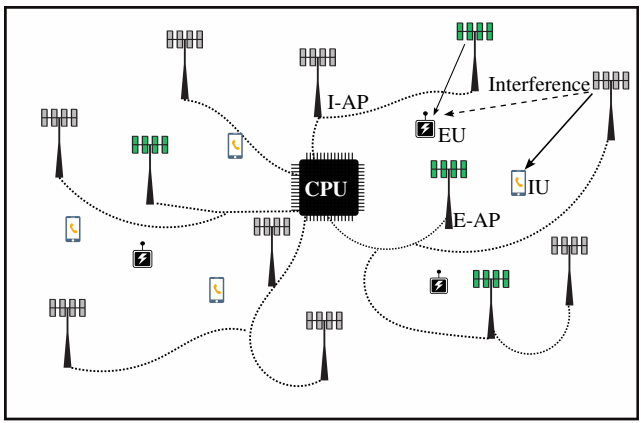
In NAFD CF-mMIMO systems, a general optimization framework for joint UE selection and transceiver design has been established, where DL UEs are equipped with time splitting~\cite{Xinjiang:TWC:2021} and power splitting~\cite{Yang:SYSJ:2022} receivers.  QoS requirements of both DL and UL UEs, fronthaul capacity constraints, energy harvesting constraint of DL UEs were considered and the superiority of the NAFD over the FD and HD designs under different setups and requirements was revealed.  Note that the EE maximization problem in NAFD, under
constraints on the UL and DL UEs' QoS requirements, fronthaul load, EH requirements and the transmit power of APs,z and UEs was studied in~\cite{Xia:TVT:2023}.   

\subsubsection{WPCN}
Wang \ettall~\cite{Wang:JIOT:2020} considered a CF-mMIMO based IoT, in which some active sensors transmit signals to APs using the harvested energy during the DL WPT. The UL and DL power control coefficients were jointly optimized to minimize the total energy consumption under the given SINR constraints. Co-located mMIMO and small-cell IoT were discussed as special cases, and the superiority of CF-mMIMO over them was shown.  Subsequently, the authors extended their work and proposed a long-term scheduling and power control optimization framework to maximize the minimum time-average achievable rate while maintaining the battery state of each sensor higher than a predefined level~\cite{Wang:IoT:2021}. By using a Lyapunov optimization approach, the transmission mode (energy harvesting or data transmission), the sensor state (active or inactive), and the UL and DL power control coefficients were jointly determined for each time slot. 
Table~\ref{tabel:WPT} shows a summary of the major related works on WPT-enabled CF-mMIMO systems.

\begin{table*}
	\centering
	\caption{\label{tabel:WPT}Summary of WIPT-enabled CF-mMIMO Literature}
	\vspace{-0.6em}
	\small
\begin{tabular}{|l|m{0.6cm}|m{0.4 cm}|m{0.70 cm}|m{0.70cm}|m{1.3cm}|m{1.2cm}|m{0.8cm}|>{\centering\arraybackslash}m{6cm}|}
	\hline
	\multirow{2}{*}{\textbf{Ref.}} & \multicolumn{2}{c|}{\textbf{Transmission}} & 
	\multicolumn{2}{c|}{\textbf{Topology}} &\multicolumn{2}{c|}{\textbf{SWIPT }}  
&\multirow{2}{*}{\textbf{WPCN}} &\multirow{2}{*}{\textbf{Technical Contribution}} \\

	\cline{2-7}
	\centering
	&\textbf{UL}  &\textbf{DL}   &\textbf{HD} &\textbf{NAFD}  &\textbf{Time switching} &\textbf{Power splitting} & & \\
\hline
	\hline
	\centering
	 	~\cite{Shrestha:GC:2018} &\centering\checkmark   &\centering\checkmark  &\centering\checkmark &\centering- &\centering\checkmark &\centering- &\centering-& UL/DL energy-rate trade-off analysis \\
	\hline	

  ~\cite{Femenias:TCOM:2021}   &\centering\checkmark   &\centering\checkmark  &\centering\checkmark &\centering- &\centering\checkmark & &\centering-& DL and UL power control design to maximize the minimum of the weighted UL SINR \\
	\hline

  ~\cite{Mohammadi:GC2023}       &\centering-   &\centering\checkmark  &\centering\checkmark &\centering- &\centering\checkmark &\centering- &\centering- & AP operation mode assignment to support EUs and IUs at the same time over the same frequency  \\
	\hline

  ~\cite{Demir:TWC:2021}       &\centering\checkmark   &\centering-  &\centering\checkmark &\centering- &\centering\checkmark &\centering- &\centering- & Joint power control and LSFD weight design to maximize the minimum guaranteed SE  \\
	\hline
 
     ~\cite{Diluka:TGCN:2021}  &\centering\checkmark   &\centering\checkmark &\centering\checkmark &\centering- &\centering\checkmark &\centering\checkmark &\centering-& Quantifying the max-min fairness optimal energy-rate trade-offs \\
	\hline

 ~\cite{Zhang:IoT:2022}&\centering-  &\centering\checkmark  &\centering\checkmark  &\centering- &\centering\checkmark &\centering- &\centering-&Design an accelerated projected gradient-based max-min power control policy to provide uniform harvested energy and achievable rate \\
	\hline	
 
 ~\cite{Alageli:TIFS:2020}&\centering-   &\centering\checkmark  &\centering\checkmark &\centering- &\centering- &\centering- &\centering-&Power control design against internal active Eve \\
	\hline 
 
	~\cite{Wang:JIOT:2020} &\centering\checkmark   &\centering-  &\centering\checkmark &\centering- &\centering- &\centering- &\centering\checkmark &UL and DL power control design for energy consumption minimization under QoS constraints \\
	\hline	

 	~\cite{Wang:IoT:2021} &\centering\checkmark   &\centering-  &\centering\checkmark &\centering- &\centering- &\centering- &\centering\checkmark & Long-term scheduling (transmission mode and sensor state) and power control optimization \\
	\hline	

   ~\cite{Xinjiang:TWC:2021}  &\centering\checkmark   &\centering\checkmark &\centering- &\centering\checkmark &\centering\checkmark & &\centering-& Joint UE selection and transceiver design, aiming at maximizing UL and DL achievable rate  \\
	\hline	
 
   ~\cite{Yang:SYSJ:2022}    &\centering\checkmark  &\centering\checkmark &\centering- &\centering\checkmark &\centering- &\centering\checkmark &\centering-&Joint UE selection and transceiver design, aiming at minimizing the total transmission power \\
	\hline
 
 ~\cite{Xia:TVT:2023}        &\centering\checkmark   &\centering\checkmark &\centering- &\centering\checkmark &\centering- &\centering\checkmark &\centering-& EE maximization under UL and DL QoS requirements, fronthaul load, and EH requirements  \\
	\hline 

 \end{tabular}
\vspace{0em}
\end{table*}

\subsection{Case Study and Discussion}
Consider a CF-mMIMO system  under TDD operation, where $M$ APs serve $K_d$ IUs and $L$ EUs with EH capabilities in the DL, as shown in Fig.~\ref{fig:WPT_CFmMIMO}. Information and energy transmissions take place simultaneously and within the same frequency band. The AP operation mode selection approach is designed according to the network requirements, determining whether an AP is dedicated to information or energy transmission. The IUs receive information from a group of the APs (I-APs), while the EUs harvest energy from the remaining APs (E-APs). The binary variables $a_m$, $m=1,\ldots, M$, are used to determine the operation mode selection for each AP. That is, $a_m=1$ indicates that AP $m$ operates as an I-AP, while $a_m=0$ indicates that AP $m$ operates as an E-AP.  During the training phase, which lasts for a duration of $\tauup$ symbols, all APs estimate their channels toward all IUs and EUs. The remaining $(\tau_c-\tauup)$ symbols are allocated for simultaneous WPT and WIT toward EUs and IUs, respectively. The EUs utilize  the harvested energy to transmit pilots and data. The transmitted signal from AP $m$ is  
\begin{align}
    \qx_{m} = \sqrt{a_m} \qx_{\mathtt{I},m} +\sqrt{(1-a_m)}\qx_ {\mathtt{E},m},
\end{align}
where $\qx_{\mathtt{I},m}=\sqrt{\rho_d}\sum_{k=1}^{K_d}\sqrt{\etamkI}\wimk \xik$ and $\qx_ {\mathtt{E},m}=\sqrt{\rho_d}\sum_{\ell=1}^{L}  \sqrt{\etamlE}\weml \xel$; $\wimk \in \C^{N\times 1}$ and $\weml\in \C^{N\times 1}$ are the precoding vectors for IU $k\in\K_d=\{1,\ldots,K_d\}$ and EU $\ell\in\mathcal{L}=\{1,\ldots,L\}$, respectively, with $\Ex\big\{\big\Vert\wimk\big\Vert^2\big\}=1$ and $\Ex\big\{\big\Vert\weml\big\Vert^2\big\}=1$. Note that AP $m$ can only transmit either $\qx_{\mathtt{I},m}$ or $\qx_ {\mathtt{E},m}$, depending on its assigned operation mode. Let $\etamkI$ and $\etamlE$ denote the DL power control coefficients chosen to satisfy the power constraint at each AP, i.e., $a_m\Ex\big\{\big\Vert \qx_{\mathtt{I},m}\big\Vert^2\big\}+(1-a_m)\Ex\big\{\big\Vert \qx_{\mathtt{E},m}\big\Vert^2\big\}\leq \rho_d$. The PPZF scheme is utilized at the APs, where local PZF precoding is deployed at the I-APs and protective MRT is used at the E-APs. The principle behind using MRT at E-APs is that MRT is the optimal beamformer for power transfer, maximizing the harvested energy when the number of antennas is large \cite{almradi2016performance}.

Different design goals can be considered to assign AP operation modes. For example, AP operation mode selection vectors ($\aaa$) along with power control coefficients ($\ETAI = [\eta_{m1}^{\mathtt{I}}, \ldots, \eta_{mK_d}^{\mathtt{I}}]$, $ \ETAE = [\eta_{m1}^{\mathtt{E}}, \ldots, \eta_{mL}^{\mathtt{E}}]$) can be optimized to maximize the average sum harvested energy, subject to minimum power requirements at the EUs, per-IU SE constraints, and transmit power at each APs. The optimization problem is mathematically formulated as
\vspace{-0.1em}
\begin{subequations}\label{P:SHE:max}
	\begin{align}
		\underset{\aaa, \ETAI, \ETAE}{\max}\,\, \hspace{0.5em}&
		\sum\nolimits_{\ell\in \mathcal{L}} 
  \Ex\left\{\Phi_{\ell}\big(\aaa,\ETAI, \ETAE\big)\right\}
		\\
		\mathrm{s.t.} \,\,
		\hspace{0.5em}& \Ex\left\{\Phi_{\ell}\big(\aaa,\ETAI, \ETAE\big)\right\} \geq \Gamma_{\ell},~\forall \ell\in\mathcal{L},\\
		& \mathrm{SE}_k (\aaa,  \ETAI, \ETAE)  \geq \SEQoS,~\forall k \in \K_d,\label{eq:SE:ct1}\\
			&\sum\nolimits_{k=1}^{K_d}
        {\etamkI}\leq a_m,
        ~\forall m\in\MM,\label{eq:etamkI:ct1}\\
       &\sum\nolimits_{\ell=1}^{L}
        {\etamlE}\leq 1-a_m,~\forall m\in\MM,\label{eq:etamellE:ct1}\\
		& a_m \in\{ 0,1\},\label{eq:am:ct1}
		\end{align}
\end{subequations}
where $\mathrm{SE}_k (\aaa,  \ETAI, \ETAE)$  denotes the DL SE for IU $k$ in $\bsHz$ given in~\eqref{eq:SINE:PPZF} at the top of the next page;
\begin{figure*}
\begin{align}~\label{eq:SINE:PPZF}
    \mathrm{SE}_k (\aaa,  \ETAI, \ETAE) \!=\!\frac{\tau_d}{\tau_c} \log_2\left(
    \!\frac{
                  \rho \big(N-K_d\big)\Big(\sum_{m\in \MM}\sqrt{ a_m\etamkI \gamuemk}  \Big)^2
                 }
                 { \rho \sum_{m\in \MM}
                 \sum_{k'\in\K_d}\!\!
  a_m\etamkpI 
  \Big(\betamkue\!-\!\gamuemk\Big)
                  \! + \!
                 \rho
                 \sum_{m\in \MM}
                 \sum_{\ell\in\mathcal{L}}
   { (1\!-\!a_m)\etamlE} \Big(\betamkue\!-\!\gamuemk\Big)
                   \!+\!  1}\right).
\end{align}
	\hrulefill
	\vspace{-4mm}
\end{figure*}
$\SEQoS$ is the minimum SE required by the $k$-th IU; $\Gamma_{\ell}$ is the minimum required harvested energy at EU $\ell$; $\Phi_\ell\big(\aaa,  \ETAE,\ETAI\big) = \frac{\Psi_{\ell}\big(\mathrm{E}_{\ell}(\aaa,  \ETAE, \ETAI)\big) - \phi \Omega }{1-\Omega}, ~\forall \ell$, is the total harvested energy at EU $\ell$,
where  $\phi$ is the maximum output DC power, $\Omega=\frac{1}{1 + \exp(\xi \chi)}$ is a constant to guarantee a zero input/output response, while $\xi$ and $ \chi$ are constant related parameters that depend on the circuit and  $\Psi\big(\mathrm{E}_{\ell}(\aaa,  \ETAE, \ETAI)\big)$  is the traditional logistic function, 
 \vspace{0.1em}
  \begin{align}~\label{eq:PsiEl}
     \Psi_{\ell}\big(\mathrm{E}_{\ell}(\aaa,  \ETAE,  \ETAI)\big) &\!=\!\frac{\phi}{1 \!+ \!\exp\Big(-\xi\big(\mathrm{E}_{\ell}(\aaa, \ETAE,\ETAI)\!-\! \chi\big)\Big)},
 \end{align}
where $\mathrm{E}_{\ell}(\aaa, \ETAE,\ETAI)$ denotes the received RF energy at EU $\ell$, $\forall \ell\in\mathcal{L}$~\cite{Boshkovska:CLET:2015}. The solution for the optimization problem~\eqref{P:SHE:max} was provided in~\cite{Mohammadi:GC2023}.

\begin{figure}[t]
	\centering
	\vspace{-0.8em}
	\includegraphics[width=0.47\textwidth]{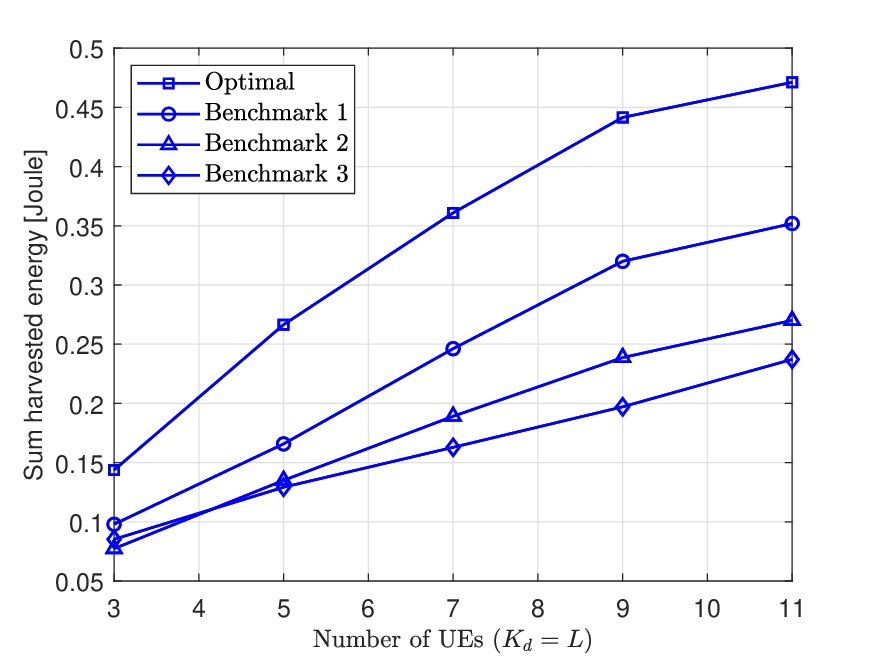}
	\vspace{-0.7em}
	\caption{Average of the sum harvested energy ($N=10$, $M=50$, $\Gamma_{\ell}=100$ $\mu$Joule,  and $\SEQoS = 1$ bit/s/Hz).}
	\vspace{0.4em}
	\label{fig:HE}
\end{figure}


Figure~\ref{fig:HE} shows the average sum harvested power achieved by the proposed scheme and the benchmark schemes as a function of the number of UEs. The optimal result achieved by solving~\eqref{P:SHE:max} is labeled as \textit{Optimal}. For comparison, three benchmark system designs are studied: i) \textit{Benchmark 1}, at which the operation mode selection parameters for the APs, denoted as ($\aaa$), are randomly assigned, while the power control coefficients ($\ETAI$, $\ETAE$) are optimized under the same SE requirement constraints as in the optimal case, ii) \textit{Benchmark 2}, in which the APs' operation mode selection parameters ($\aaa$) are randomly assigned, and no power control is performed at the APs, i.e., $\etamkI = \frac{1}{K_d}$, $\forall k\in\mathcal{K}_d$  and $\etamlE = \frac{1}{L}$, $\forall \ell\in\mathcal{L}$, iii) \textit{Benchmark 3}, wherein all APs are used for DL WIT and WPT over orthogonal time frames of equal length $(\tau_c-\tauup)/2$. It is observed that the optimal scenario achieves superior performance, while Benchmark 1 consistently outperforms the other benchmarks. These results illustrate the significance of AP mode selection and power control design within the network.  

\subsection{Future Research Directions}
Possible future direction areas in WIPT-enabled CF-mMIMO networks are considering  EE-related optimization problems with joint AP clustering and UE scheduling, and radio resource management to cope with practical challenges, such as limited-capacity fronthaul links between the APs and CPU. Moreover, NAFD CF-mMIMO networks can be efficiently developed to support separate ERs and IRs at the same time. To this end,  part of the APs can be assigned for WPT to the ERs and the remaining APs can be scheduled to serve UL and DL data transmission from information UEs. EE-related optimization problems can be developed to design the AP's mode of operation (energy transfer/DL data transfer/UL data receive) subject to different  QoS requirement levels at the energy and information receivers.

\section{Cell-free Massive MIMO and Millimeter Wave Communication}~\label{sec:mmwave}
MmWave communication over the $30$-$300$ GHz spectrum to support the emerging  bandwidth-demanding services, such as augmented/virtual reality and AI,  is one of the promising technologies for the 6G systems, which can be used as complement to the current sub-$6$ GHz band~\cite{Rappaport:ACCESS:2013,Heath:JSTSP.2016,Xiao:JSAC:2017}. While the wireless communication distance is severely restricted, owing to serious path-loss attenuation and high sensitivity to blockage, the short wavelength of mmWave frequencies, on the other hand, allows for a dense packing of large antenna arrays for highly directional beamforming. From these aspects, the synergistic integration between the mmWave and CF-mMIMO with deployment of large number of antennas at the APs envisions a significant improvement in the system performance in terms of SE and network connectivity~\cite{Guo:TWC:2021,Jafri:TVT:2023,Nguyen:TVT:2022,Femenias:TWC:2022,He:CLET:2023}.

Nonetheless, operating over mmWave frequencies brings up new challenges for CF-mMIMO, which need to be well understood before its roll-out. Considering the fact that mmWave communications can offer higher data rates than conventional sub $6$-GHz communications, implementation of mmWave CF-mMIMO systems entails high capacity fronthaul links to transfer data between the APs and CPU. This is due to the fact that in DL (UL), all UEs' data should be transmitted from the CPU (APs) to the APs (CPU).  Moreover, channel estimation is perceived as an extremely challenging task for mmWave CF-mMIMO systems, due to the large communication bandwidth and strongly spatially correlated channels from the APs to the UEs due to the close location of the APs~\cite{Kim:TVT:2022}. Furthermore, due to the sparsity of the mmWave channels in the spatial domain, the number of operating simultaneous connections becomes limited~\cite{Wang:JSAC:2017}. Finally, the power consumption of  mmWave CF-mMIMO systems, may grow dramatically as network densification, high capacity fronthaul links, mMIMO antenna arrays at APs and associated complicated signal processing tasks for beamforming and channel estimation demand an escalated amount of energy. 

\begin{figure}[t]
    \def\svgwidth{240pt} 
    \fontsize{8}{7}\selectfont 
    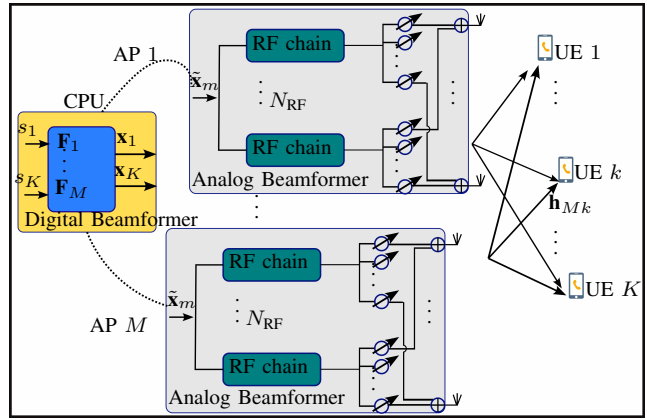 \vspace{0mm}
    \caption{Illustration of hybrid beamforming in cell-free mmWave MIMO systems.}\vspace{0mm} \label{fig:mmwave_CFmMIMO}
    \vspace{2em}
\end{figure}

\vspace{-1em}
\subsection{Literature Review}
In this subsection, we summarize the most notable contributions that have dealt with the aforementioned challenges and the development of mmWave CF-mMIMO systems. Moreover, the corresponding key design insights are provided.

\subsubsection{Hybrid Beamforming}
Deployment of the conventional fully-digital transceiver architecture in mmWave mMIMO systems imposes a high cost and power consumption, due to the need of per-antenna RF chains and  ADCs~\cite{Liang:WCL:2014,Han:MCOM:2015,Sohrabi:JSTSP:2016}. To make mMIMO at mmWave practically feasible, \emph{hybrid beamforming}, i.e., a cascaded combination of high-dimensional analog-precoder with low-dimensional digital precoder, has been proposed in the literature as a substitute of full-dimensional digital precoders~\cite{Liang:WCL:2014,Han:MCOM:2015,Sohrabi:JSTSP:2016}. Consider the cell-free mmWave mMIMO system shown in Fig.~\ref{fig:mmwave_CFmMIMO}, where each AP is equipped with $N_{\text{RF}}$ RF chains and $N$ antennas, where $N_{\text{RF}}\leq K$. Let $s_k$, $k=1,\ldots,K$, denote the transmit signal for UE $k$. Then, by using the digital beamforming at the CPU, $\qx_{i}=\sum_{k=1}^{K} \qf_{i,k} s_k \in \mathbb{C}^{N_{\text{RF}}\times 1}$ is precoded for the AP $i$, with $\qf_{i,k} \in \mathbb{C}^{N_{\text{RF}}\times 1}$ is the digital beamformer utilized for the transmission between the AP $i$ and the UE $k$ and $\qF_i = [\qf_{i,1},\ldots,\qf_{i,K}]\in \mathbb{C}^{N_{\text{RF}}\times K}$, for $i=1,\ldots,M$. The analog beamformer at AP $i$ is denoted by $\qF_{\text{RF}_i} \in\mathbb{C}^{N\times N_{\text{RF}}}$ and realized by
a phase-shifter network. By applying the analog beamformer, the received signal at UE $k$ can be expressed as
\begin{align} 
y_{k}&= \underbrace {\sum\nolimits_{m =1}^{M}  \mathbf {h}^{H}_{mk}\mathbf {F}_{\mathrm {RF}_{m}} \mathbf {f}_{{mk}} s_{k}}_{\textrm {desired signal}} + 
\\&+\underbrace {\sum\nolimits_{l=1, l \neq k}^{K} \sum\nolimits_{m = 1}^{M}  \mathbf {h}^{H}_{mk}\mathbf {F}_{\mathrm {RF}_{m}} \mathbf {f}_{{m,l}} s_{l}}_{\textrm {inter-user interference}} + n_{k},
\end{align}
where $n_k\sim \mathcal{CN}(0,1)$.

Given the sparse nature of the mmWave channels, codebook-based hybrid beamforming designs are commonly used, where the column of the analog-beamformer are selected from specific candidate vectors, such as the array response vector of the channel~\cite{Ayach:TWC:2014,Alkhateeb:TWC:2015}. Similar to the fully-digital sub-$6$ GHz wireless systems, digital precoding is used to multiplex independent data streams and to mitigate interference. ZF, MMSE, and MRT/maximum ratio combining (MRC) digital filters are preferred in baseband processing of the mmWave communication systems~\cite{Long:JSTSP:2018}. In mmWave multi-user scenarios, a digital precoder can provide higher degrees-of-freedom than an analog beamformer and hence can be used for interference cancellation among different UEs.

Hybrid beamforming design has been so far mainly addressed for point-to-point communication and classic cellular systems~\cite{Han:MCOM:2015,Sohrabi:JSTSP:2016,Ayach:TWC:2014,Alkhateeb:TWC:2015,Long:JSTSP:2018}, while fewer studies~\cite{Guo:TWC:2021,Jafri:TVT:2023,Nguyen:TVT:2022,Femenias:TWC:2022} have been carried out in the space of mmWave CF-mMIMO. More specifically, Guo~\ettall~\cite{Guo:TWC:2021} proposed a hybrid precoding scheme with two heuristic statistical-CSI based analog-beamforming designs to minimize the simplified sum-mean-square-error objective function for mmWave cell-free and UC mMIMO systems. Moreover, a theoretical analysis on the impact of imperfect state information of beams was presented, from both instantaneous and ergodic aspects, to verify the robustness of the proposed hybrid scheme.   Jafri~\ettall~\cite{Jafri:TVT:2023} developed optimal hybrid beamformers for broadcast, unicast and multicast mmWave CF-mMIMO systems in the presence of interuser/intergroup interference. Moreover, they developed a successive UL hybrid beamforming scheme that maximizes the SINR.

With the vision of reducing the complexity of the beamforming design in a centralized manner, Nguyen~\ettall~\cite{Nguyen:TVT:2022} studied hybrid beamforming design for UL CF-mMIMO systems and proposed two designs, namely decentralized and semi-centralized hybrid beamformer design. In the former case, analog and digital beamformes are designed locally at each AP. In contrast, in the latter design, analog beamformers are generated at the CPU, using the global CSI received from all APs and the digital beamformers are designed locally at each AP. Decentralized beamforming design can achieve approximately the same achievable rate as semi-centralized one, with substantially lower computational complexity and no CSI exchange between the APs and CPU.

On the other side, the beam squint caused by the spatial-wideband effect deteriorates the system performance. Moreover, the application of numerous RF chains increases the power consumption and reduces the global EE. The beam squint occurs when receiving different resolvable versions of the same electromagnetic wave from the same propagation path with different propagation delays, due to high bandwidth, very high carrier frequency and large-scale antenna array. In~\cite{Femenias:TWC:2022}, the impact of beam squint on the design of statistical CSI-based hybrid beamformers is addressed in the context of wideband mmWave CF-mMIMO orthogonal frequency division modulation networks. He~\ettall~\cite{He:CLET:2023} focused on  designing energy-efficient wideband hybrid precoders utilizing low-resolution phase shifters for fully connected and subarray-based phase shifter architectures. To achieve this, they  developed a beam squint compensation algorithm employing an iterative heuristic Gram–Schmidt approach. This algorithm facilitates the design of wideband hybrid precoder with the maximal number of RF chains. Leveraging the proposed hybrid precoders, the number of RF chains was optimized as a shortest-path problem.

\begin{table*}
	\centering
	\caption{\label{tabel:MWAVe} Summary of mmWave CF-mMIMO Literature}
	\vspace{-0.6em}
	\small
\begin{tabular}{|l|m{0.9cm}|m{0.9cm}|m{0.70cm}|m{0.70cm}|>{\centering\arraybackslash}m{10cm}|}
	\hline
	\multirow{2}{*}{\textbf{Ref.}} & \multicolumn{2}{c|}{\textbf{CSI}} & 
	\multicolumn{2}{c|}{\textbf{Transmission}} & \multirow{2}{*}{\textbf{Technical Contribution}}\\

	\cline{2-5}
	\centering
	  &\textbf{Stat.}  &\textbf{Instant.} &\textbf{UL} &\textbf{DL}  & \\
\hline
	\hline
	\centering

~\cite{Guo:TWC:2021} &\centering-   &\centering\checkmark &\centering- &\centering\checkmark & Hybrid precoding to minimize the simplified sum-mean-square-error \\
	\hline	

~\cite{Jafri:TVT:2023} &\centering-   &\centering\checkmark &\centering\checkmark &\centering\checkmark &Hybrid beamforming design for broadcast, unicast and multicast scenarios\\
	\hline

~\cite{Nguyen:TVT:2022} &\centering-   &\centering\checkmark  &\centering\checkmark &\centering- &Decentralized and semi-centralized hybrid beamformer design \\
	\hline


~\cite{Femenias:TWC:2022} &\centering\checkmark   &\centering- &\centering- &\centering\checkmark &Beam squint-aware hybrid beamforming design \\
	\hline 

~\cite{He:CLET:2023} &\centering-  &\centering\checkmark  &\centering- &\centering\checkmark &Energy efficient wideband hybrid
precoding design using low-resolution phase shifters based on a resilient beam squint compensation method\\
	\hline
 
~\cite{Kim:TVT:2022} &\centering-   &\centering\checkmark &\centering- &\centering\checkmark &UL channel estimation and DL precoding design with low-capacity fronthaul links and low-resolution ADC/DACs \\
	\hline
 
 ~\cite{Alonzo:TGCN:2019} &\centering-   &\centering\checkmark &\centering\checkmark &\centering\checkmark & EE maximization via DL power control with fixed hybrid beamforming \\
	\hline	
 
 ~\cite{Wang:TCOM:2022} &\centering-  &\centering\checkmark  &\centering- &\centering\checkmark &Joint UE association, hybrid beamforming, and fronthaul compression design \\
	\hline	
 
 ~\cite{Femenias:ACCESS:2019} &\centering-   &\centering\checkmark &\centering\checkmark &\centering\checkmark &Performance analysis in the presence of capacity-constrained fronthaul links
  \\
	\hline	

  ~\cite{Wang:WCL:2022} &\centering\checkmark   &\centering- &\centering- &\centering\checkmark & Joint beamforming and AP-UE association  \\
	\hline	

   ~\cite{li2023network} &\centering-   &\centering\checkmark &\centering\checkmark &\centering\checkmark &Optimizing NAFD system under fronthaul compression capacity constraints\\
	\hline	
 
  ~\cite{Morales:ACCESS:2020} &\centering\checkmark   &\centering- &\centering\checkmark &\centering\checkmark &Developing energy efficient AP sleep mode techniques \\
	\hline	

~\cite{Yetis:JOPC:2021} &\centering-   &\centering\checkmark &\centering- &\centering\checkmark &Joint analog beam selection and digital filter design via  ML algorithms \\
	\hline

~\cite{Buzzi:TWC:2022} &\centering-   &\centering\checkmark &\centering\checkmark &\centering\checkmark &UC AP-UE association and beam alignment \\
	\hline

\end{tabular}
\vspace{-1em}
\end{table*}
\subsubsection{AP Association/Activation and Fronthaul Quantization}
To achieve further energy consumption and reduce the amount of fronthaul data transmission, UC clustering or AP-UE association concept and/or AP (de)activation with hybrid beamforming can be deployed in mmWave communication systems~\cite{Alonzo:TGCN:2019, Wang:TCOM:2022,Femenias:ACCESS:2019,Wang:WCL:2022, li2023network, Morales:ACCESS:2020}. AP selection schemes, in which each UE chooses and connects to only a subset of APs, reduce the power consumption caused by the fronthaul links. Alonzo~\ettall~\cite{Alonzo:TGCN:2019} investigated UC CF-mMIMO systems at mmWave frequencies, considering the training-based channel estimation and DL/UL data transmission phases. The problem of DL power control was formulated to maximize the EE of the network, in the presence of hybrid analog-digital partial ZF beamforming at the APs and simple $0-1$ beamforming architecture at the UEs. Wang~\ettall~\cite{Wang:TCOM:2022} concentrated on joint UE association, hybrid beamforming, and fronthaul compression design with the aid of UL training. They considered a CF-mMIMO system based on cloud radio access network, where multiple RRHs are distributed to communicate with UEs via analog beamforming, and connected to a centralized baseband unit through fronthaul links which executes digital beamforming. The baseband unit optimizes the digital beamforming and fronthaul compression based on the training results and through the weighted sum-rate maximization and max-min fairness design criteria. Feminias and Riera-Palou~\cite{Femenias:ACCESS:2019} developed an analytical framework for the performance analysis of the mmWave CF-mMIMO systems, considering hybrid precoders and assuming the availability of capacity-constrained fronthaul links connecting the APs and the CPU. They proposed a  UE selection algorithm for the cases in which the number of active UEs in the networks is greater than the number of RF chains at a particular AP. Wang~\ettall~\cite{Wang:WCL:2022} studied the problem of joint beamforming and AP-UE association to maximize the minimum average received signal power among all UEs, by considering statistical-CSI.

To provide simultaneous UL and DL transmissions, Li~\ettall~\cite{li2023network} studied NAFD mmWave CF-mMIMO networks, considering DAC quantization and fronthaul compression. They proposed to maximize the weighted UL and DL sum rate by jointly optimizing the power allocation of both the transmitting remote antenna units and UL UEs and the variances of the DL and UL fronthaul compression noise. Finally, the superiority of the NAFD over NAFD co-time co-frequency FD cloud radio access network in the cases of practical limited-resolution DACs was quantified.

Another line of research has established that a fraction of AP can be dynamically (de)activated in response to variations in the UE locations and traffic demands, thereby improving the EE. Morales~\ettall~\cite{Morales:ACCESS:2020} proposed energy efficient AP sleep mode techniques for mmWave CF-mMIMO networks that are able to capture the inhomogeneous nature of spatial traffic distribution in the networks. 

\subsubsection{Beam Selection and Beam Alignment}
Analog beamforming in mmWave is usually designed through the beam alignment process, based on a codebook of different patterns. The aim of this process is to find the angular directions of an active link with sufficiently high signal strength between the receiver and transmitter, without explicit channel estimation~\cite{Heath:JSTSP.2016}.  The assignment of the same beam to multiple UEs in the network results in beam conflict in mmWave networks, which can significantly reduce the sum-rate of the network~\cite{Sun:TWC:2019}. In CF-mMIMO systems, the complexity of the beam alignment increases as the number of active links between the APs and UEs becomes large. This, on the other hand, increases the probability of beam conflict. To address this issue, Yetis~\ettall~\cite{Yetis:JOPC:2021} proposed low-cost joint designs of analog beam selection and digital filters. The input-output mapping functions of the beam selection decisions of the joint designs were efficiently approximated via supervised  ML algorithms. Buzzi~\ettall~\cite{Buzzi:TWC:2022} developed a protocol to estimate for each UE, the strongest path from the surrounding APs, and perform UC AP-UE association. Two beam alignment algorithms were proposed to enable UEs to  discriminate the signals coming from different APs. 

\subsubsection{Channel Estimation}
Channel estimation is a critical module for configuring the hybrid beamformers. Analog beamformers make the implementation of the conventional channel estimation methods difficult in mmWave communication systems. More specifically, the channel measured in the digital baseband is a function of the analog beamformer and, thus, the channel matrix cannot be directly measured~\cite{Heath:JSTSP.2016}. Furthermore,  due to the large number of elements in the transmit array and also high communication  bandwidth, long training sequences are needed if the classical channel estimation methods are employed. By using compressive adaptation techniques, which leverage mmWave channel spatial sparsity,  the estimation of the channel can be obtained from a small set of RF measurements. We refer the interested reader to~\cite{Heath:JSTSP.2016} for more details on the channel estimation of mmWave channels. 

In CF-mMIMO systems, due to the large number of APs, the distances between the APs are small. Therefore, the channels between the APs and UEs become strongly correlated, which makes the channel estimation more complicated. Moreover, limited-capacity fronthaul must be taken into consideration, when the channel estimation schemes are designed.  Kim~\ettall~\cite{Kim:TVT:2022} investigated the  UL channel estimation and DL precoding design in mmWave CF-mMIMO systems with low-capacity fronthaul links and low-resolution ADC/DACs. They optimized  the codebook design problem associated with the fronthaul compression in the UL channel estimation phase, where the goal is to minimize the channel estimation error. 

\subsection{Future Research Directions}
Utilizing FD transceivers in mmWave systems has the potential to accelerate the next generation of wireless networks, offering increased SE gains and reduced latency. However, realizing mmWave FD with large antenna arrays remains a challenging issue~\cite{Xiao:MWC:2017,Roberts:WMC:2021}. As an alternative, NAFD CF-mMIMO architecture can be employed in the mmWave frequency band to address the simultaneous UL and DL data transmission demands. Nevertheless, the problem of joint AP mode assignment, hybrid beamforming design, and power control becomes more  challenging compared to the conventional sub-$6$ GHz systems. Finding scalable and low-complexity approaches for SE and/or EE-related problems, such as ML-based algorithms, is a timely future research direction.

Further, the synergy of CF-mMIMO and terahertz communication ($300$ GHz–$10$ THz)~\cite{Sarieddeen:PIEEE:2021} remains largely an untapped field of future technology that has the potential to bring about drastic changes to how we live today. Out of many obstacles, channel estimation in the THz band is very challenging in CF-mMIMO networks, where accurate CSI is required in the beamforming mechanisms and for accurately directing beams to avoid misalignment issues~\cite{Sarieddeen:PIEEE:2021}. Moreover, how to design  signal processing algorithms with fronthaul-limited capacity should be investigated.  Table~\ref{tabel:MWAVe} summarizes a number of existing contributions to mmWave
CF-mMIMO networks.

In high-density mmWave CF-mMIMO networks, the probability of users entering the near-field zone of the APs increases.\footnote{The electromagnetic radiation field can generally be divided into far-field and near-field regions. The boundary between these two regions is determined by the Rayleigh distance, which is proportional to the product of the square of the array aperture and the carrier frequency. In the near-field region, we encounter spherical wavefronts, whereas far-field electromagnetic propagation is effectively approximated by planar waves~\cite{Zhang:MCOM:2023,liu2024near}.}  Different from beam steering in conventional far-field communication, where a multi-antenna transmitter sends electromagnetic signals in a specific direction, beam focusing in the near-field avails of the spherical wavefronts to concentrate the radiated energy at a specific spatial location. This involves focusing not only by angle but also by a specific depth along the direction of propagation~\cite{Zhang:MCOM:2023,liu2024near}. In this environment, the ability of mmWave CF-mMIMO to utilize near-field communications can lead to more efficient spectrum usage and improved throughput. Specifically, by leveraging beam focusing, i.e., generating focused beams in specific spatial regions, new levels of interference mitigation can be achieved in the network~\cite{Zhang:MCOM:2023}. More importantly, near-field communications in CF-mMIMO systems can enable advanced applications, such as high-precision sensing, imaging, and localization, which are particularly useful in industrial and IoT scenarios. From a security perspective, the directional nature of near-field communications can enhance privacy, making it harder for unauthorized users to intercept signals~\cite{Zhang:MCOM:2023,liu2024near}. However, harnessing such capabilities in mmWave CF-mMIMO introduces new design and signal processing challenges, presenting an interesting direction for future research.

\section{Cell-free Massive MIMO and Reconfigurable Intelligent Surfaces  }~\label{sec:RIS}
RISs are a recent technological breakthrough that holds the potential of intelligently improving the network infrastructures, making the wireless environment flexible enough to automatically adapt to the wireless scenario changes, and customizing the network according to traffic conditions and requirements~\cite{Renzo:JSAC:2020,Wu:TCOM:2021}.
A RIS comprises an array of passive~\cite{Renzo:JSAC:2020}, active~\cite{Zhang:TCOM:2023}, or hybrid (active/passive)~\cite{Nguyen2:TVT:2022,Yigit:TWC:2022} reflecting elements for reconfiguring the impinging signals. By adaptively adjusting the phase  shift of the RIS elements, radiated signals from the RIS can be constructively or destructively combined with the signals from the other paths to enhance the received signal power at the desired UE, or to mitigate/degrade the undesired signals (with application in interference management in multiuser networks or PLS perspective). 

\begin{figure}[!t]\centering \vspace{0em}
    \def\svgwidth{240pt} 
    \fontsize{8}{7}\selectfont 
     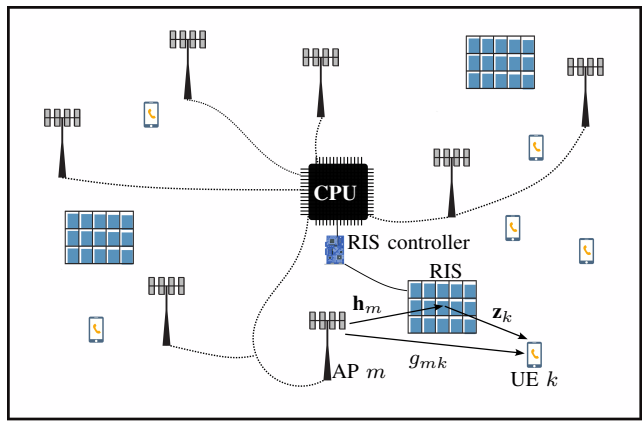 \vspace{0mm}
    \caption{Illustration of a RIS-aided CF-mMIMO network.}\vspace{0mm} \label{fig:RIS_CFmMIMO}
    \vspace{2em}
\end{figure}

Passive RISs which merely reflect the incident signal, eliminates the use of receive/transmit RF chains and operates in short range, thus, can be deployed densely with scalable cost and low energy consummation. Nevertheless,  RIS-assisted wireless communication systems, with fully passive RISs, suffer from multiplicative path attenuation~\cite{Zhang:TCOM:2023}. In scenarios with strong direct links, the capacity achieved through the RIS link is limited,  compromising the promising potential of the RIS. At the expense of additional power consumption, active RIS  have the ability to actively reflect signals with amplifications, and compensate for the large path loss of the reflected links~\cite{Long:TWC:2021,Zhang:TCOM:2023}. A hybrid relay-reflecting RIS, with only a few active elements to amplify the incident signals, leverages the advantages of both passive RIS and active FD amplify-and-forward relay to offer, not only the reflecting, but also relaying gains to the aided systems~\cite{Nguyen2:TVT:2022,Yigit:TWC:2022}.  

In scenarios of poor scattering environments or high attenuation due to the presence of large obstacles, CF-mMIMO systems fail to provide satisfactory QoS. A large-scale deployment of the APs  guarantee favorable performance and overcome these challenges, yet on the other hand, may lead to unsatisfactory EE performance due to enormous hardware cost and power dissipation~\cite{Hien:TGCN:2018}. To assist communications between the APs and UEs and providing wireless services for the blind spots, multiple low-cost energy-efficient  RISs or HR-RISs can be deployed to provide a new dimension to CF-mMIMO systems to enhance their SE and EE~\cite{van2021reconfigurable,Shi:TVT:2022,Nguyen:TWC:2022,Gan:TCOM:2022,Huang:WCL:2021,Shi:JSAC:2023,Zhang:JCCN:2021,Shi:MCOM:2022,Lyu:WCL:2023,Zhang:TSP:2021,Ma:TWC:2023,Elhoushy:WCL:2022,Hao:TIFS:2022} (cf.Fig~\ref{fig:RIS_CFmMIMO}). Nevertheless, channel acquisition and node coordination, which is formidable even for conventional CF-mMIMO networks, is typically exacerbated with the deployment of RISs. Moreover, RISs will impact the  complexity and signal processing methods and synergy between individual communication devices, bringing new challenges and opportunities for the network design.

\subsection{Literature Review}
\subsubsection{Spectral Efficiency Analysis and Optimization}
The complementary features of CF-mMIMO and RIS can be leveraged to enhance the SE of wireless systems in the harsh propagation environments. Trinh \ettall~\cite{van2021reconfigurable} developed an analytical framework for analyzing and optimizing the UL and DL transmissions of RIS-assisted  CF-mMIMO systems under spatially correlated channels and in the presence of direct links subject to the presence of blockages.  They developed an efficient channel estimation scheme to overcome the high overhead that may be associated with the estimation of the individual channels of the RIS elements. Shi \ettall~\cite{Shi:TVT:2022} provided an analysis of the UL SE in the presence of channel estimation error and under spatially correlated channels. The analysis revealed that designing the inter-distance among RIS elements as at least half wavelength reduces the detrimental effect of spatial correlation and, hence, improves the SE performance.  As a further development, Nguyen \ettall~\cite{Nguyen:TWC:2022} proposed a hybrid relay-reflecting RIS  for improving the SE of the CF-mMIMO systems, with respect to the state-of-art approaches, e.g., conventional CF-mMIMO and RIS-aided CF-mMIMO systems. Moreover, the capability of the HR-RISs to overcome the multiplicative path loss in the reflecting channels was shown. To reduce the channel estimation overhead, Gan \ettall ~\cite{Gan:TCOM:2022} applied the two-timescale transmission protocol proposed in~\cite{Zhao:TWC:2021} to the RIS-assisted CF-mMIMO systems. Transmissions are performed over time frames, consisting of several coherence interval, within which the statistical CSI remains unchanged. At the beginning of each time frame, the statistical CSI of all links is estimated, while the instantaneous CSI from APs to UEs is obtained at the beginning of each coherence interval. Then, long-term RIS phase shifts are optimized according to the statistical CSI and the short-term transmit precoding vectors at the AP are adapted to the instantaneous effective CSI with fixed phase shifts. To avoid the instantaneous CSI exchange among BSs via backhauling and reduce the computation complexity at CPU, Huang~\ettall~\cite{Huang:WCL:2021} proposed a decentralized design framework for cooperative beamforming in RIS-aided CF-mMIMO networks, which maximizes the weighted DL sum-rate via jointly optimizing the APs' digital beamformers and the RIS's analog beamformers. Shi~\ettall~\cite{Shi:JSAC:2023} derived the UL SE of RIS-assisted CF-mMIMO systems over spatially correlated fading channels, taking into account the spatial electromagnetic interference at the RIS. This study considered local MR combining at the APs and the LSFD design at the CPU (Level-3 processing). Numerical results in~\cite{Shi:JSAC:2023} indicated that the electromagnetic interference significantly degrades the performance of UEs with unsatisfactory channel conditions. Moreover, increasing the number of AP antennas pronounce the negative impact of electromagnetic interference. On the other hand, increasing the number of RIS elements is always beneficial as it provides more degrees-of-freedom to alleviate the impairment caused by electromagnetic interference.

\begin{table*}
	\centering
	\caption{\label{tabel:RIS}Summary of RIS-assisted CF-mMIMO Literature}
	\vspace{-0.6em}
	\small
\begin{tabular}{|l|>{\centering\arraybackslash}m{1cm}|>{\centering\arraybackslash}m{1cm}|m{0.95cm}|m{0.95cm}|>{\centering\arraybackslash}m{10cm}|}
\hline
\multirow{2}{*}{\textbf{Ref.}} & \multicolumn{2}{c|}{\textbf{CSI}} & 
\multicolumn{2}{c|}{\textbf{RIS Topology}} & \multirow{2}{*}{\textbf{Technical Contribution}}\\

\cline{2-5}
\centering
&\textbf{Stat.}  &\textbf{Instant.} &\textbf{Passive} &\textbf{Hybrid}  &  \\
\hline\hline\centering

~\cite{van2021reconfigurable} &\centering\checkmark       &\centering- &\centering\checkmark &\centering- & Analyzing and optimizing the UL and DL transmissions under spatially correlated channels \\
	\hline	

~\cite{Shi:TVT:2022}       &\centering\checkmark   &\centering- &\centering\checkmark &\centering- &UL SE analysis under channel estimation error and over spatially correlated channels \\
	\hline

~\cite{Nguyen:TWC:2022}     &\centering\checkmark   &\centering- &\centering- &\centering\checkmark & UL and DL SE analysis considering the MMSE estimate of the effective channels\\
	\hline	

 ~\cite{Gan:TCOM:2022}        &\centering\checkmark   &\centering- &\centering\checkmark &\centering- & Maximization of weighted sum-rate  under the two-timescale transmission protocol \\
	\hline

~\cite{Huang:WCL:2021} &\centering-       &\centering\checkmark &\centering\checkmark &\centering- & Maximization of weighted sum-rate via decentralized beamforming design for APs and RISs\\
	\hline	

~\cite{Shi:JSAC:2023}      &\centering\checkmark   &\centering- &\centering\checkmark &\centering- &  Max-min SE power control design based on UL SE results with local MR combining at the APs and LSFD at the CPU\\
	\hline

~\cite{Zhang:JCCN:2021}      &\centering-   &\centering\checkmark &\centering\checkmark &\centering- &EE optimization via joint digital and analog beamforming at the APs and RISs, respectively \\
	\hline	

~\cite{Shi:MCOM:2022}  &\centering-   &\centering\checkmark &\centering\checkmark &\centering- & Adding wireless energy scavenging module to the original RIS panel with different amalgamation of RIS and CF-mMIMO\\
	\hline	

 ~\cite{Lyu:WCL:2023}         &\centering-   &\centering\checkmark &\centering- &\centering\checkmark &Digital beamforming and hybrid RIS coefficients design for EE optimization  \\
	\hline	

 ~\cite{Jin:WCL:2023}         &\centering-   &\centering\checkmark &\centering- &\centering\checkmark & Joint transmit beamforming and RIS coefficients optimization to maximize the EE  \\
	\hline	
 
 ~\cite{Zhang:TSP:2021}       &\centering-   &\centering\checkmark  &\centering\checkmark &\centering- &Joint precoding design at the APs and RISs to maximize the weighted sum-rate of all UEs \\
	\hline	

  ~\cite{Ma:TWC:2023}     &\centering\checkmark   &\centering- &\centering\checkmark &\centering- &AP selection problem and beamforming design over mmWave channels by considering statistical CSI error model\\
	\hline	

   ~\cite{Lan:IoT:2023}
    &\centering-   &\centering\checkmark &\centering\checkmark &\centering- &AP selection as well as a joint power
control, precoding and phase shift design to maximize the SE/EE\\
	\hline	

   ~\cite{Li:TVT:2023}
    &\centering-   &\centering\checkmark &\centering\checkmark &\centering- & UE and RIS subsurface association with phase shift design  to maximize the EE\\
	\hline

~\cite{Elhoushy:WCL:2022}    &\centering-   &\centering\checkmark &\centering\checkmark &\centering- &Minimize the information leakage to active Eve while maintaining certain QoS for the legitimate UEs \\
	\hline 

  ~\cite{Hao:TIFS:2022}       &\centering-   &\centering\checkmark &\centering\checkmark &\centering- & RIS-UE matching and maximizing the weighted sum secrecy rate in PLS-based systems \\
	\hline

\end{tabular}
\vspace{-1em}
\end{table*}

\subsubsection{Energy Efficiency and Optimization}
Zhang \ettall~\cite{Zhang:JCCN:2021} considered the EE problem in RIS-assisted CF-mMIMO systems by jointly optimizing the digital beamforming at the APs and analog beamforming at the RISs, and then the impact of the transmit power, number of RIS, and  RIS size on the EE were investigated. To enhance the EE performance,   Shi~\ettall~\cite{Shi:MCOM:2022} suggested adding a wireless energy scavenging module to the original RIS panel and exploit some elements for energy reception and other elements for signal reflection. The energy harvesting elements are connected with energy storage hardware, which can store the harvested energy and support the energy consumption of other elements performing reflection. Accordingly, three different ways of amalgamation of the energy scavenging-enabled RIS and CF-mMIMO, namely centralized RIS, non-cooperative distributed RIS, and cooperative distributed RIS, have been studied. Jin~\ettall ~\cite{Lyu:WCL:2023} suggested the application of hybrid RISs, with a few active elements capable of amplifying the incident signal, and studied the digital beamforming and hybrid RIS coefficients design. Lyu \ettall~\cite{Jin:WCL:2023} studied the EE enhancement achieved by deploying multiple hybrid RISs into a DL CF-mMIMO system. They formulated an optimization problem under the minimum rate constraint and optimized the digital beamforming and hybrid RIS coefficients using a block coordinate descent  based iterative algorithm. The simulation results in~\cite{Jin:WCL:2023} indicated that, when incorporating a limited number of active elements in a RIS, hybrid RIS outperforms both passive and active RISs in terms of EE.

\subsubsection{RIS and/or AP Selection}
Zhang and Dai~\cite{Zhang:TSP:2021} addressed the problem of joint precoding design at the APs and RISs to maximize the weighted sum-rate of all UEs to improve the network capacity. To avoid the huge overhead and delay caused by the channel estimation process, a two-timescale scheme has been proposed, where each UE is matched with several well-performed RISs at the beginning of a large timescale. Therefore, only the CSI of the selected RISs are required for each UE. Ma \ettall~\cite{Ma:TWC:2023} carried out a study on active and passive beamforming design for the RIS-aided mmWave CF-mMIMO system, by considering the statistical CSI error model for all channels that follows the circularly symmetric complex Gaussian distribution. To cope with the heavy communication and computational costs in the system, caused by channel estimation overhead, an AP selection problem was formulated as a binary integer quadratic programming problem and then solved via a relaxed linear approximation algorithm. Lan \ettall~\cite{Lan:IoT:2023} focused on the SE/EE optimization of a RIS-aided UC CF-mMIMO system, where a certain number of APs are selected to serve each UE. To address the channel estimation challenge in this scenario, the authors verified the feasibility of deploying extra APs in the vicinity of the RISs to obtain the approximate RIS-UE channels. They proposed a joint power control, precoding and phase shift iterative algorithm to maximize the SE/EE subject to power constraints of the APs and the phase constraint of RISs. Li~\ettall~\cite{Li:TVT:2023}  developed a joint UE and RIS subsurface association algorithm, which allows each RIS to support multiple UEs at the same time. Moreover, to maximize the EE, the phase shift matrix of RISs and the transmission power of APs were jointly optimized based on Riemannian product manifolds.

\subsubsection{Secrecy Enhancement}   
The potential of RIS in improving the secrecy capacities offered in CF-mMIMO in the presence of an active Eve was pursued by Elhoushy \ettall~\cite{Elhoushy:WCL:2022}. In a PLS-based multi RIS CF-mMIMO system, Hao \ettall~\cite{Hao:TIFS:2022} looked into the matching problem among RISs and UEs~\cite{Zhang:TSP:2021} to discard some of the RISs with small contribution to the UE's secrecy and accordingly to reduce the channel estimation overhead. Then, a mixed integer non-linear programming problem was formulated to maximize the weighted sum secrecy rate and solved via linear conic relaxation. Table~\ref{tabel:RIS} provides a summary of the existing works in the space of RIS-assisted CF-mMIMO.

\subsection{Case Study and Discussion}
Consider a RIS-assisted CF-mMIMO system, where $M$ APs connected to a CPU serve $K$ UEs on the same time and frequency resource. Our system is a special case of the system model in Fig.~\ref{fig:RIS_CFmMIMO}, where the APs are assumed to have a single antenna and the communication is assisted by one single RIS that comprises $\Nris$ engineered scattering elements that can modify the phases of the incident signals. The phase shift matrix of the RIS is denoted by $\boldsymbol{\Phi}=\{e^{j\theta_1},\ldots,e^{j\theta_{\Nris}}\}$ where $\theta_n\in[-\pi, \pi]$ is the phase shift applied by the $n$th element of the RIS. The aggregated channel between the AP $m$ and UE $k$ is denoted by $u_{mk}=g_{mk} + \qh_m^H \boldsymbol{\Phi} \qz_k$, where $g_{mk}\sim\mathcal{CN}(0,\beta_{mk})$ is the direct channel between the UE $k$ and the AP $m$; $\qh_m\in\mathbb{C}^{\Nris\times 1}$ is the channel between the AP $m$ and the RIS; and $\qz_k\in\mathbb{C}^{\Nris\times 1}$ is the channel between the RIS and the UE $k$. By incorporating a realistic channel model to account for the spatial correlation among the RIS elements,  $\qh_m\sim\mathcal{CN}(\boldsymbol{0},\qR_m)$ and $\qz_k\sim\mathcal{CN}(\boldsymbol{0},\tilde{\qR}_k)$, where $\qR_m\in \mathbb{C}^{\Nris\times \Nris}$ and $\tilde{\qR}_k\in \mathbb{C}^{\Nris\times \Nris}$ are the spatial covariance matrices, as described in~\cite{Emil:WCL:2021}.

To estimate the desired channel from UE $k$, AP $m$ projects the received training signal onto the corresponding pilot sequence to obtain $y_{pmk} = \sqrt{\rho_p\tauup} u_{mk} + \sum_{k'\in\mathcal{P}_k\setminus K} \sqrt{\rho_p\tauup} u_{mk'} + w_{pmk}$, where $\rho_p$ is the normalized SNR of each pilot symbol and $w_{pmk}\sim\mathcal{CN}(0,1)$ is the additive noise at the AP $m$. By applying the linear MMSE estimation method, the estimate  of the aggregated channel can be obtained as $\hat{u}_{mk} =c_{mk} y_{pmk}$, where $c_{mk} = \frac{\sqrt{\rho_p\tauup} \delta_{mk}}{{\rho_p\tauup}  \sum_{k'\in\mathcal{P}_k}  \delta_{mk'} + 1}$ with $\delta_{mk} \triangleq \beta_{mk} + \trace(\pmb {\Theta }_{mk})$, where $\pmb {\Theta }_{mk}\triangleq\boldsymbol{\Phi}^H \qR_m \boldsymbol{\Phi}\tilde{\qR}_k$. Moreover, let $e_{mk} = u_{mk} - \hat{u}_{mk}$ be the channel estimation error. The variance of the channel estimate, $\gamma_{mk} =\Ex\{\vert u_{mk} \vert^2\}$ is $\sqrt{\rho_p\tauup} \delta_{mk} c_{mk}$, while the variance of the channel estimation error $\Ex\{\vert e_{mk} \vert^2\} = \delta_{mk}-\gamma_{mk}$. Since the quality of the channel estimation highly depends on the phase shift design at the RIS, $\boldsymbol{\Phi}$ can be optimized to minimize the total normalized mean square error, $\mathtt{NMSE}_{mk} = \frac{\Ex\{\vert e_{mk} \vert^2\}}{\Ex\{\vert u_{mk} \vert^2\}}$, obtained from all the UEs and all the APs, as follows
\begin{subequations}~\label{eq:NMSE}
    \begin{align}
&\underset {\{ \theta _{n} \} }{\mathrm {min}} \sum\nolimits_{m=1}^{M} \sum\nolimits _{k=1}^{K} \mathrm {NMSE}_{mk} \\&\mathrm {S.t} ~\,\,-\pi \leq \theta _{n} \leq \pi, \quad \forall n.
\end{align}
\end{subequations}
\begin{figure}[t]
	\centering
	\vspace{-0.25em}
	\includegraphics[width=0.47\textwidth]{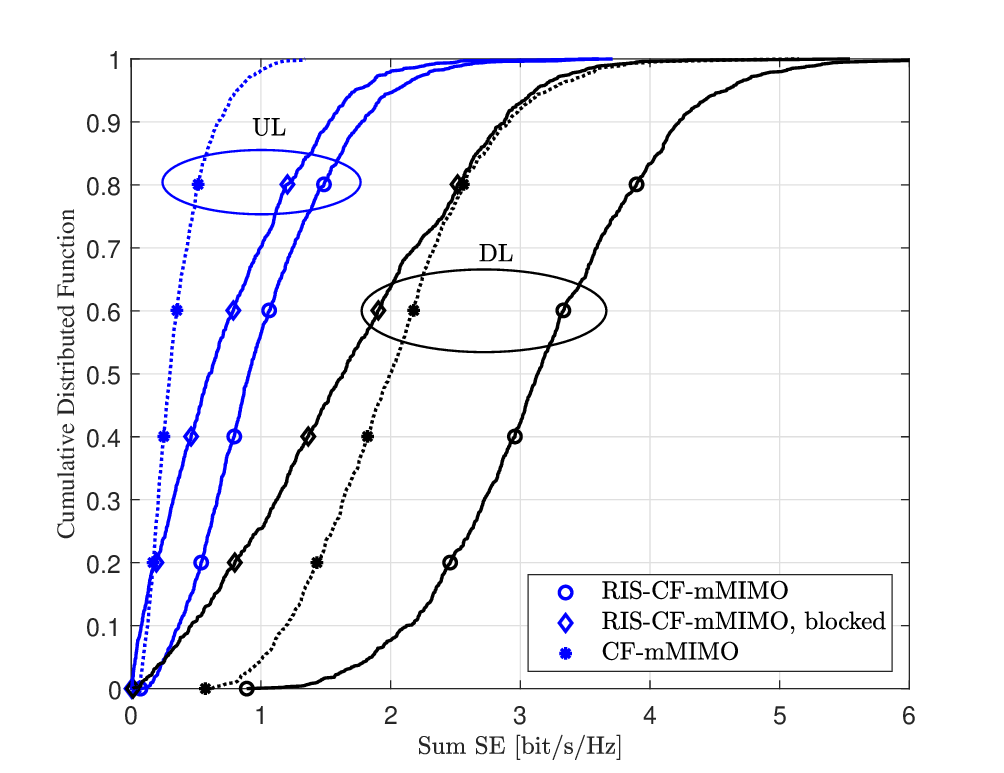}
	\vspace{-0.7em}
	\caption{CDF of the sum SE with $M=100$, $\Nris=900$, $K=10$,  $\tau_c=200$, $\tauup=5$, $p_{\mathtt{direct}}=0.2$,  $p_p=0.1$ W, and $p_d=0.5$ W.}
	\vspace{1em}
	\label{figRIS}
\end{figure}

The optimization problem~\eqref{eq:NMSE} is a fractional program, whose globally-optimal
solution is not simple to be obtained for an RIS with a large number of independently tunable elements. Nevertheless, the results in~\cite{van2021reconfigurable} indicate that in the absence of the direct link, the equal phase shift design is optimal. 

\begin{figure*}
\begin{align}
\SINR_{\ul k}&=
\frac{\rho_{u} \varsigma_{k} \left ({\sum_{m=1}^{M} \gamma_{mk} }\right)^{2}}{%
\splitfrac{\textstyle 
            \rho_{u} \sum _{k'=1}^{K} \sum _{m=1}^{M} \varsigma_{k'} \gamma _{mk} \delta _{mk'} +\,  \tauup \rho_{p}\rho_{u} \sum_{k' =1 }^{K} \sum _{k'' \in \mathcal {P}_{k}} \sum _{m=1}^{M} \sum _{m'=1}^{M} \varsigma_{k'} c_{mk}c_{m'k} \trace(\pmb {\Theta }_{mk'} \pmb {\Theta }_{m'k''})}
          {\textstyle 
           + \sum _{m=1}^{M} \gamma _{mk} + \tauup \rho_{p}\rho _{u} \sum _{k' \in \mathcal {P}_{k}} \sum _{m=1}^{M} \varsigma_{k'} c_{mk}^{2} \mathrm {tr}(\pmb {\Theta }_{mk'}^{2}) +\, \tauup \rho_{p}\rho _{u} \sum _{k' \in \mathcal {P}_{k} \setminus \{ k\} } \varsigma_{k'} \left ({\sum _{m=1}^{M}c_{mk} \delta _{mk'} }\right)^{2}} },~\label{eq:SINRRISu}\\
\SINR_{\dl k} &= 
\frac{\rho _{d} \left({\sum _{m=1}^{M} \sqrt {\eta _{mk}} \gamma _{mk} }\right)^{2}}{%
\splitfrac{\textstyle 
            \rho _{d} \!\sum _{k' =1}^{K}\! \sum _{m=1}^{M}\! \eta _{mk'} \gamma _{mk'} \delta _{mk} \!+ \!\tauup\rho_{p}  \rho _{d} \, \sum _{k' =1}^{K} \!\sum _{k'' \in \mathcal {P}_{k'}}\! \sum _{m=1}^{M}\! \sum _{m'=1}^{M}\! \sqrt {\eta _{mk'} \eta _{m'k'}} c_{mk'} c_{m'k'} \,\trace(\pmb {\Theta }_{mk} \pmb {\Theta }_{m'k''})}
          {\textstyle 
          \!+\!\tauup\rho_{p}\rho_{d} \sum_{k' \in \mathcal {P}_{k} } \sum_{m=1}^{M} \eta _{mk'} c_{mk'}^{2} \mathrm {tr}(\pmb {\Theta }_{mk}^{2})\!+\!\, \tauup\rho_{p}  \rho_{d} \sum_{k' \in \mathcal {P}_{k} \setminus \{ k\}} \left ({\sum_{m=1}^{M} \sqrt {\eta _{mk'} } c_{mk'} \delta _{mk} }\right)^{2}\!+\! 1}}~\label{eq:SINRRISd}.
\end{align}
  	\hrulefill
	\vspace{-4mm}
  \end{figure*}

Figure~\ref{figRIS} shows the CDF of the sum UL and sum DL SE, defined as $\SE_{\ul} = \frac{\tau_{u}}{\tau_{c}}  \sum_{k=1}^{K}   \log _{2}  (1 + \SINR_{\ul k} )$ and $\SE_{\dl} = \frac{\tau_{d}}{\tau_{c}} \sum_{k=1}^{K}\log _{2} \left ({1 + \SINR_{\dl k} }\right)$, where, $\SINR_{\ul k}$,  and $\SINR_{\dl k}$ are given in~\eqref{eq:SINRRISu} and~\eqref{eq:SINRRISd}, respectively, on the top of the next page~\cite{{van2021reconfigurable}}. To quantify the impact of RIS, we consider a scenario in which the direct links, represented by $g_{mk}$, are assumed to be unblocked with a certain probability. Therefore, we  model the large-scale fading
coefficient $\beta_{mk}$ as $\beta_{mk} = \bar{\beta}_{mk}a_{mk}$, where $\bar{\beta}_{mk}$ accounts for the path loss due to the transmission distance and the shadow fading according to the three-slope propagation model in~\cite{Hien:cellfree} and binary variable $a_{mk}$ accounts
for the probability that the direct links are unblocked. More specifically, $a_{mk}=1$ with a probability $p_{\mathtt{direct}}$ and otherwise $a_{mk}=0$. We consider two benchmarks: i) a CF-mMIMO system without RIS deployment (CF-mMIMO), and ii) an RIS-assisted CF-mMIMO system with blocked direct links (RIS-CF-mMIMO, blocked). In the latter case, all direct links are blocked with unit probability, and as a result, UL and DL transmissions are performed exclusively through the RIS. From Fig.~\ref{figRIS}, we observe that by using RIS in the network, both the UL and DL SE of the network are significantly improved compared to the benchmarks.

\vspace{-1em}
\subsection{Future Research Directions}
Looking ahead, some open problems are still worth further investigation. For example, in some specific scenarios, how to optimize the placement of RISs is an interesting research direction. Moreover, the integration of different type of RISs (passive, active, and hybrid) into WIPT-enabled CF-mMIMO networks, with the aim of overall SE and/or EE enhancement, is a promising research direction that can be applied to different architectures including UC, FD, and NAFD. For example, active RISs can be deployed and the CPU can decide which RISs to be used to assist WPT or WIT, while the overall power budget can be optimally shared between the APs and active RISs, to avoid extra power consumption. Furthermore, investigating the potential of a more general class of RIS, called beyond diagonal RIS (BD-RIS)~\cite{Hongyu:TWC:2023}, in CF-mMIMO systems is an interesting research direction. The non-diagonal scattering matrix of BD-RIS significantly affects the analysis and design of CF-mMIMO systems. It is still unclear whether integrating BD-RIS into CF-mMIMO systems provides more performance gains compared to conventional RIS~\cite{hua2024cell}. 

To mitigate the half-space coverage limitation of conventional RISs, a simultaneous transmitting and reflecting RIS (STAR-RIS)~\cite{Xu:CLET:2021} has been recently proposed. An interesting application scenario of the STAR-RIS is to assist the simultaneous UL and DL transmissions in FD and NAFD CF-mMIMO networks. This paradigms opens several issues for future research, including the transmission power control for FD and HD-mode APs, the operation mode selection for STAR-RIS and APs, and the channel estimation at APs.  As an initial attempt, Papazafeiropoulos~\ettall~\cite{Papazafeiropoulos:TVT:2023} developed an analytical framework for the study of the DL of a STAR-RIS-assisted CF-mMIMO system by using statistical CSI, where the aggregate APs-UEs channels, based on imperfect CSI, are considered
with MR beamforming for information decoding. 

Recent efforts to improve connectivity and achieve higher data rates involve increasing the aperture size of the transceivers and using extremely high-frequency band. Therefore, the concept of extremely large-scale RIS has emerged~\cite{Liu:TCOM:2023}, where large-scale reflecting elements are deployed at the RIS to compensate for the severe multiplicative fading effect in the cascaded channel. Nevertheless, by increasing the size of the RIS, near-field signal propagation will become more dominant, leading to a significant transformation in the electromagnetic field structure. Therefore, it is essential to consider investigating the performance of CF-mMIMO systems under near-field propagation conditions. 

\section{Emerging Application Scenarios}~\label{sec:other}
In this section, we shed light on the new opportunities in emerging technologies and network architectures and highlight  open challenges, trends, and opportunities to  enable the widespread use of CF-mMIMO.
\subsection{Ultra-Reliable and Low-Latency Communications}
URLLC is a pivotal technique for supporting next generation industrial IoT devices such as autonomous vehicles and robots~\cite{Bennis:PRI:2018}. For these applications, most devices use short packets (on the order of 100 bits) to transmit information with low latency (within hundreds of microseconds) and with a reliability no smaller than $99.999\%$. Under such stringent latency requirements, harnessing time diversity becomes infeasible. Moreover, the use of frequency diversity poses difficulties, because the regulations set by standardization do not allow UEs to distribute a coded packet over the noncontagious frequency resources. Space diversity, on the other hand, can be leveraged to meet the demanding reliability requirements of URLLC. Although mMIMO offers a substantial degree of spatial diversity, it is prone to significant variations in path loss and inter-cell interference. CF-mMIMO presents a solution to mitigate this problem.  

\subsubsection{Literature review} The pioneering work by Nasir~\ettall~\cite{Nasir:TWC:2021} proposed a new class of CB for URLLC to maintain the low computational complexity for its design while requiring only local CSI for its transmit implementation. Both the Shannon rate of UEs (in the long blocklength) and the URLLC rate (in the short blocklength) were optimized. Furthermore, this study developed improper Gaussian signaling to improve both the Shannon function rate, URLLC rate, and EE. By using the UC approach, Peng~\ettall~\cite{Peng:TVT:2023} derived lower bounds on the achievable downlink data rate with imperfect CSI for the MRT, FZF, and local ZF precoding schemes under finite channel blocklength. Then, they maximized the weighted sum rate by jointly optimizing the pilot power and the transmission power, while considering the minimal requirements of decoding error probability and data rate. Shi~\ettall~\cite{Shi:TVT:2023} considered the path-following algorithm-based precoding design in CF-mMIMO systems for URLLC in the centralized and decentralized fashion. While centralized design is performed at the CPU, the decentralized approach involves partitioning the APs into distinct cooperative clusters without overlaps.  Within each cluster, the APs share the data and instantaneous CSI in each cluster to design the precoding vectors, thereby mitigating the computational complexity. Lancho~\ettall~\cite{Lancho:TWC:2023} provided a general framework for characterizing (in numerical way) the packet error probability in CF-mMIMO architectures supporting URLLC services. Moreover, upper bounds of the UL and DL decoding error probabilities have been derived in~\cite{Lancho:TWC:2023} by using the saddlepoint method to support URLLC. The findings indicate that in the URLLC regime, it is beneficial to minimize the average distance between the UEs and APs by densifying the AP deployment. 

Zeng~\ettall~\cite{Zeng:IoT:2023} introduced the concept of energy-efficient massive URLLC in CF-mMIMO systems, which aims to provide communication services with high reliability, low latency, low energy consumption, and massive access. In this study, the general $\kappa-\mu$ shadowed fading model was considered along with MR combining at the APs, followed by simple LSFD detection (with weighting coefficient $\chi_{mk} = \beta_{mk}/\sum_{k=1}^K \beta_{mk}$ for UE $k$ at AP $m$) at the CPU for UL data detection. 


\vspace{-0.8em}
\subsection{Unmanned Aerial Vehicles}
UAVs have recently found numerous promising applications, including photography, traffic control, packet delivery, and telecommunication. The inclusion of UAVs into wireless networks, especially their deployment as flying APs/BSs in cellular networks, has been a compelling research focus over the past decade~\cite{Mozaffari:tuts:2019}. These airborne APs offer a viable alternative to terrestrial counterparts, providing advantages in coverage, cost, and deployment flexibility. A significant benefit of using UAVs is their ability to be repositioned across different zones to meet dynamic network connectivity demands. 

In CF-mMIMO networks, UAVs can function as either APs~\cite{Diaz:TWC:2023,Vilor:TWC:2023,LinlinLTCOM:2024} or UEs~\cite{Andrea:JOP:2020,Elwekeil:TCOM:2023,Yong:IoT:2024}. Compared to terrestrial cell-free networks, their aerial counterparts present major differences and new challenges: \textit{i)} The ground-to-air and air-to-ground links are dominated by LoS, which can be better modeled via a Rician channel model. This results in a larger channel coherence bandwidth relative to ground networks, with the phase of the LoS component adding another layer of uncertainty due to its stochastic nature~\cite{Vilor:TWC:2023}; \textit{ii)} Deployment of large arrays becomes infeasible, but smaller directional antennas can be mounted onboard~\cite{Guo:TCOM:2020}; \textit{iii)} The fronthaul connecting the flying APs/UEs to the rest of the network (CPU) becomes wireless with limited capacity; \textit{iv)} Scenarios involving high-mobility UAVs result in fast-varying spatial channels.

\subsubsection{Literature review}
Exploring the potential of UAVs as flying BSs/APs represents an attractive avenue of research, as they can provide superior coverage and can be deployed quickly and on-demand~\cite{Vilor:TWC:2023,Diaz:TWC:2023,LinlinLTCOM:2024}. Diaz-Vilor~\ettall~\cite{Vilor:TWC:2023} considered the UL of fully and partially centralized CF-mMIMO networks with linear MMSE and MRC receivers,  where the UAVs function as flying APS. Two different Ricean channel models were considered: 1) with known phase and 2) with random phase in the LoS component. Moreover, the UAV deployment problem for different receiver architectures was investigated to maximize the minimum UE SE. The authors extended their works in~\cite{Diaz:TWC:2023} and investigated the challenges related to the wireless nature of access and fronthaul links. They formulated and solved the optimization subproblems involving the 3D deployment of the UAVs, UE transmit powers, UAV transmit powers, and the joint optimization of these subproblems, all in the presence of imperfect wireless fronthaul links. Xu~\ettall \cite{LinlinLTCOM:2024} investigated a scenario with both aerial and terrestrial APs. The limited capacity of both wired and wireless fronthaul links, along with the constraints imposed by individual QoS requirements of users, were considered as design criteria for developing the system.

UAVs can also function as UEs in the network. For example, D’Andrea~\ettall~\cite{Andrea:JOP:2020} analyzed and compared the UL and DL SE results for conventional and UC CF-mMIMO systems in the presence of the UAV communications with both UAV and legacy terrestrial UEs. Several power control designs were developed, including one maximization of the minimum of the SEs across the UEs. Zheng~\ettall~\cite{Zheng:JSAC:2021} investigated the UL SE of UAV communication with CF-mMIMO system, where the UAV is energized via WPT with RF signals from APs. The superiority of CF-mMIMO systems over cellular mMIMO in terms of downlink harvested energy and UL SE was also demonstrated. Elwekeil~\ettall~\cite{Elwekeil:TCOM:2023} proposed two power control schemes, optimizing the sum rate and the worst UE’s rate for both UL and DL, in UC CF-mMIMO networks that support URLLC applications for both ground UEs and UAVs. Chen~\ettall~\cite{Yong:IoT:2024} explored pilot assignment and power control challenges for ensuring secure UAV communications in a UC CF-mMIMO network. The network scenario involved numerous distributed APs simultaneously serving multiple UAVs and terminal UEs, all contending with the presence of a UAV acting as an eavesdropper capable of executing pilot spoofing attacks.

\subsubsection{Future research directions}
In the presence of UAVs, the coherence bandwidth of the air-to-ground channels is greater when compared to the ground networks~\cite{Matolak:TVT:2017}. Therefore, a larger number of orthogonal pilot dimensions are available, allowing for effective mitigation of pilot contamination. However, the phase of the LoS component, typically represented deterministically to account for the receiver's phase lock loop's tracking and adjustments, may have to be modelled stochastically to account for drift. This introduces new challenges during the channel estimation phase. Moreover, deployment of small directional antennas onboard, necessitates the development of novel precoding designs. Furthermore, the performance and operational duration of UAVs is fundamentally constrained by the limited onboard energy, while offering the following functionalities:  (i) wireless communication and (ii) UAV movement control~\cite{Zeng:MCOM:2016}. Thus, WPT-enabled UAV communication and EE emerge as primary focal points in the design of CF-mMIMO systems.     
Finally, when fixed-wing UAVs are used as UEs in the network for aerial surveillance, they introduce fast-varying spatial channels due to their high speed.\footnote{In the literature, rotary-wing UAVs are frequently utilized as flying APs~\cite{Diaz:TWC:2023,Vilor:TWC:2023,LinlinLTCOM:2024}. These drones have the ability to hover in the air and perform vertical take-off and landing, rendering them more stable and suitable for indoor areas. Additionally, their corresponding channels exhibit less rapid changes. However, they are subject to higher energy restrictions, slower speeds, and lower capacity compared to fixed-wing UAVs.} To tackle this challenge, one solution is to employ orthogonal time frequency space (OTFS) modulation in these networks. OTFS modulation has already been investigated for terrestrial CF-mMIMO networks in~\cite{Mohammad:OTFS}, and exploring its integration into aerial CF-mMIMO networks with high-mobility UAVs represents an interesting future direction.

\subsection{Cell-Free Massive MIMO and Machine Learning}~\label{sec:cell-free learning}
CF-mMIMO systems involve numerous APs/antennas and UEs, resulting in a high-dimensional complex (non-linear) optimization challenge. Traditional optimization methods frequently encounter difficulties handling this complexity, whereas ML techniques thrive in such high-dimensional spaces. ML methods have the ability to efficiently process and analyze vast datasets in real-time, making them ideal for large-scale deployments and effectively mitigating scalability concerns in these networks. Furthermore, ML techniques enable real-time decision-making for resource management, performance optimization, and low-latency communication. These aspects are prerequisites for the designed algorithms in future wireless networks, including CF-mMIMO networks. ML algorithms can effectively meet these stringent requirements.

For wireless resource allocation and designing encoders and decoders, supervised learning is typically used due to its fast convergence and high-quality output, though it requires a large amount of human-labeled data, increasing data processing complexity. Unsupervised learning is suitable for challenges like UE association, hybrid multiple access UE grouping, and detecting malicious UE attacks, as it can deduce patterns from abundant unlabeled data without human guidance, though the data may not fully reflect real-world scenarios. RL discovers optimal actions in uncertain environments through trial-and-error, learning from feedback without direct supervision, but it requires significant resources and sometimes lacks clear physical explanations. Despite these limitations, RL and deep reinforcement learning (DRL) have been successfully applied in areas, such as UAV communications, autonomous driving, mobile edge computing, and wireless caching placement, allowing network entities to develop optimal decision-making policies through interactions with dynamic and uncertain environments~\cite{Quang:TVT:2021,Eryani:JSAC:2021}.

\textit{Distributed learning:} Solving problems centrally at the CPU enables efficient resource allocation, but it requires a comprehensive understanding of the entire network state for effective decision-making. For example, data from the users' side—such as computing demands, channel conditions, and maximum tolerable task execution deadlines—must be gathered and processed. Then, allocation decisions must be relayed back to the users within a strict delay tolerance, which results in significant overhead and additional delay due to the two-way information exchange. Moreover, future wireless networks including CF-mMIMO networks will involve multiple network entities (multiple agents) and, thus, single-agent RL may not be efficient. This is because the policy of the network entity, learned by the single-agent RL, does not consider the impact of the policies of other network entities.  This may cause non-stationary issues or reduce the learning efficiency. To address these challenges, the concept of mobile edge computing can be leveraged, which involves processing data at edge nodes instead of the central cloud. As the computational capability of mobile devices is growing significantly, it is now feasible to push network computation even further to the mobile device level. Therefore, distributed solution approaches, based on multi-agent reinforcement learning (MARL)~\cite{Tianxu:TuT:2022} and federated learning, can be applied to CF-mMIMO networks, efficiently leveraging the network's distributed nature. Distributed learning not only addresses the challenges posed by the increasing number of devices in wireless networks, such as an IoT network, but also effectively fulfills the requirements of safeguarding the UE privacy and ensuring data security.

\subsubsection{Literature review}
In recent years, there has been significant attention towards harnessing the advantages of ML for the advancement of the CF-mMIMO systems. In particular, ML techniques have been employed for different purposes including channel estimation~\cite{Jin:TVT:2019,Athreya:WCL:2020}, power control~\cite{Bashar:JSAC:2020,Zhao:ACCESS:2021,Zaher:TWC:2023,Zhang:TVT:2023,hao2023user}, beamforming design~\cite{Hojatian:CLET:2022,Eryani:JSAC:2021,Eryani:TCOM:2022}, and UE clustering within NOMA CF-mMIMO systems~\cite{Quang:TVT:2021,Eryani:JSAC:2021}.

Inspired by the huge computational complexity of classical least-squares and MMSE estimators, a channel estimation framework based on the denoising convolutional neural network was proposed by Jin~\ettall~in~\cite{Jin:TVT:2019}. Athreya~\ettall~\cite{Athreya:WCL:2020} proposed a framework that leveraged cascade deep learning for inter-pilot interpolation and TDD reciprocity calibration in orthogonal frequency division multiplexing based CF-mMIMO systems.

Bashar~\ettall~\cite{Bashar:JSAC:2020} concentrated their attention on the power control design for UL transmissions in CF-mMIMO systems with limited fronthaul. In this considered situation, a LSF DL-based power control scheme was designed to allocate the power control coefficients for both quantize-and-forward and combine-quantize-and-forward schemes. The main goal of this study was to train a neural network to determine optimal transmit power levels at the UEs using LSF coefficients as input. This optimization was based on results obtained from a large dataset of randomly chosen small-scale fading coefficients. A deep convolutional neural network was exploited to determine a mapping from the LSF coefficients and the optimal power through solving the sum rate maximization problem using the quantized channel. Zhao~\ettall~\cite{Zhao:ACCESS:2021}  introduced two power allocation methods for DL transmission in CF-mMIMO based on DRL: the deep Q-network and the deep deterministic policy gradient. These methods offer the advantage of lower computational complexity compared to the WMMSE, while still maintaining competitive DL sum-SE performance. Unlike traditional supervised learning, which requires a large training dataset generated through a computationally complex algorithm~\cite{EMIL:MSP:2020}, in DRL, training is conducted through trial-and-error interactions with the environment, relying on rewards obtained during these interactions. 
Zhang~\ettall~\cite{Zhang:TVT:2023} used unsupervised learning to implement low-complexity power control strategies for UL and DL data transmission in CF-mMIMO. The proposed learning algorithms rely on LSF coefficients and take into account the pilot contamination effects. More specifically, the proposed learning algorithm learns to map between power coefficients and LSF coefficients and avoids the prohibitive computational complexity of the conventional power allocation methods. Hao~\ettall~\cite{hao2023user}  proposed an accelerated projected gradient algorithm for joint UE association and power control in a DL UC CF-mMIMO system with local PPZF processing.    

Two unsupervised deep neural network architectures were proposed in~\cite{Hojatian:CLET:2022} to perform decentralized hybrid beamforming in FDD-based CF-mMIMO. By drawing analogies between an UL cell-free network and a quasi-neural network, and borrowing the idea of backpropagation algorithm, the authors in~\cite{Wang:TCOM:2023} introduced a distributed learning approach for UL CF-mMIMO beamforming. This scheme achieves multi-AP cooperation, while requiring no information exchange between the APs. The proposed algorithm outperforms Level-3 processing, even though no explicit CSI is required, and each AP only needs to pass the scalar sequences (product of the received signal and local beamforming vectors) to the CPU over the fronthaul. The authors in~\cite{Eryani:TCOM:2022} introduced a DL CF-mMIMO network architecture based on dynamic partitioning (which partitions the network into a set of independent cell-free subnetworks) along with a hybrid analog-digital DL beamforming method by using DRL techniques. Through a DRL-cum-convex optimization model, joint network partitioning, analog beamsteering, and digital beamforming is designed to maximize the per-UE transmission rate, while inter-subnetwork interference and intra-UE interference are mitigated. It was shown that the performance enhancement over the conventional all-digital counterpart becomes more significant as the number of network partitions increases. 

In the space of NOMA CF-mMIMO networks, ML-based algorithms have been proposed for AP/UE clustering. For example,  Le~\ettall~\cite{Quang:TVT:2021} developed two unsupervised ML-based UE clustering algorithms with the objective of maximizing the sum-rate, which outperform the distance-based and Jaccard-based UE clustering schemes. To reduce the  computational complexity and processing delay for signal detection at the CPU, Al-Eryani~\ettall~\cite{Eryani:JSAC:2021} proposed a hybrid DRL model, with the objective of implementing joint AP clustering and beamforming design in NOMA CF-mMIMO, aiming to maximize either the sum-rate or the minimum rate among all UEs. 

Distributed learning has been leveraged in various scenarios in CF-mMIMO networks. Zaher~\ettall~\cite{Zaher:TWC:2023} developed a fully distributed feedforward deep neural network for each AP to approximate per-AP normalized power coefficients and the total transmit power from the AP. Local information at each AP is used as input, while during the training stage, the deep neural network incorporated network-wide solutions for sum SE and proportional fairness as labeled output. Researchers have also investigated the use of CF-mMIMO to support federated learning in wireless environments~\cite{vu20TWC,Sifaou:TWC:2023,Qin:WCL:2024}. Vu~\ettall~\cite{vu20TWC} formulated a mixed timescale stochastic nonconvex optimization problem aimed at minimizing the training time of a federated learning process. Their results confirmed that CF-mMIMO offers the lowest training time when compared to cell-free time division multiple access mMIMO and collocated mMIMO. To alleviate the communication overhead of federated learning over CF-mMIMO networks, Sifaou~\ettall~\cite{Sifaou:TWC:2023} proposed an over-the-air federated learning approach. This method allows clients to send their local updates simultaneously, enabling the CPU to directly obtain the sum from the received signal. As a result, the communication resources required to reach convergence do not scale with the number of clients.

Furthermore, multi-agent ML approaches have been explored in several studies~~\cite{Zhang:TWC:2023,Tilahun:TVT:2023,Zhu:TVT:2024}. Zhang \textit{et al.}~\cite{Zhang:TWC:2023} investigated SWIPT-enabled CF-mMIMO networks with power splitting receivers and NOMA. They designed a ML-based approach to address the mixed combinatorial and non-convex optimization problem. First, UE clustering is optimized using a K-means based method. Then, the power control coefficients and power splitting ratios are jointly optimized using a multi-agent deep Q-network method, which is executed in a distributed manner. Tilahun \textit{et al.}~\cite{Tilahun:TVT:2023} developed a distributed approach based on a cooperative MARL framework, to address the joint communication and computing resource allocation problem in a CF-mMIMO-enabled mobile edge computing network. The objective was to minimize the total energy consumption of the UEs while meeting ultra-low delay constraints. Each UE was implemented as a learning agent to make joint resource allocation decisions relying solely on local information during execution. Zhu \textit{et al.}~\cite{Zhu:TVT:2024} introduced a distributed MARL-based method, incorporating fuzzy logic, for joint precoding and phase shift design in RIS-aided CF-mMIMO systems. The presented method leverages parallel computing to reduce the computational time, making it highly suitable for deployment in large-scale networks.

\subsubsection{{Future research directions}}
ML-based methods in CF-mMIMO systems face several challenges that warrant further investigation. Firstly, they are often developed and evaluated in simulated environments, which cannot effectively resemble the  complexities of real-world scenarios. Consequently, there can be discrepancies between the simulated and actual performance when deployed in practical CF-mMIMO systems, leading to unexpected behavior and performance degradation. Secondly, these methods may struggle to be generalized across different CF-mMIMO scenarios or adapt to dynamic environments. They are typically trained on specific datasets or scenarios, and their performance may deteriorate when confronted with unseen or novel conditions. Ensuring the robustness and generalization of learning-based methods in CF-mMIMO systems is crucial, particularly in scenarios with heterogeneous network conditions or varying user behaviors. Finally,  many learning-based methods, especially those based on deep learning or RL, require significant computational resources during both training and inference phases. In CF-mMIMO systems, where real-time processing and decision-making are crucial, the computational overhead introduced by learning-based methods may be prohibitive. This can lead to practical challenges in implementing these methods in CF-mMIMO systems, especially in resource-constrained environments or applications where low latency is critical.

ML-based algorithms can be utilized in NAFD CF-mMIMO networks to reduce the complexity of the joint AP mode assignment and AP clustering in UC scenarios. Zhu~\ettall~\cite{Zhu:Access:2022} proposed a load-aware dynamic mode selection scheme for the APs, aiming to maximize the UL-DL sum-rate of the network while considering the per-user traffic load. They investigated both centralized Q-learning and distributed multi-agent Q-learning algorithms with varying complexities, demonstrating that the former is more suitable for real-world applications due to its smaller storage unit and lower complexity. Sun~\ettall~\cite{Sun:JSYS:2023} proposed a “preallocation—optimization” mechanism for AP duplex mode optimization in NAFD CF-mMIMO systems.  The preallocation part involves a network load prediction algorithm based on autoregressive integrated moving average model, ensuring accurate load forecasting and efficient preallocation of resource blocks. In the optimization part,  DRL-based and  hierarchical RL-based AP duplex mode optimization algorithm were developed to solve the multi-objective optimization problem of AP mode optimization. Another direction is to leverage AI techniques to control the trajectories of multiple UAVs to efficiently cover vehicles in dynamic cell-free vehicular networks, where communication infrastructure is not available or severely damaged~\cite{Samir:JMC:2021}. Furthermore, to meet the real-time processing constraints and mitigate the high fronthaul overhead in RIS-aided CF-mMIMO systems, ML can be effectively employed. Chen~\ettall~\cite{Chen:TWC:2023} proposed a fully distributed ML
algorithm where each AP determines its own beamforming vectors using a graph
neural network based on its local CSI. The
RIS reflection coefficients can be determined by only one
of the APs.

\subsection{ Integrated Sensing and Communication}~\label{sec:CF-ISAC}
Integrating sensing functions into communication systems is expected to be a cornerstone of 6G and future communication technologies~\cite{Tataria:PROC:2021}. Next-generation wireless networks require both high-quality connectivity and highly accurate sensing capabilities. By efficiently sharing the hardware and wireless resources, the communication infrastructure can integrate sensing functions at minimal cost, potentially repurposing sensing frequency bands for wireless communication~\cite{Liu:JSAC:2022}. These sensing capabilities can unlock a wide range of applications in security, healthcare, and traffic management. While the design of various aspects of integrated sensing and communication (ISAC) systems has recently attracted growing research interest, most prior work has primarily focused on scenarios with single infrastructure/AP for ISAC~\cite{Liu:JSAC:2022}. In practice, however, multiple infrastructures/APs will operate within the same geographical region, frequency band, and time, causing mutual interference in both sensing and communication functions. Nevertheless, by enabling coordination among these distributed APs, both the communication and sensing performance can be significantly improved. Consequently, this leads to the development of CF-mMIMO ISAC systems, where distributed APs collaboratively serve the same set of communication users and sense the same targets. CF-mMIMO yields great potential in radar/sensing applications due to: \textit{i)} The cooperation of multiple APs enhances the localization and tracking accuracy, especially for fluctuating targets; \textit{ii)} Echo signals can be coherently and jointly processed, increasing detection probability and achieving robust parameter estimation; \textit{iii)} It offers seamless connections, making it ideal for low-latency target detection scenarios.

\subsubsection{Literature review}
Network ISAC was introduced in~\cite{Zhang:VMG:2021}, wherein multiple distributed BSs in cloud radio access networks (C-RANs) are enabled to cooperate in performing both distributed radar sensing and coordinated wireless communications. Advanced coordinated multipoint transmission/reception techniques were leveraged to mitigate or even utilize co-channel interference among different communication users, while also properly controlling the interference between sensing and communication signals. Given the design challenges in MIMO networks and the potential of CF-mMIMO to overcome these challenges, recent works have investigated the integration of ISAC and CF-mMIMO from different aspects. 
Behdad~\ettall~\cite{Behdad:GC:2022} studied the DL communication and multi-static sensing in a C-RAN assisted CF-mMIMO system, where a set of APs jointly serve the UEs and optionally steer a beam towards the target, with known location. A maximum a posteriori ratio test detector was derived to detect the target using signals received at remaining APs. In~\cite{Sakhnini:GC:2022}, a CF-mMIMO-based radar system was implemented in the UL by dedicating a fixed set of APs for radar probing signal transmission, while the remaining APs received a combination of radar echoes and communication signals. The impact of user interference on the system was managed by power control using linear interference constraints based on an average radar SINR expression. 
In~\cite{Cao:JSYS:2023}, vector orthogonal frequency
division multiplexing integrated signals were proposed and investigated for CF-mMIMO ISAC systems. These signals demonstrated high SE for communication system, superior detection resolution for sensing application, and accurate latency/Doppler estimation results. Mao~\ettall~\cite{Mao:TWC:2024} studied the impact of the uncertainty of the target locations on the propagation of wireless signals during both UL and DL phases, and  derived the main statistics of the MIMO channel estimation error. A fundamental performance metric, termed communication-sensing region, was introduced and optimized to capture the trade-off between the communication and sensing functionalities. In~\cite{elfiatoure2024multiple}, a distributed implementation for ISAC supported by CF-mMIMO was proposed, featuring a dynamic AP operation mode selection strategy that determines the allocation of APs for DL information transmission or radar sensing.

\subsection{Miscellaneous Topics}
With the advancement in other emerging technologies, CF-mMIMO is progressively being employed as a crucial component, driven by its significant benefits.  Inspired by the WPT efficiency of  CF-mMIMO, which is due to the significantly lower average distance between the energy sources and receiver, the coexistence of cell-free and symbiotic backscatter communication was investigated in~\cite{dai:TCOM:2023,Ataeeshojai:IoT:2023}. The underlaid backscatter devices utilize ambient “legacy” radio signals from the cell-free network as both a harvested energy source and a carrier on which to modulate data. Expanding these studies to encompass mmWave communication presents an interesting avenue for future research.  Content caching stands out as an another prevalent method within CF-mMIMO systems, aimed at enhancing the quality of communications. Backhaul/fronthaul loads can be reduced by avoiding fetching duplicate trending contents. Therefore, applications, such as video streaming, that require low latency and high throughput can be enabled~\cite{Wang:COMMG:2014}. Chen~\ettall~\cite{Chen:ICC:2021} proposed a greedy caching strategy to minimize energy consumption in a CF-mMIMO network, assuming that the content popularity and the number of requests at each AP is known. Chuang~\ettall~\cite{Chuang:TVT:2023} provided a DRL framework to address UE association and content caching jointly for the EE maximization problem in a CF-mMIMO network and under unknown content popularity.  

Ensuring communication for a diverse range of UEs with varying mobility profiles stands as an another fundamental goal of upcoming wireless networks. Specifically, both 5G and 6G networks are anticipated to deliver reliable transmissions for high-speed trains, and to introduce new services or improvement for vehicular communications in intelligent transportation systems. In such scenarios, a wireless channel is rapidly time varying, thus Doppler shifts can be much larger than those in traditional cellular networks. To leverage the capabilities of CF-mMIMO in such scenarios, certain factors need to be taken into account, along with the exploration of new solutions. Zheng~\ettall~\cite{Jiakang:JSAC:2022} investigated the UL SE of CF-mMIMO rthogonal frequency division modulation systems for high-speed train communications. Under these circumstances, they indicated that CF-mMIMO systems exhibit a lower susceptibility to Doppler frequency offsets compared to small cell and cellular systems. Mohammadi~\ettall~\cite{Mohammad:OTFS} studied the UL and DL SE of the CF-mMIMO systems with orthogonal time-frequency space modulation (OTFS), taking into account the impact of channel estimation error.  We recall that OTFS modulation was proposed recently to address high mobility-related issues~\cite{Raviteja:TWC:2018}. Additionally, the authors have extended their work to the scenario where UEs use the MMSE-SIC  to detect the received signal~\cite{Mohammadi:WCL:2023}. They developed a power control algorithm to ensure max-min fairness among the UEs. 

In recent years, antennas have transcended their traditional function as basic signal conduits, transforming into dynamic, adaptable, and intelligent components. Modern antennas can actively shape, steer, and manage data flow to fulfill the intricate requirements of today's wireless communication systems. This transformation is largely attributed to the integration of metamaterials into antenna design. Metamaterial antennas boast distinctive structures for radiation elements, surpassing the half-wavelength limitations inherent in traditional antenna production. This breakthrough allows for an increased number and/or placement options of antenna elements within a given array size, thereby enhancing beam-steering capabilities and introducing a new paradigm for interference mitigation and enhancement of transmit/receive spatial diversity. In this context, holographic and fluid antennas stand out as notable designs~\cite{Huang:WC:2020,Shojaeifard:IPC:2022}. Deploying these architectures in cellular and satellite communications, they demonstrate significant potential for reducing complexity and cost while upholding performance standards. Nevertheless, their integration into CF-mMIMO systems has yet to be thoroughly explored. Specifically, their potential for managing intra-user/AP interference in NAFD CF-mMIMO networks and dual-function systems such as SWIPT/ISAC-enabled CF-mMIMO could be leveraged to enhance network design requirements.

\section{Conclusion}~\label{sec:conc}
We have provided a brief understanding on CF-mMIMO networks. Then, we reviewed CF-mMIMO research contributions combined with state-of-art technologies and promising ones for 5G wireless networks and beyond, including non-orthogonal transmission, PLS, energy harvesting, mmWave communication,  RIS, URLLC, and UAV-aided communication.  We delved into the motivation behind integrating these technologies into CF-mMIMO networks and discuss the associated design challenges. To enhance comprehension, we presented several case studies that illustrate the core concepts. Finally, we identified various research gaps, laying the foundation for future research directions.

\bibliographystyle{IEEEtran}
\bibliography{IEEEabrv,references}


\begin{IEEEbiography}[{\includegraphics[width=1in,height=1.25in,clip,keepaspectratio]
{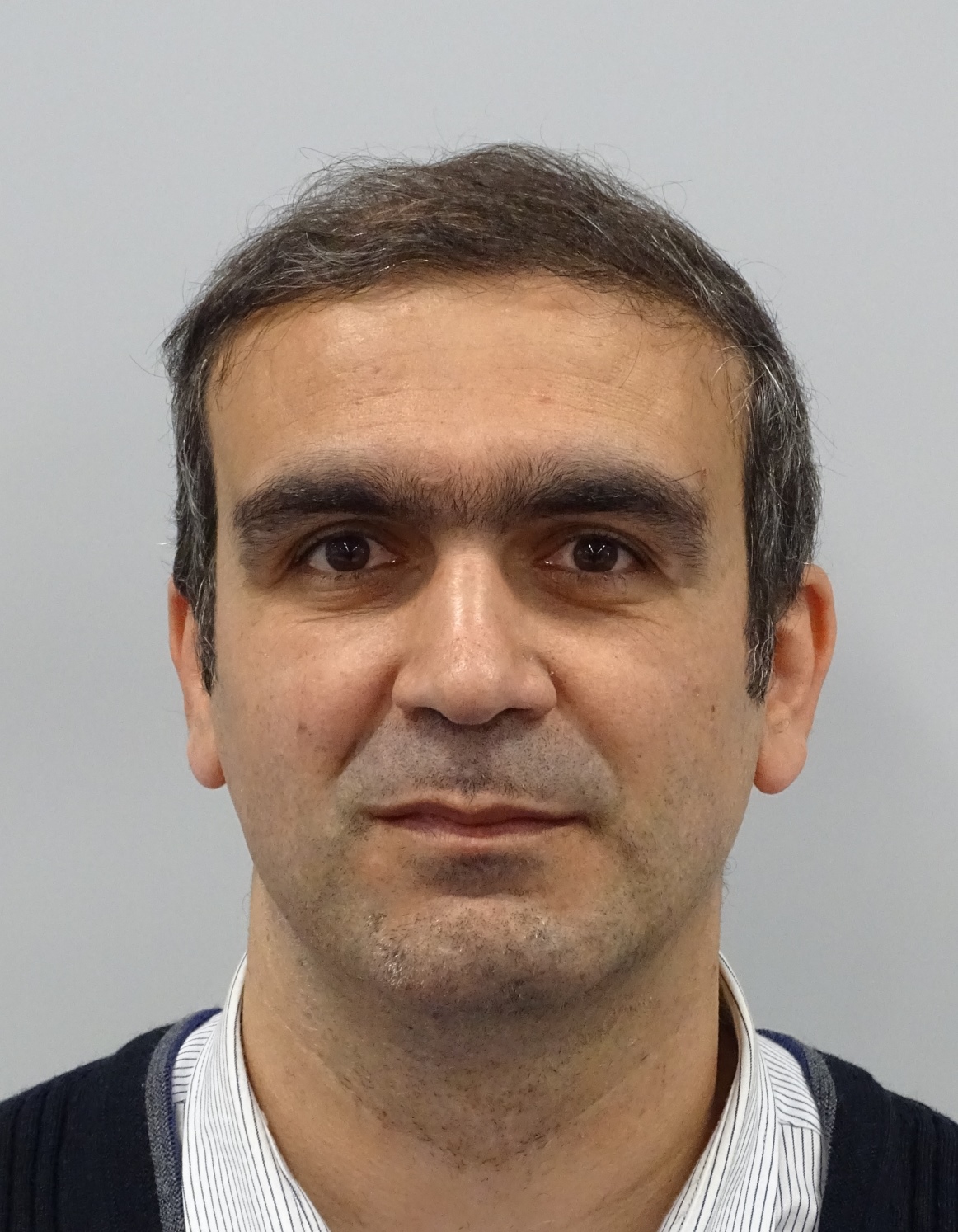}}]
{Mohammadali Mohammadi} (Senior Member, IEEE) is currently a Lecturer at the Centre for Wireless Innovation (CWI), Queen’s University Belfast, U.K. He previously held the position of Research Fellow at CWI from 2021 to 2024. His research interests include signal processing for wireless communications, cell-free massive MIMO, wireless power transfer, OTFS modulation, reconfigurable intelligent surfaces, and full-duplex communication. He has published more than 70 research papers in accredited international peer reviewed journals and conferences in the area of wireless communication. He has co-authored two book chapters, ``Full-Duplex Non-orthogonal Multiple Access Systems," invited chapter in Full-Duplex Communication for Future Networks (Springer-Verlag, 2020) and ``Full-Duplex wireless-powered communications", invited chapter in Wireless Information and Power Transfer: A New Green Communications Paradigm (Springer-Verlag, 2017). He was a recipient of the Exemplary Reviewer Award for IEEE Transactions on Communications in 2020 and 2022, and IEEE Communications Letters in 2023. He has been a member of Technical Program Committees for many IEEE conferences, such as ICC, GLOBECOM, and VTC.
\end{IEEEbiography}

\begin{IEEEbiography}[{\includegraphics[width=1in,height=1.25in,clip,keepaspectratio]{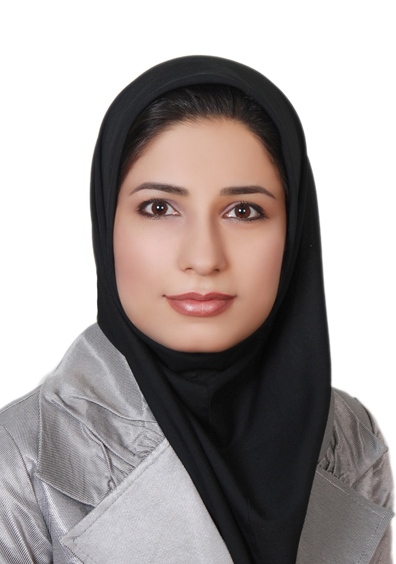}}]{Zahra Mobini}
 received the B.S. degree in electrical engineering from Isfahan University of Technology, Isfahan, Iran, in 2006, and the M.S and Ph.D.
degrees, both in electrical engineering, from the M. A. University of Technology and K. N. Toosi University of Technology, Tehran, Iran, respectively. From November 2010 to November 2011, she was a Visiting Researcher at the Research School of Engineering, Australian National University, Canberra, ACT, Australia. She is currently  a Post-Doctoral Research Fellow at the Centre for Wireless Innovation (CWI), Queen's University Belfast (QUB). Before joining QUB,  she was an Assistant and then Associate Professor with the Faculty of Engineering, Shahrekord University, Shahrekord, Iran (2015-2021). 
Her research interests include physical-layer security, massive  MIMO, cell-free massive  MIMO, full-duplex communications, and resource management and optimization. She has co-authored many research papers in wireless communications. She has actively served as the reviewer for a variety of IEEE journals,  such as TWC, TCOM, and TVT.

\end{IEEEbiography}

\begin{IEEEbiography}[{\includegraphics[width=1in,height=1.25in,clip,keepaspectratio]{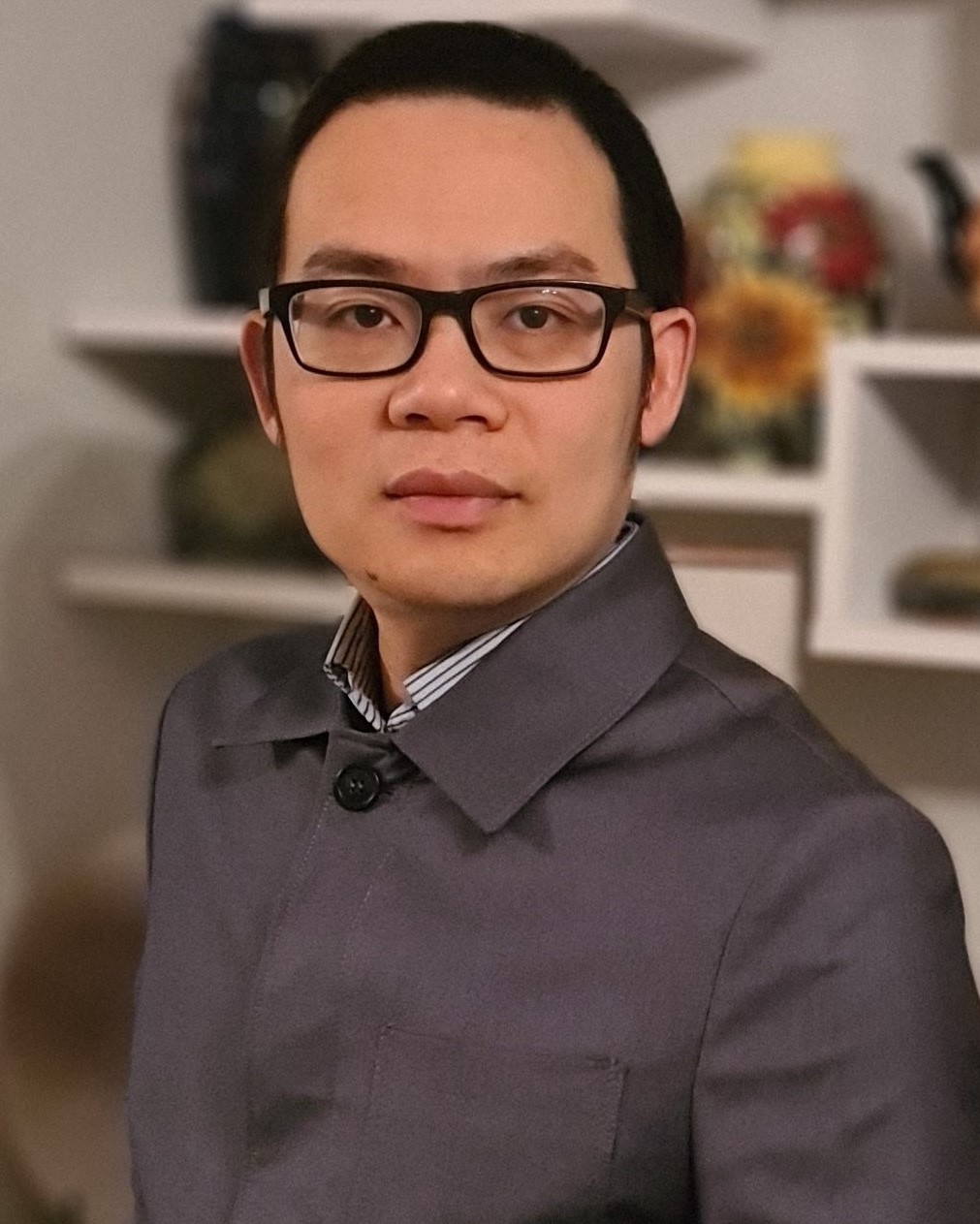}}]
{Hien Quoc Ngo} is currently a Reader with Queen's University Belfast, U.K. His main research interests include massive MIMO systems, cell-free massive MIMO, reconfigurable intelligent surfaces, physical layer security, and cooperative communications. He has co-authored many research papers in wireless communications and co-authored the Cambridge University Press textbook \emph{Fundamentals of Massive MIMO} (2016).

He received the IEEE ComSoc Stephen O. Rice Prize in 2015, the IEEE ComSoc Leonard G. Abraham Prize in 2017, the Best Ph.D. Award from EURASIP in 2018, and the IEEE CTTC Early Achievement Award in 2023. He also received the IEEE Sweden VT-COM-IT Joint Chapter Best Student Journal Paper Award in 2015. He was awarded the UKRI Future Leaders Fellowship in 2019. He serves as the Editor for the IEEE Transactions on Wireless Communications, IEEE Transactions on Communications, the Digital Signal Processing, and the Physical Communication (Elsevier). He was an editor of the IEEE Wireless Communications Letters, a Guest Editor of IET Communications, and a Guest Editor of IEEE ACCESS in 2017.
\end{IEEEbiography}

\begin{IEEEbiography}[{\includegraphics[width=1in,height=1.35in,clip,keepaspectratio]{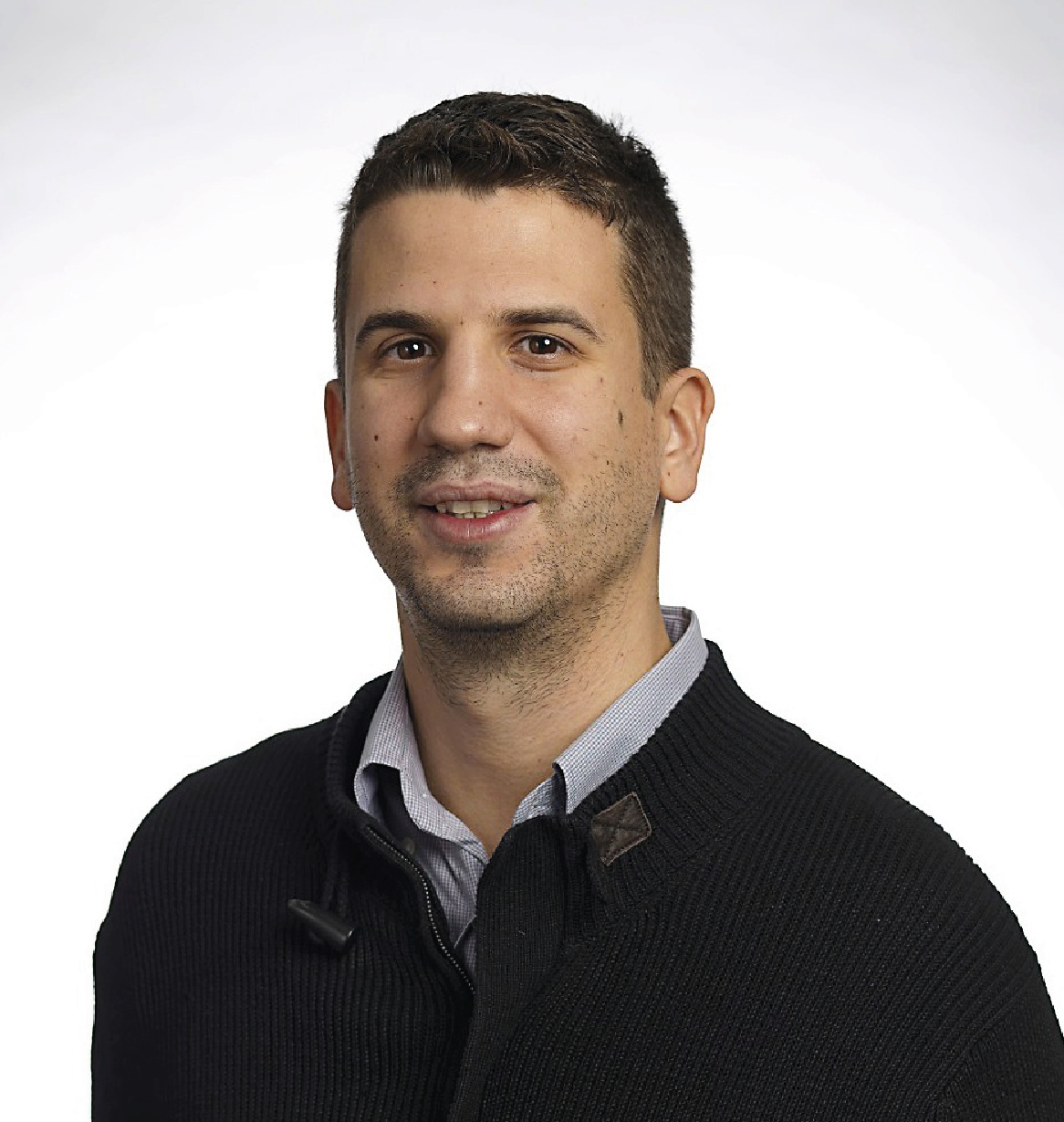}}]
{Michail Matthaiou}(Fellow, IEEE) obtained his Ph.D. degree from the University of Edinburgh, U.K. in 2008. He is currently a Professor of Communications Engineering and Signal Processing and Deputy Director of the Centre for Wireless Innovation (CWI) at Queen’s University Belfast, U.K. He has also held research/faculty positions at Munich University of Technology (TUM), Germany and Chalmers University of Technology, Sweden.
His research interests span signal processing for wireless communications, beyond massive MIMO, intelligent reflecting surfaces, mm-wave/THz systems and deep learning for communications.

Dr. Matthaiou and his coauthors received the IEEE Communications Society (ComSoc) Leonard G. Abraham Prize in 2017. He currently holds the ERC Consolidator Grant BEATRICE (2021-2026) focused on the interface between information and electromagnetic theories. To date, he has received the prestigious 2023 Argo Network Innovation Award, the 2019 EURASIP Early Career Award and the 2018/2019 Royal Academy of Engineering/The Leverhulme Trust Senior Research Fellowship. His team was also the Grand Winner of the 2019 Mobile World Congress Challenge. He was the recipient of the 2011 IEEE ComSoc Best Young Researcher Award for the Europe, Middle East and Africa Region and a co-recipient of the 2006 IEEE Communications Chapter Project Prize for the best M.Sc. dissertation in the area of communications. He has co-authored papers that received best paper awards at the 2018 IEEE WCSP and 2014 IEEE ICC. In 2014, he received the Research Fund for International Young Scientists from the National Natural Science Foundation of China. He is currently the Editor-in-Chief of Elsevier Physical Communication, a Senior Editor for \textsc{IEEE Wireless Communications Letters} and \textsc{IEEE Signal Processing Magazine}, and an Area Editor for \textsc{IEEE Transactions on Communications}. He is an IEEE and AAIA Fellow.
\end{IEEEbiography}

\end{document}